\documentclass{article}

\usepackage[preprint]{neurips_2023}

\usepackage{microtype}
\usepackage{graphicx}
\usepackage{subfigure}
\usepackage{booktabs} % for professional tables

\usepackage[hypertexnames=false, bookmarks=true, bookmarksnumbered=true]{hyperref}
\usepackage{amsmath}
\usepackage{amssymb}
\usepackage{mathtools}
\usepackage{amsthm}

\usepackage[utf8]{inputenc} % allow utf-8 input
\usepackage[T1]{fontenc}    % use 8-bit T1 fonts
\usepackage{url}            % simple URL typesetting
\usepackage{booktabs}       % professional-quality tables
\usepackage{amsfonts}       % blackboard math symbols
\usepackage{nicefrac}       % compact symbols for 1/2, etc.
\usepackage{microtype}      % microtypography
\usepackage{colortbl}
\usepackage{xcolor}         % colors

\usepackage{algorithm}
\usepackage{algorithmic}
\usepackage{amsmath}
\usepackage{amssymb}
\usepackage{mathtools}
\usepackage{amsthm}
\usepackage{multicol}

\usepackage{graphicx}
\usepackage{enumitem}
\usepackage{subfigure}
\usepackage{bm}
\usepackage{array}
\usepackage{indentfirst}

\usepackage{lipsum}		% Can be removed after putting your text content
\usepackage{natbib}
\usepackage{appendix}
\usepackage{import}
\usepackage{bbding}
\usepackage{threeparttable}
\usepackage{dsfont}
\usepackage{float}
\usepackage{tikz}

\usepackage{mathtools}

\usepackage[capitalize,noabbrev]{cleveref}
\hypersetup{
    colorlinks=true,
    linkcolor=red,
    urlcolor=green,
    citecolor=blue,
}
\usepackage{autonum}

\usepackage{listings}
\usepackage{xcolor}
\usepackage{wrapfig}

\newcommand{\ie}{{i.e.}}           % i.e.

\theoremstyle{plain}
\newtheorem{theorem}{Theorem}[section]

\newtheorem{lemma}[theorem]{Lemma}

\theoremstyle{definition}

\theoremstyle{remark}

\crefname{theorem}{Theorem}{Theorems}
\crefname{proposition}{Proposition}{Propositions}
\crefname{lemma}{Lemma}{Lemmas}
\crefname{corollary}{Corollary}{Corollaries}
\crefname{definition}{Definition}{Definitions}
\crefname{assumption}{Assumption}{Assumptions}
\crefname{remark}{Remark}{Remarks}

\crefname{equation}{equation}{equations}
\Crefname{equation}{Equation}{Equations}

\DeclareMathOperator*{\argmin}{arg\,min}
\newcommand\blfootnote[1]{%
  \begingroup
  \renewcommand\thefootnote{}\footnote{#1}%
  \addtocounter{footnote}{-1}%
  \endgroup
}

\title{Efficient Last-Iterate Convergence in Solving Games}

\author {
    \textbf{Linjian Meng}\textsuperscript{\rm 1{$\dagger$}},
    \textbf{Youzhi Zhang}\textsuperscript{\rm 2{$\dagger$}}, 
    Zhenxing Ge\textsuperscript{\rm 1},
    Shangdong Yang\textsuperscript{\rm 3},
    Tianyu Ding\textsuperscript{\rm 4}, \\
    \textbf{Wenbin Li}\textsuperscript{\rm 1},
    \textbf{Tianpei Yang}\textsuperscript{\rm 1},
    \textbf{Bo An}\textsuperscript{\rm 5},
    \textbf{Yang Gao}\textsuperscript{\rm 1}
    \\
    \textsuperscript{\rm 1} National Key Laboratory for Novel Software Technology, Nanjing University\\
    \textsuperscript{\rm 2} Centre for Artificial Intelligence and Robotics, Hong Kong Institute of Science \& Innovation, CAS\\
    \textsuperscript{\rm 3} Jiangsu Key Laboratory of Big Data Security and Intelligent Processing, Nanjing\\ University of Posts and Telecommunications\\
    \textsuperscript{\rm 4} Microsoft Corporation\\
    \textsuperscript{\rm 5} School of Computer Science and Engineering, Nanyang Technological University\\
    menglinjian@smail.nju.edu.cn, youzhi.zhang@cair-cas.org.hk, zhenxingge@smail.nju.edu.cn, \\
    sdyang@njupt.edu.cn, tianyuding@microsoft.com, liwenbin@nju.edu.cn, tianpei.yang@nju.edu.cn\\
    boan@ntu.edu.sg, gaoy@nju.edu.cn \\
}

\begin{document}

\maketitle

\begin{abstract}
To establish last-iterate convergence for Counterfactual Regret Minimization (CFR) algorithms in learning a Nash equilibrium (NE) of extensive-form games (EFGs), recent studies reformulate learning an NE of the original EFG as learning the NEs of a sequence of (perturbed) regularized EFGs. Consequently, proving last-iterate convergence in solving the original EFG reduces to proving last-iterate convergence in solving (perturbed) regularized EFGs. However, the empirical convergence rates of the algorithms in these studies are suboptimal, since they do not utilize Regret Matching (RM)-based CFR algorithms to solve perturbed EFGs, which are known the exceptionally fast empirical convergence rates. Additionally, since solving multiple perturbed regularized EFGs is required, fine-tuning across all such games is infeasible, making parameter-free algorithms highly desirable. In this paper, we prove that CFR$^+$, a classical parameter-free RM-based CFR algorithm, achieves last-iterate convergence in learning an NE of perturbed regularized EFGs. Leveraging CFR$^+$ to solve perturbed regularized EFGs, we get Reward Transformation CFR$^+$ (RTCFR$^+$). Importantly, we extend prior work on the parameter-free property of CFR$^+$, enhancing its stability, which is crucial for the empirical convergence of RTCFR$^+$. Experiments show that RTCFR$^+$ significantly outperforms existing algorithms with theoretical last-iterate convergence guarantees.
\end{abstract}

\blfootnote{$^\dagger$ Equal Contribution.}

\vspace{-0.45cm}
\section{Introduction}\label{sec:Introduction}
Extensive-form games (EFGs) are a foundational model for capturing interactions among multiple agents and sequential events. They are widely applied in simulating real-world scenarios, such as medical treatment~\citep{sandholm2015steering}, security games~\citep{lisy2016counterfactual}, and recreational games~\citep{ brown2019superhuman}. A common goal to address EFGs is to learn a Nash equilibrium (NE), where no player can unilaterally improve their payoff by deviating from the equilibrium.

To learn an NE in EFGs, recent research often employs regret minimization algorithms~\citep{zhao2022no-regret}. Among regret minimization algorithms, Counterfactual Regret Minimization (CFR) algorithms are the most widely used ones for learning an NE in real-world EFGs~\citep{bowling2015heads,moravvcik2017deepstack,brown2018superhuman,brown2019superhuman,perolat2022mastering}. They typically use Regret Matching (RM) algorithms~\citep{hart2000simple,gordon2006no,lanctot2009monte,johanson2012finding,lanctot2013monte,tammelin2014solving,brown2019solving,farina2021faster,zhang2022equilibrium,xu2022autocfr,farina2023regret,xu2024minimizing} as the local regularizer, since RM algorithms usually exhibit a faster empirical convergence rate than other local regret minimizers, such as Online Mirror Descent (OMD)~\citep{nemirovskij1983problem}. For convenience, we refer to the CFR algorithms that employ RM algorithms and OMD algorithms as local regularizers as RM-based CFR algorithms and OMD-based CFR algorithms, respectively.

However, most of regret minimization algorithms, including CFR algorithms, typically only achieve average-iterate convergence and their strategy profile may diverge or cycle, even in normal-form games (NFGs)~\citep{bailey2018multiplicative, mertikopoulos2018cycles}, a special form of EFGs where each player has only one information set (infoset). Average-iterate convergence implies that the averaging of strategies is necessary, which increases computational and memory overhead. Additionally, when strategies are parameterized via function approximation, a new approximation function must be trained to represent the average strategy, resulting in further representation errors. Consequently, algorithms with last-iterate convergence to NE, which ensures that the sequence of strategy profiles converges to the the set of NEs, are preferable as they avoid such averaging.

To establish last-iterate convergence for CFR algorithms, recent studies~\citep{perolat2021poincare,perolat2022mastering,liu2022power} employ the Reward Transformation (RT) framework, which (i) transforms the task of learning an NE of the original EFG into learning the NEs of a sequence of (perturbed) regularized EFGs and (ii) ensures the sequence of the NEs of these (perturbed) regularized EFGs converges to the set of NEs of the original EFG. Therefore, to ensure last-iterate convergence in learning an NE of the original EFG, it is sufficient to establish last-iterate convergence in learning an NE of (perturbed) regularized EFGs. Unfortunately, previous studies only establish last-iterate convergence in learning an NE of (perturbed) regularized EFGs for OMD-based CFR algorithms, incurring a poor empirical convergence rate to the set of NEs of the original EFG, as illustrated in our experiments.

To improve the empirical convergence rate, we propose Reward Transformation CFR$^+$ (RTCFR$^+$), utilizing CFR$^+$~\citep{tammelin2014solving}, a classical parameter-free RM-based CFR algorithm, to solve perturbed regularized EFGs. RTCFR$^+$ is inspired by two observations: (i) RM-based CFR algorithms usually outperform OMD-based CFR algorithms, and (ii) parameter-free algorithms, implying no parameters need to be tuned~\citep{grand2021conic}, are desirable to solve multiple perturbed regularized EFGs as fine-tuning across all perturbed regularized EFGs is infeasible. Notably, the parameter in CFR algorithms typically refers to the step sizes. Based on the RT framework, if CFR$^+$ has last-iterate convergence guarantee in learning an NE of perturbed regularized EFGs, then RTCFR$^+$ has last-iterate convergence guarantee in learning an NE of the original EFG. Unfortunately, it remains unknown whether CFR$^+$ achieves the parameter-free (i.e., holds for any step sizes) last-iterate convergence in learning an NE of perturbed regularized EFGs. It motivates a key question:
\begin{center}
\vspace{-0.25cm}
\textit{
Does CFR$^+$ have parameter-free last-iterate convergence guarantee \\ in learning an NE of perturbed regularized EFGs?
}
\end{center}
\vspace{-0.25cm}
To answer this question, we first provide the non-parameter-free (w.r.t. the step sizes) last-iterate convergence of CFR$^+$, \ie, for any initial accumulated counterfactual regrets, CFR$^+$ achieves last-iterate convergence in learning an NE of perturbed regularized EFGs when the step size exceeds a positive constant. We then extend this non-parameter-free result to establish the parameter-free result, \ie, CFR$^+$ achieves last-iterate convergence for any initial accumulated counterfactual regrets and step sizes. Note that our parameter-free result holds for any initial accumulated counterfactual regrets—not just the zero initialization in previous works~\citep{farina2021faster}—enhancing the stability of CFR$^+$~\citep{farina2023regret}, which is critical for the empirical convergence of RTCFR$^+$ in solving the original EFG. Without our parameter-free result, RTCFR$^+$ fails to converge to the set of NEs of original EFG! To the best of our knowledge, this is the first parameter-free last-iterate convergence guarantee for RM-based CFR algorithms in learning an NE of perturbed regularized EFGs. As a consequence, based on the convergences of the RT framework and CFR$^+$, RTCFR$^+$ achieves last-iterate convergence in learning an NE of the original EFG.

{Specifically, we propose novel techniques to overcome the challenges in the above two steps of the proof.} First, the primary challenge in proving the non-parameter-free result is that the smoothness of the instantaneous counterfactual regrets—the key property used in prior works~\citep{liu2022power} to establish the last-iterate convergence of CFR algorithms—cannot be leveraged, since RM algorithms update within the cone of the strategy space while the final output lies in the strategy space itself. To address this, we exploit the fact that an NE represents a best response to others at each infoset in perturbed EFGs. More specifically, this fact allows a term—related to the accumulated counterfactual regrets and the utility obtained by deviating from an NE of perturbed EFGs—can be added to enable smoothness of the instantaneous counterfactual regrets can be leveraged to ensure that the cumulative squared distance between the iterated strategy profiles and the NE of perturbed regularized EFGs remains bounded by a constant across all iterations, thereby guaranteeing last-iterate convergence.
{Second, the} main challenge of proving our parameter-free result is that the property used in prior proofs of the parameter-free property of CFR$^+$—the strategy sequence produced by CFR$^+$ remains invariant across different step sizes—holds only when the initial accumulated counterfactual regrets are zero~\citep{farina2021faster}. We address this by leveraging the linearity of the projection in CFR$^+$ alongside our non-parameter-free convergence result that holds for any initial accumulated counterfactual regrets. In particular, we use this linearity of projection to show that for any initial accumulated counterfactual regrets and step sizes, there exists an alternative choice of these parameters that yields an identical strategy profile sequence. By then applying our non-parameter-free result to this alternative setting, we establish that the resulting strategy profile sequence converges to the set of NEs of perturbed regularized EFGs, thus proving the parameter-free last-iterate convergence result.

Experimental evaluations across nine instances from five standard EFG benchmarks---Kuhn Poker, Leduc Poker, Goofspiel, Liar's Dice, and Battleship---demonstrate that RTCFR$^+$ significantly outperforms existing algorithms with theoretical last-iterate convergence guarantees.

% \vspace{-0.35cm}
\section{Preliminaries}\label{sec:Preliminaries}
\textbf{Extensive-form games (EFGs).} EFG is a common used model for modeling tree-form sequential decision-making problems. An EFG can be formulated as $G=\{\mathcal{N}, \mathcal{H}, P, A, \mathcal{I}, \{u_i\}\}$.  {Here,} $\mathcal{N}$ is the set of players. $\mathcal{H}$ is the set of all possible history sequences. The set of leaf nodes is denoted by $\mathcal{Z}$. For each history $h\in \mathcal{H}$, the function $P(h)$ represents the player {acting at} node $h${, and} $A(h)$ denotes the actions available at node $h$. To {account for} private information, the nodes for each player $i$ are partitioned into a collection $\mathcal{I}_i$, {referred to as} information sets\ (infosets). For any {infoset} $I \in \mathcal{I}_i$, histories $h,h'\in I$ are indistinguishable to player $i$. Thus, $P(I)=P(h)$, $A(I)=A(h), \forall h \in I$. The notation $\mathcal{I}$ denotes $\mathcal{I} = \{ \mathcal{I}_i | i \in \mathcal{N}\}$. We also use $C_i(I,a)$ to denote the set of infosets that belongs to $i$ and will counter after executing $a \in A(I)$ at infoset $I \in \mathcal{I}_i$. The notations $A_{max}$ and $C_{max}$ denote $\max_{I \in \mathcal{I}} |A(I)|$ and $ \max_{i \in \mathcal{N}, I \in \mathcal{I}_i, a \in A(I)} C_i(I,a)$, respectively. For each leaf node $z$, there is a pair $(u_0(z), u_1(z)) \in [-1,\ 1]$ which denotes the payoffs for the min player\ (player 0) and the max player\ (player 1), respectively. We define $H$ as the maximum number of actions taken by all players along any path from the root to a leaf node. In two-player zero-sum EFGs, $u_0(z) = -u_1(z), \forall z \in \mathcal{Z}$.

\textbf{Sequence-form strategy.} A sequence is an {infoset}-action pair $(I,a)$, where $I \in \mathcal{I}$ is an {infoset} and $a$ is an action belonging to $A(I)$. Each sequence identifies a path from the root node to the {infoset} $I${, selecting the action $a$ along this path}. The set of sequences for player $i$ is denoted by $\Sigma_i$. The last sequence encountered on the path from {the root node} $r$ to $I$ is denoted by $\rho_I$ ($\rho_I \in \Sigma_i$). In other words, $\forall i \in \mathcal{N}, I \in \mathcal{I}_i$, $I \in C_i(\rho_I)$. {A sequence}-form strategy for player $i$ is a non-negative vector $\bm{x}_i$ indexed over the set of sequences $\Sigma_i$. For each sequence $q=(I,a) \in \Sigma_i$, $\bm{x}_i(q)$ is the probability that player $i$ reaches the sequence $q$ {when following} the strategy $\bm{x}_i$. We formulate the sequence-form strategy space as a treeplex~\citep{hoda2010smoothing}. Let $\bm{\mathcal{X}}_i$ denote the set of sequence-form strategies for player $i$. We use $\bm{x}_i(I)=[\bm{x}_i(I,a)|a \in A(I)]$ to denote the slice of a given strategy $\bm{x}_i$ corresponding to sequences belonging to {infoset} $I$, where $\bm{x}_i(I,a)$ is value of $\bm{x}_i$ at the sequence $(I,a)$. We assume that $\forall i \in \mathcal{N}$ and $\bm{x}_i \in \bm{\mathcal{X}}_i$, $\Vert \bm{x}_i \Vert_1\leq D$.

\textbf{Nash equilibrium (NE).} NE describes a rational behavior where no player can benefit by unilaterally deviating from the equilibrium. For any player, her strategy is the best-response to the strategies of others. From the sequence-form strategy framework, learning an NE of EFGs is represented by
\begin{equation}\label{eq:BSPP}
% \small
% \setlength\abovedisplayskip{2pt}
% \setlength\belowdisplayskip{2pt}
    \min_{{\bm{x}}_0 \in \bm{\mathcal{X}}_0}\max_{{\bm{x}}_1 \in \bm{\mathcal{X}}_1} {\bm{x}}_0^\text{T} \bm{A} {\bm{x}}_1,
\end{equation}
where $\bm{A}$ is the payoff matrix. We use $\small \bm{\mathcal{X}}$ and $\bm{\mathcal{X}}^{*}$ to denote $\small \times_{i \in \mathcal{N}} \bm{\mathcal{X}}_i$ and the set of NEs, respectively.

\textbf{Behavioral strategy.} While sequence-form strategy suffices to depict the strategy within EFGs, to illustrate the counterfactual regret minimization framework~\citep{zinkevich2007regret,farina2019online}, behavioral strategy is necessary. This strategy $\sigma_i$ is defined on each inforset. For any inforset $I \in \mathcal{I}_i$, the probability for the action $a \in A(I)$ is denoted by $\sigma_i(I,a)$. We use $\sigma_i(I) = [\sigma_i(I,a)|a \in A(I)] \in \Delta^{|A(I)|}$ to denote the strategy at inforset $I$, where $\Delta^{|A(I)|}$ is a $(|A(I)|-1)$-dimension simplex. If all players follow the strategy profile $\sigma=\{ \sigma_0, \sigma_1\}$ and reaches inforset $I$, the reaching probability is denoted by $\pi^{{\sigma}}(I)$. The contribution of $i$ to this probability is $\pi_{i}^{\sigma}(I)$ and $\pi_{-i}^{\sigma}(I)$ for other than $i$, where $-i$ denotes the players other than $i$. As analyzed in \citet{von1996efficient}, $\forall i \in \mathcal{N}, I \in \mathcal{I}_i, a \in A(I), \bm{x}_i \in \bm{\mathcal{X}}_i$, $\bm{x}_i(I,a) = \pi^{\sigma}_i(I) \sigma_i(I,a)$.

\textbf{Perturbed extensive-form games (Perturbed EFGs).} This game is a variant of the original EFG. Specifically, the strategy space of each infoset $I \in \mathcal{I}$ in a $\gamma$-perturbed EFG is a $\gamma$-perturbed simplex $\Delta^{|A(I)|}_{\gamma}$, rather than the standard simplex $\Delta^{|A(I)|}$ used in the original EFG, where $\gamma > 0$ is a constant. Formally, for any $\hat{\sigma}_{i}(I) \in  \Delta^{|A(I)|}_{\gamma}$ and $a \in A(I)$, the constraint $\gamma \leq \hat{\sigma}_i(I,a) \leq 1$ holds, where $i = P(I)$. For convenience, we denote the set of sequence-form strategies for player $i$ in the $\gamma$-perturbed EFGs as $\bm{\mathcal{X}}^{\gamma}_i$. In $\gamma$-perturbed EFGs with $\gamma > 0$, any behavioral strategy $\hat{\sigma}_{i}$, with $\hat{\sigma}_{i}(I) \in  \Delta^{|A(I)|}_{\gamma}$ for all $i \in \mathcal{N}$ and $I \in \mathcal{I}_i$, can be uniquely mapped to a sequence-form strategy $\hat{\bm{x}}_i \in \bm{\mathcal{X}}^{\gamma}_i$, and vice versa. Specifically, $\forall i \in \mathcal{N}, I \in \mathcal{I}_i$, $\hat{\sigma}_i(I) = \hat{\bm{x}}_i(I)/\hat{\bm{x}}_i(\rho_I) \geq \gamma$. Similarly, we use the notation $\bm{\mathcal{X}}^{\gamma}$ and $\bm{\mathcal{X}}^{*,\gamma}$ to denote the joint strategy space $\times_{i \in \mathcal{N}} \bm{\mathcal{X}}^{\gamma}_i$ and the set of NEs of $\gamma$-perturbed EFGs, respectively.

\textbf{Learning an NE via regret minimization algorithms.} 
For any sequence of strategies $\bm{x}_i^1, \cdots, \bm{x}_i^T$ of player $i$, player $i$'s regret is $R^{T}_i = \max_{\bm{x}_i \in \bm{\mathcal{X}}_i}\sum^{T}_{t=1}  \langle \bm{\ell}^{t}_i, \bm{x}^{t}_i - \bm{x}_i\rangle$, where $\bm{\ell}^{t}_i$ is the loss for player $i$ at iteration $t$. Regret minimization algorithms are algorithms ensuring $R^{T}_i$ grows sublinearly. To learn an NE of EFGs via regret minimization algorithms, we set $\bm{\ell}^{t}_i = \bm{\ell}^{\bm{x}^t}_i$ with $\bm{\ell}^{\bm{x}}_0 = \bm{A}\bm{x}_1$ and $\bm{\ell}^{\bm{x}}_1 = -\bm{A}^{\text{T}} \bm{x}_0$. If all players follow regret minimization algorithms, then the average strategy converges to the set of NEs in two-player zero-sum EFGs. In this paper, we assume that, $\forall \bm{x}, \bm{x}^{\prime} \in \bm{\mathcal{X}}$, $
% \begin{equation}\label{eq:L and P}
% \small
% \setlength\abovedisplayskip{0pt}
% \setlength\belowdisplayskip{0pt}
    \Vert \bm{\ell}^{\bm{x}} - \bm{\ell}^{\bm{x}^{\prime}}\Vert_1 \leq L\Vert \bm{x} - \bm{x}^{\prime} \Vert_1\ \text{and}\ \Vert \bm{\ell}^{\bm{x}} \Vert_1 \leq P$,
% \end{equation}
where $\bm{\ell}^{\bm{x}} = [\bm{\ell}^{\bm{x}}_i | i \in \mathcal{N}]$, as well as $L > 0$ and $P >0$ are game-dependent constants.

\textbf{Counterfactual regret minimization (CFR) framework.} This framework~\citep{zinkevich2007regret, farina2019online} is designed to solve EFGs by decomposing the global regret $R^{T}_i$ into local regrets at each infoset, allowing for independent minimization within each infoset, rather than directly minimizing global regret. This approach has led to the development of several superhuman Game AIs~\citep{bowling2015heads,moravvcik2017deepstack,brown2018superhuman,brown2019superhuman,perolat2022mastering}. Formally, for player $i$, given the observed loss when all players follow $\bm{x} \in \bm{\mathcal{X}}_i$ is $\bm{\ell}^{\bm{x}}_i$, the CFR framework computes the counterfactual values at each infoset $I \in \mathcal{I}_i$ according to
\begin{equation}\label{eq: counterfactual values}
\small
\setlength\abovedisplayskip{0pt}
\setlength\belowdisplayskip{0pt}
    \bm{v}^{\bm{x}}_i(I,a) = -\bm{\ell}_i^{\bm{x}}(I,a) + \sum_{I^{\prime} \in C_i(I,a)} \langle \bm{v}^{\bm{x}}_i(I^{\prime}), \sigma_i(I^{\prime}) \rangle
\end{equation}
where $\bm{\ell}_i^{\bm{x}}(I,a)$ is the value of $\bm{\ell}_i^{\bm{x}}$ at the sequence $(I,a)$, $\bm{v}^{\bm{x}}_i(I^{\prime}) = [\bm{v}^{\bm{x}}_i(I^{\prime},a^{\prime})|a^{\prime} \in A(I^{\prime})]$, and $\sigma_i$ represents the behavioral strategy of player $i$ corresponds to $\bm{x}_i$. \citet{farina2019online} demonstrate that
\begin{equation}\label{eq:regret is upper bounded by counterfactual regret}
\small
\setlength\abovedisplayskip{0pt}
\setlength\belowdisplayskip{0pt}
\begin{aligned}
    R^{T}_i = & \max_{\bm{x}_i \in \bm{\mathcal{X}}_i} \sum^{T}_{t=1} \langle \bm{\ell}^{t}_i, \bm{x}^{t}_i - \bm{x}_i \rangle
    \leq  \sum_{I \in \mathcal{I}_i} \max_{\sigma_i(I)} \sum^{T}_{t=1} \langle \bm{v}^{t}_i(I), \sigma_i(I) - \sigma^t_i(I) \rangle,
\end{aligned}
\end{equation}
where $\bm{v}^{t}_i(I) = \bm{v}^{\bm{x}^t}_i(I)$ and $\sigma^t_i$ is the behavioral strategy of player $i$ corresponds to $\bm{x}^t_i$. It indicates that minimizing the local regret $\max_{\sigma_i(I)} \sum^{T}_{t=1} \langle \bm{v}^{t}_i(I), \sigma_i(I) - \sigma^t_i(I) \rangle$ at $I \in \mathcal{I}_i$ contributes to minimizing the global regret $R^{T}_i$.

\textbf{Blackwell approachability framework.} RM algorithms are come from this framework whose core insight lies in reframing the problem of regret minimization within $\bm{\mathcal{Z}}$ as regret minimization within $\text{cone}(\bm{\mathcal{Z}})$~\citep{abernethy2011blackwell,grand2021conic,chakrabarti2024extensive}. Specifically, a regret minimization algorithm is instantiated in $\text{cone}(\bm{\mathcal{Z}})$, where its output at iteration $t$ is $\bm{\theta}^t_i$. This corresponds to the strategy $\bm{x}^t_i = \bm{\theta}^t_i/\langle \bm{\theta}^t_i, \bm{1} \rangle$ within $\bm{\mathcal{Z}}$. Given the loss $\bm{\ell}^t_i$ at iteration $t$, the algorithm observes the transformed loss $\bm{m}^t_i = \langle \bm{\ell}^t_i, \bm{x}^t_i \rangle \bm{1} - \bm{\ell}^t_i$ and subsequently generates $\bm{\theta}^{t+1}_i$. The main advantage of this framework is its capacity to develop parameter-free algorithms. More details are provided below.

\textbf{Regret Matching (RM).} 
To minimize local regret within each infoset, CFR algorithms commonly employ local regret minimizers based on RM~\citep{hart2000simple,gordon2006no,bowling2015heads,farina2021faster,farina2023regret,xu2022autocfr,xu2024minimizing,cai2023lastiterate}, which show strong empirical convergence rate and are typically parameter-free. RM is a traditional algorithm grounded in Blackwell approachability framework. It corresponds to an instantiation of FTRL~\citep{shalev2007primal}, a classical regret minimization algorithm, in the cone of the simplex, $\mathbb{R}^{d}_{\geq 0} = \{\bm{y} \in \mathbb{R}^{A(I)} \mid \bm{y} \geq \bm{0}\}$~\citep{farina2021faster}. At iteration $t$ and infoset $I \in \mathcal{I}_i$, the update rule of RM is
\begin{equation}\label{eq:update-rule-RM-GF}
\small
\begin{aligned}
     & {\bm{\theta}}^{t+1}_I \in \argmin_{{\bm{\theta}}_I \in \mathbb{R}^{|A(I)|}_{\geq 0}} \left\{  \langle -\sum_{\tau=1}^t \bm{m}^{t}_i(I), \bm{\theta}_I\rangle + \frac{1}{\eta} \psi (\bm{\theta}_I) \right\},\ \
     &\sigma^{t+1}_i(I) = \frac{{\bm{\theta}}^{t+1}_I}{\langle {\bm{\theta}}^{t+1}_I, \bm{1} \rangle},
\end{aligned}
\end{equation}
where $i = P(I)$, $\eta > 0$ is the step size, $\bm{m}^{t}_i(I) = -\langle \bm{v}_i^{t}(I), \sigma^{t}_i(I)\rangle \bm{1} + \bm{v}_i^{t}(I)$ represents the instantaneous counterfactual regret, ${\bm{\theta}}^{t}_I$ is the accumulated counterfactual regret, and $\psi(\cdot) = \Vert \cdot \Vert^2_2/2$ is the quadratic regularizer. When ${\bm{\theta}}^{1}_I = 0$, $\forall \eta > 0$, the sequence $\{ \sigma^{1}_i(I), \sigma^{2}_i(I), \dots, \sigma^{t}_i(I), \dots\}$ remains unchanged~\citep{farina2021faster}. Thus, RM is parameter-free. When integrated with the CFR framework, yields vanilla CFR~\citep{zinkevich2007regret}, a parameter-free CFR algorithm.

\textbf{Regret Matching$^+$ (RM$^+$).} 
\citet{tammelin2014solving} proposes RM$^+$, a variant of RM that typically exhibits a faster empirical convergence rate than RM. RM$^+$ is also a algorithm grounded in Blackwell approachability framework. It corresponds to an OMD instantiated in the cone of the simplex~\citep{farina2021faster}. Formally, at each iteration $t$ and infoset $I \in \mathcal{I}_i$, RM$^+$ updates the strategy via
\begin{equation}\label{eq:update-rule-RM+-GF}
\small
\begin{aligned}
     & {\bm{\theta}}^{t+1}_I \in \argmin_{{\bm{\theta}}_I \in \mathbb{R}^{|A(I)|}_{\geq 0}} \left\{  \langle - \bm{m}^{t}_i(I), \bm{\theta}_I\rangle + \frac{1}{\eta} D_{\psi} (\bm{\theta}_I, {\bm{\theta}}^{t}_I) \right\},\ \
     &\sigma^{t+1}_i(I) = \frac{{\bm{\theta}}^{t+1}_I}{\langle {\bm{\theta}}^{t+1}_I, \bm{1} \rangle},
\end{aligned}
\end{equation}
where $i = P(I)$, $\eta > 0$ is the step size, $\bm{m}^{t}_i(I)$ is as defined in (\ref{eq:update-rule-RM-GF}), and $D_{\psi}(\bm{u}, \bm{v}) = \psi(\bm{u}) - \psi(\bm{v}) - \langle \nabla\psi(\bm{v}), \bm{u} - \bm{v} \rangle$ is the Bregman divergence associated with $\psi(\cdot)$. As with RM, if ${\bm{\theta}}^{1}_I = 0$, for all the step size $\eta > 0$, the output sequence $\{ \sigma^{1}_i(I), \sigma^{2}_i(I), \dots, \sigma^{t}_i(I), \dots\}$ remains unchanged~\citep{farina2021faster}. Combining RM$^+$ with the CFR framework yields CFR$^+$~\citep{tammelin2014solving}, which is also a parameter-free CFR algorithm.

\section{Problem Statement}\label{sec:Problem Statement}
% \vspace{-0.25cm}

To establish the last-iterate convergence for CFR algorithms, \citet{perolat2021poincare,perolat2022mastering,liu2022power} use the RT framework, transforming the task of learning an NE of the original EFG into finding the NEs of a sequence of (perturbed) regularized EFGs and ensuring the sequence of the NEs of these (perturbed) regularized EFGs converges to the set of NEs of the original EFG. Therefore, establishing last-iterate convergence in learning an NE the original EFG reduces to establishing last-iterate convergence in learning an NE of (perturbed) regularized EFGs. In this paper, we consider the following perturbed regularized EFG:
\begin{equation}\label{eq:BSPP-perturbed regularized}
% \small
% \setlength\abovedisplayskip{2pt}
% \setlength\belowdisplayskip{2pt}
\begin{aligned}
    \min_{{\hat{\bm{x}}}_0 \in \bm{\mathcal{X}}^{\gamma}_0} \max_{{\hat{\bm{x}}}_1 \in \bm{\mathcal{X}}^{\gamma}_1} {\hat{\bm{x}}}_0^{\text{T}} \bm{A} {\hat{\bm{x}}}_1 + \mu D_{\psi}({\hat{\bm{x}}}_0, \bm{r}_0) - \mu D_{\psi}({\hat{\bm{x}}}_1, \bm{r}_1),
\end{aligned}
\end{equation}
where $\gamma > 0$ and $\mu > 0$ are constants, and $\bm{r} = [\bm{r}_0;\bm{r}_1] \in \bm{\mathcal{X}}$ is the reference strategy profile. The NE of this perturbed regularized EFG is unique and denoted by $\hat{\bm{x}}^{*,\gamma,\mu,\bm{r}}$ or $\hat{\sigma}^{*,\gamma,\mu,\bm{r}}$. By continuously decreasing the value of $\gamma$ and updating $\bm{r}$ to $\hat{\bm{x}}^{*,\gamma,\mu,\bm{r}}$, from the studies by \citet{abe2023slingshot,bernasconi2024learning}, the sequence of the NEs of the perturbed regularized EFGs converges to the set of NEs of the original EFG. Consequently, achieving the last-iterate convergence for solving (\ref{eq:BSPP-perturbed regularized}) implies achieving the last-iterate convergence for solving (\ref{eq:BSPP}). This paper refrain from investigating the RT framework and its convergence as these have been thoroughly investigated in other studies~\citep{perolat2021poincare,liu2022power,abe2023slingshot,bernasconi2024learning,wang2025magnetic}. 

% \footnote{While intuitively the sequence of the NEs of the perturbed regularized EFGs converges to the set of the extensive-form perfect equilibrium of the original EFG, \citet{bernasconi2024learning} show that it may only converge to the set of NEs of the original EFG (the text around the Proposition 4.1 of \citet{bernasconi2024learning})}

The introduction of perturbation and regularization ensures the smoothness of counterfactual values and the strongly monotonicity, respectively. The smoothness is $
% \begin{equation}
\small
% \setlength\abovedisplayskip{2pt}
% \setlength\belowdisplayskip{2pt}
%     \begin{aligned}
        \Vert \bm{v}^{{\hat{\sigma}}}_i(I) - \bm{v}^{{\hat{\sigma}}^{\prime}}_i(I) \Vert_1 \leq O(\Vert {\hat{\bm{x}}}- {\hat{\bm{x}}}^{\prime} \Vert_1),\ \forall {\hat{\bm{x}}}, {\hat{\bm{x}}}^{\prime} \in \bm{\mathcal{X}}^{\gamma},\
%     \end{aligned}
% \end{equation}
$where ${\hat{\sigma}}$ and ${\hat{\sigma}}^{\prime}$ are the behavioral strategy profiles associated with ${\hat{\bm{x}}}$ and $ {\hat{\bm{x}}}^{\prime}$, respectively. The strongly monotonicity indicates that $
% \begin{equation}
\small
% \setlength\abovedisplayskip{2pt}
% \setlength\belowdisplayskip{2pt}
%     \begin{aligned}
        O(\langle \bm{\ell}^{{\hat{\bm{x}}}} - \bm{\ell}^{\bm{x}^{\prime}}, {\hat{\bm{x}}} - {\hat{\bm{x}}}^{\prime} \rangle )\geq \Vert {\hat{\bm{x}}} - {\hat{\bm{x}}}^{\prime} \Vert_2^2,\ \forall {\hat{\bm{x}}}, {\hat{\bm{x}}}^{\prime} \in \bm{\mathcal{X}}^{\gamma}.
%     \end{aligned}
% \end{equation}
$

Although some works have investigated the last-iterate convergence of CFR algorithms for solving perturbed regularized EFGs~\citep{liu2022power}, their CFR algorithms do not use RM-based algorithms as the local regret minimizer. The absence of RM-based algorithms leads to significantly weaker empirical performance in terms of last-iterate convergence than traditional RM-based average-iterate convergence CFR algorithms, as illustrated in our experiments. In addition, as solving multiple perturbed regularized EFGs is required, fine-tuning across all perturbed regularized EFGs is computationally infeasible. Consequently, parameter-free algorithms, implying no parameters need to be tuned~\citep{grand2021conic}, are desirable. Based on these observations, we propose Reward Transformation CFR$^+$ (RTCFR$^+$), utilizing CFR+ ~\citep{tammelin2014solving}, a classical parameter-free RM-based CFR algorithm, to solve perturbed regularized EFGs defined in (\ref{eq:BSPP-perturbed regularized}) (details of RTCFR$^+$ are in \Cref{sec:Our Method}). Unfortunately, it remains unknown whether CFR$^+$ achieves the parameter-free (i.e., holds for any step sizes) last-iterate convergence in solving (\ref{eq:BSPP-perturbed regularized}). Therefore, our objective is to demonstrate the parameter-free last-iterate convergence for CFR$^+$ in solving (\ref{eq:BSPP-perturbed regularized}). More discussions about the related works are in \Cref{sec:Related Work}.

% \vspace{-0.25cm}
\section{Last-Iterate Convergence of CFR$^+$ in Solving Perturbed Regularized EFGs}\label{sec:Our Method}
Now, we show that CFR$^+$ exhibits last-iterate convergence for solving the perturbed regularized EFGs defined in (\ref{eq:BSPP-perturbed regularized}). Before introducing the last-iterate convergence of CFR$^+$, we first extend CFR$^+$ to perturbed EFGs as the original CFR$^+$ algorithm is only designed for the case where $\gamma=0$. Specifically, we (i) first update the accumulated counterfactual regrets within the original simplex’s cone while ensuring strategy outputs lie within the perturbed simplex by mixing the non-perturbed strategy formed by the accumulated counterfactual regrets with the uniform vector, then (ii) compute the instantaneous counterfactual regrets using the non-perturbed strategy and the counterfactual values observed through following the output perturbed strategy. This enables the use of the strong monotonicity to establish last-iterate convergence in learning an NE of the perturbed regularized EFGs in (\ref{eq:BSPP-perturbed regularized}), as shown in (\ref{eq:Sec-4-3}).
Formally, the update rule of CFR$^+$ for learning an NE of the perturbed regularized EFGs in (\ref{eq:BSPP-perturbed regularized}) at iteration $t$ and infoset $I \in \mathcal{I}_i$ is
\begin{equation}
\small
\setlength\abovedisplayskip{3pt}
\setlength\belowdisplayskip{3pt}
% \thinmuskip=0mu
% \medmuskip=0mu
% \thickmuskip=0mu
% \spaceskip=0pt
\begin{aligned}
     & {\bm{\theta}}^{t+1}_I \in \argmin_{{\bm{\theta}}_I \in \mathbb{R}^{|A(I)|}_{\geq 0}} \left\{  \langle - \hat{\bm{m}}^{t}_i(I), \bm{\theta}_I\rangle + \frac{1}{\eta} D_{\psi} (\bm{\theta}_I, {\bm{\theta}}^{t}_I) \right\},\ \sigma^{t+1}_i(I) = \frac{{\bm{\theta}}^{t+1}_I}{\langle {\bm{\theta}}^{t+1}_I, \bm{1} \rangle},\\
    & \hat{\sigma}^{t+1}_i(I) = (1 - \alpha_I)\sigma^{t+1}_i(I) + \gamma \bm{1}, \ \alpha_{I} = \gamma|A(I)|,\\
    & \hat{\bm{m}}^{t}_i(I) = \hat{\bm{v}}^{t}_i(I) - \langle \hat{\bm{v}}^{t}_i(I), \sigma^{t}_i(I)\rangle \bm{1}, \\
    & \hat{\bm{v}}^{t}_i(I,a) = -\hat{\bm{\ell}}_i^{t}(I,a) + \sum_{I^{\prime} \in C_i(I,a)} \langle \hat{\bm{v}}^{t}_i(I^{\prime}) , \hat{\sigma}^{t}_i(I^{\prime}) \rangle, \\
    & \hat{\bm{\ell}}_0^{t} = \bm{A} \hat{\bm{x}}^{t}_1 + \mu \nabla \psi(\hat{\bm{x}}^{t}_0) - \mu \nabla \psi({\bm{r}}_0),\ \hat{\bm{\ell}}_1^{t} = - \bm{A}^{\text{T}} \hat{\bm{x}}^{t}_0 + \mu \nabla \psi(\hat{\bm{x}}^{t}_1) - \mu \nabla \psi({\bm{r}}_1),\\
\end{aligned}
\label{eq:update-rule-MPCFR}
\end{equation}
where $\eta > 0$ is the step size and $\hat{\bm{x}}^{t}_i(I) = {\pi}^{\hat{\sigma}^{t}}_i(I) \hat{\sigma}^{t}_i(I)$. The second line in (\ref{eq:update-rule-MPCFR}) is related to mixing the non-perturbed strategy $\sigma^{}$ with the uniform vector $\bm{1}$, and the third line is related to constructing the instantaneous counterfactual regrets $\hat{\bm{m}}^{t}_I$ using non-perturbed strategy $\sigma^{t}_i$ derived from accumulated counterfactual regrets $\bm{\theta}^{t}_I$ and counterfactual values $\hat{\bm{v}}^{t}_i$ observed through following the perturbed strategy $\hat{\sigma}^{t}_i$.

\begin{theorem}
[Proof is in \Cref{sec:prf:thm:convergence results of our algorithm}]
Assuming all players follow the update rule of CFR$^+$ with any $\bm{\theta}^{1}_I \in \mathbb{R}^{|A(I)|}_{\geq 0}$ and $\eta > 0$, the strategy profile $\hat{\bm{x}}^{t}$ converges to the set of NEs of the perturbed regularized EFGs defined in (\ref{eq:BSPP-perturbed regularized}) with any $\gamma > 0$ and $\mu >0$. 
\label{thm:convergence results of our algorithm}
\end{theorem}

\textbf{Proof sketch of \Cref{thm:convergence results of our algorithm}.}
Our proof consists of two steps. Firstly, we establish the non-parameter-free last-iterate convergence; that is, for all $\bm{\theta}^{1}_I \in \mathbb{R}^{|A(I)|}_{\geq 0}$, the last-iterate convergence of CFR$^+$ in solving (\ref{eq:BSPP-perturbed regularized}) holds when $\eta$ exceeds a certain constant. The principal challenge is that the smoothness of the instantaneous counterfactual regrets cannot be used since RM algorithms update within the cone of the strategy space, $\text{cone}(\Delta^{A(I)})$, whereas the final output lies in the strategy space, $\Delta^{A(I)}$. We address this challenge by leveraging the fact that an NE is a best response to other strategies at each infoset in perturbed EFGs, as shown in the text around (\ref{eq:sec-4-0-2}), (\ref{eq:Sec-4-3}), and \Cref{lem:add term is positive}. Secondly, we derive the parameter-free convergence result, namely, that the last-iterate convergence of CFR$^+$ holds for all $\bm{\theta}^{1}_I \in \mathbb{R}^{|A(I)|}_{\geq 0}$ and $\eta > 0$. The main challenge here is that the property used in previous proofs of the parameter-free property—that the strategy sequence produced by CFR$^+$ is invariant w.r.t. different step sizes $\eta > 0$—holds only when $\bm{\theta}^{1}_I= \bm{0}$. We overcome this by exploiting the linearity of the projection in CFR$^+$ and the fact that our non-parameter-free last-iterate convergence of CFR$^+$ holds for all $\bm{\theta}^{1}_I \in \mathbb{R}^{|A(I)|}_{\geq 0}$, as presented in the second paragraph following \Cref{lem:add term is positive}. The details of our proof sketch is shown in the following.

\begin{lemma}
[Adapted from the proof of Lemma 4 in \citet{farina2021faster}]
\label{lem:update-rule-MPCFR-inequality}
Assuming all players follow the update rule of CFR$^+$, then for any ${\bm{\theta}}_I \in \mathbb{R}^{|A(I)|}_{\geq 0}$, we have
\begin{equation}
\small
\setlength\abovedisplayskip{3pt}
\setlength\belowdisplayskip{3pt}
\begin{aligned}
    & D_{\psi}({\bm{\theta}}_I, {\bm{\theta}}^{t+1}_I) - D_{\psi}({\bm{\theta}}_I, {\bm{\theta}}^{t}_I) 
    \leq  \eta \langle \hat{\bm{m}}^{t}_i(I), {\bm{\theta}}^{t+1}_I - {\bm{\theta}}_I \rangle - D_{\psi}({\bm{\theta}}^{t+1}_I, {\bm{\theta}}^{t}_I).
\end{aligned}
\end{equation}
\end{lemma}

By applying \Cref{lem:update-rule-MPCFR-inequality} with ${\bm{\theta}}_I = \sigma_i^{*,\mu,\gamma,\bm{r}}(I) = {(\hat{\sigma}_i^{*,\mu,\gamma,\bm{r}}(I) - \gamma \bm{1})}/{(1 - \alpha_I)} \in \Delta^{|A(I)|}$, we get
\begin{equation}\label{eq:sec-4-0}
\small
% \setlength\abovedisplayskip{3pt}
% \setlength\belowdisplayskip{3pt}
% \thinmuskip=0mu
% \medmuskip=0mu
% \thickmuskip=0mu
% \spaceskip=0pt
    \begin{aligned}
        & \eta \langle \hat{\bm{m}}^{t}_i(I), \sigma_i^{*,\mu,\gamma,\bm{r}}(I) - \bm{\theta}^{t+1}_I \rangle \leq  D_{\psi}(\sigma_i^{*,\mu,\gamma,\bm{r}}(I), {\bm{\theta}}^{t}_I) - D_{\psi}(\sigma_i^{*,\mu,\gamma,\bm{r}}(I), {\bm{\theta}}^{t+1}_I) - D_{\psi}({\bm{\theta}}^{t+1}_I, {\bm{\theta}}^{t}_I).
    \end{aligned}
\end{equation}
Also, we define 
\begin{equation}\label{eq:sec-4-0-0}
\small
\thinmuskip=0mu
\medmuskip=0mu
\thickmuskip=0mu
\spaceskip=0pt
    \begin{aligned}
    & \hat{\bm{m}}^{*,\mu,\gamma,\bm{r}}_i(I) = \hat{\bm{v}}^{*,\mu,\gamma,\bm{r}}_i(I) - \langle \hat{\bm{v}}^{*,\mu,\gamma,\bm{r}}_i(I), \sigma^{*,\mu,\gamma,\bm{r}}_i(I)\rangle \bm{1}, \\   
    & \hat{\bm{v}}^{*,\mu,\gamma,\bm{r}}_i(I) = -\hat{\bm{\ell}}_i^{*,\mu,\gamma,\bm{r}}(I,a) + \sum_{I^{\prime} \in C_i(I,a)} \langle \hat{\bm{v}}^{*,\mu,\gamma,\bm{r}}_i(I^{\prime}) , \hat{\sigma}_i^{*,\mu,\gamma,\bm{r}}(I^{\prime}) \rangle,\\
    & \hat{\bm{\ell}}_0^{*,\mu,\gamma,\bm{r}} = \bm{A} \hat{\bm{x}}^{*,\mu,\gamma,\bm{r}}_1 + \mu \nabla \psi(\hat{\bm{x}}^{*,\mu,\gamma,\bm{r}}_0) - \mu \nabla \psi({\bm{r}}_0),\ \hat{\bm{\ell}}_1^{*,\mu,\gamma,\bm{r}} = - \bm{A}^{\text{T}} \hat{\bm{x}}^{*,\mu,\gamma,\bm{r}}_0 + \mu \nabla \psi(\hat{\bm{x}}^{*,\mu,\gamma,\bm{r}}_1) - \mu \nabla \psi({\bm{r}}_1).
    \end{aligned}
\end{equation}
Then, adding $\eta \langle - \hat{\bm{m}}^{*,\mu,\gamma,\bm{r}}_i(I), \bm{\theta}^{t+1}_I - \bm{\theta}^{t}_I \rangle$ to each hand side of (\ref{eq:sec-4-0}), we can get
\begin{equation}\label{eq:sec-4-0-2}
\small
% \setlength\abovedisplayskip{3pt}
% \setlength\belowdisplayskip{3pt}
% \thinmuskip=0mu
% \medmuskip=0mu
% \thickmuskip=0mu
% \spaceskip=0pt
    \begin{aligned}
        & \eta \langle \hat{\bm{m}}^{t}_i(I), \sigma_i^{*,\mu,\gamma,\bm{r}}(I) - \bm{\theta}^{t}_I \rangle - \eta^2 \frac{\Vert \hat{\bm{m}}^{t}_i(I) - \hat{\bm{m}}^{*,\mu,\gamma,\bm{r}}_i(I) \Vert^2_2}{2} \\
        % \leq  & D_{\psi}(\sigma_i^{*,\mu,\gamma,\bm{r}}(I), {\bm{\theta}}^{t}_I) + \eta \langle - \hat{\bm{m}}^{*,\mu,\gamma,\bm{r}}_i(I), {\bm{\theta}}^{t}_I \rangle
        % - D_{\psi}(\sigma_i^{*,\mu,\gamma,\bm{r}}(I), {\bm{\theta}}^{t+1}_I) - \eta \langle - \hat{\bm{m}}^{*,\mu,\gamma,\bm{r}}_i(I), {\bm{\theta}}^{t+1}_I \rangle \\
        % & + \eta \langle \hat{\bm{m}}^{t}_i(I) - \hat{\bm{m}}^{*,\mu,\gamma,\bm{r}}_i(I), \bm{\theta}^{t+1}_I - \bm{\theta}^{t}_I \rangle - D_{\psi}({\bm{\theta}}^{t+1}_I, {\bm{\theta}}^{t}_I) \\
        \leq  & D_{\psi}(\sigma_i^{*,\mu,\gamma,\bm{r}}(I), {\bm{\theta}}^{t}_I) + \eta \langle - \hat{\bm{m}}^{*,\mu,\gamma,\bm{r}}_i(I), {\bm{\theta}}^{t}_I \rangle
        - D_{\psi}(\sigma_i^{*,\mu,\gamma,\bm{r}}(I), {\bm{\theta}}^{t+1}_I) - \eta \langle - \hat{\bm{m}}^{*,\mu,\gamma,\bm{r}}_i(I), {\bm{\theta}}^{t+1}_I \rangle. 
    \end{aligned}
\end{equation}
In OMD algorithms~\citep{sokota2022unified}, the addition of the term $\eta \langle - \hat{\bm{m}}^{*,\mu,\gamma,\bm{r}}_i(I), \bm{\theta}^{t+1}_I - \bm{\theta}^{t}_I \rangle$ is not required to exploit the smoothness of the instantaneous counterfactual regrets. However, this term is necessary to prove the last-iterate convergence of CFR$^+$. This step is crucial in our proof, and to the best of our knowledge, no prior work has proposed a similar approach.

\begin{lemma}
[Proof is in \Cref{subsec:proof:lem:sum of counterfactual regret}]
For any $\bm{x}, \bm{x}^{\prime} \in \bm{\mathcal{X}}$, $\bm{\ell} \in \mathbb{R}^{|\mathcal{X}|}$, $i \in \mathcal{N}$, $\mu \geq 0$, and $\gamma \geq 0$,
\begin{equation}
\small
\setlength\abovedisplayskip{3pt}
\setlength\belowdisplayskip{3pt}
\langle \bm{\ell}_i, \bm{x}_i - \bm{x}^{\prime}_i \rangle = \sum_{I \in \mathcal{I}_i} \pi^{\sigma^{\prime}}_i(I) \langle -\bm{v}^{\sigma}_i(I,a), \sigma_i(I) - \sigma^{\prime}_i(I) \rangle,
\end{equation}
where $\bm{v}^{\sigma}_i(I,a) = -\bm{\ell}_i (I,a) + \sum_{I^{\prime} \in C_i(I,a)} \langle \bm{v}^{\sigma}_i(I^{\prime}) , \sigma_i (I^{\prime}) \rangle$, as well as $\sigma$ and $\sigma^{\prime}$ are the behavioral strategy profiles associated with $\bm{x}$ and $ \bm{x}^{\prime}$, respectively.
\label{lem:sum of counterfactual regret}
\end{lemma}

Combining (\ref{eq:sec-4-0-2}) with \Cref{lem:sum of counterfactual regret}, and setting $\zeta_I = (1-\alpha_I) \beta_I$ with $ \beta_I = {\pi}^{\hat{\sigma}^{*,\mu,\gamma,\bm{r}}}_i(I)$, we have
\begin{equation}\label{eq:Sec-4-1}
\small
\setlength\abovedisplayskip{3pt}
\setlength\belowdisplayskip{3pt}
\thinmuskip=-0.5mu
\medmuskip=-0.5mu
\thickmuskip=-0.5mu
\spaceskip=-0.5pt
    \begin{aligned}
        & 
        \eta \sum_{t=1}^T \sum_{i \in \mathcal{N}} \langle \hat{\bm{\ell}}^{t}_i, \hat{\bm{x}}^{t}_i - \hat{\bm{x}}^{*,\mu,\gamma,\bm{r}}_i \rangle 
        - \sum_{t=1}^T \sum_{i \in \mathcal{N}} \sum_{I \in \mathcal{I}_i} \eta^2 \frac{\Vert \hat{\bm{m}}^{t}_i(I) - \hat{\bm{m}}^{*,\mu,\gamma,\bm{r}}_i(I) \Vert^2_2}{2} \\
        \leq & 
        \sum_{i \in \mathcal{N}} \sum_{I \in \mathcal{I}_i} 
        \zeta_I \left(
         D_{\psi}(\sigma_i^{*,\mu,\gamma,\bm{r}}(I), {\bm{\theta}}^{1}_I) + \eta \langle - \hat{\bm{m}}^{*,\mu,\gamma,\bm{r}}_i(I), {\bm{\theta}}^{1}_I \rangle
        - D_{\psi}(\sigma_i^{*,\mu,\gamma,\bm{r}}(I), {\bm{\theta}}^{T+1}_I) - \eta \langle - \hat{\bm{m}}^{*,\mu,\gamma,\bm{r}}_i(I), {\bm{\theta}}^{T+1}_I \rangle 
        \right).
    \end{aligned}
\end{equation}
% The term $\hat{\bm{x}}^{*,\mu,\gamma,\bm{r}}_i$ in the left-hand side of (\ref{eq:Sec-4-1}) can be any $\bm{x}_i \in \bm{\mathcal{X}}_i$. Therefore, we can minimize the regret within each infoset to minimize the regret across the entire game.

\vspace{-0.25cm}
By using the strongly monotonicity ($O(\sum_{t=1}^T \sum_{i \in \mathcal{N}} \langle \hat{\bm{\ell}}^{t}_i, \hat{\bm{x}}^{t}_i - \hat{\bm{x}}^{*,\mu,\gamma,\bm{r}}_i \rangle) \geq \Vert \hat{\bm{x}}^{t}_i - \hat{\bm{x}}^{*,\mu,\gamma,\bm{r}}_i \Vert^2_2$)
% (details are in \Cref{sec:prf:thm:convergence results of our algorithm})
% , we obtain
% \begin{equation}\label{eq:Sec-4-2}
% \small
% \setlength\abovedisplayskip{3pt}
% \setlength\belowdisplayskip{3pt}
% \thinmuskip=0mu
% \medmuskip=0mu
% \thickmuskip=0mu
% \spaceskip=0pt
%     \begin{aligned}
%         & 
%         \mu \eta \sum_{t=1}^T \Vert \hat{\bm{x}}^{t}- \hat{\bm{x}}^{*,\mu,\gamma,\bm{r}} \Vert^2_2 
%         - \sum_{t=1}^T \sum_{i \in \mathcal{N}} \sum_{I \in \mathcal{I}_i} \eta^2 \frac{\Vert \hat{\bm{m}}^{t}_i(I) - \hat{\bm{m}}^{*,\mu,\gamma,\bm{r}}_i(I) \Vert^2_2}{2} \\
%         \leq & 
%         \sum_{i \in \mathcal{N}} \sum_{I \in \mathcal{I}_i} 
%         \zeta_I \left(
%          D_{\psi}(\sigma_i^{*,\mu,\gamma,\bm{r}}(I), {\bm{\theta}}^{1}_I) + \eta \langle - \hat{\bm{m}}^{*,\mu,\gamma,\bm{r}}_i(I), {\bm{\theta}}^{1}_I \rangle
%         - D_{\psi}(\sigma_i^{*,\mu,\gamma,\bm{r}}(I), {\bm{\theta}}^{T+1}_I) - \eta \langle - \hat{\bm{m}}^{*,\mu,\gamma,\bm{r}}_i(I), {\bm{\theta}}^{T+1}_I \rangle 
%         \right).
%     \end{aligned}
% \end{equation}
% Then, according to the 
and the smoothness of instantaneous counterfactual regrets (  $\Vert \hat{\bm{m}}^{t}_i(I) - \hat{\bm{m}}^{*,\mu,\gamma,\bm{r}}_i(I) \Vert^2_2 \leq O(\Vert \hat{\bm{x}}^{t}_i - \hat{\bm{x}}^{*,\mu,\gamma,\bm{r}}_i \Vert^2_2)$) (see details in \Cref{sec:prf:thm:convergence results of our algorithm}), we get
\begin{equation}\label{eq:Sec-4-3}
\small
\setlength\abovedisplayskip{3pt}
\setlength\belowdisplayskip{3pt}
% \thinmuskip=0mu
% \medmuskip=0mu
% \thickmuskip=0mu
% \spaceskip=0pt
    \begin{aligned}
        & 
        \mu \eta \sum_{t=1}^T \Vert \hat{\bm{x}}^{t}- \hat{\bm{x}}^{*,\mu,\gamma,\bm{r}} \Vert^2_2 
        - \sum_{t=1}^T \eta^2 C_0 \Vert \hat{\bm{x}}^{t}- \hat{\bm{x}}^{*,\mu,\gamma,\bm{r}} \Vert^2_2 
        \leq  
        \sum_{i \in \mathcal{N}} \sum_{I \in \mathcal{I}_i} 
        \zeta_I \bigg(
         D_{\psi}(\sigma_i^{*,\mu,\gamma,\bm{r}}(I), {\bm{\theta}}^{1}_I) \\ & \quad \quad \quad \quad + \eta \langle - \hat{\bm{m}}^{*,\mu,\gamma,\bm{r}}_i(I), {\bm{\theta}}^{1}_I \rangle 
        - D_{\psi}(\sigma_i^{*,\mu,\gamma,\bm{r}}(I), {\bm{\theta}}^{T+1}_I) - \eta \langle - \hat{\bm{m}}^{*,\mu,\gamma,\bm{r}}_i(I), {\bm{\theta}}^{T+1}_I \rangle 
        \bigg),
    \end{aligned}
\end{equation}
where $C_0 = |\mathcal{I}|A_{max}^2 \left( 4 ( L + \mu )^2 + 20 (P+2\mu D)^2{C^2_{max}}/{\gamma^{2H}} \right)$. Note that the form of smoothness we adopt differs from that commonly used in OMD algorithms~\citep{sokota2022unified}, where smoothness typically takes the form $\Vert \hat{\bm{m}}^{t}_i(I) - \hat{\bm{m}}^{t+1}_i(I) \Vert^2_2 \leq O(\Vert \hat{\bm{x}}^{t} - \hat{\bm{x}}^{t+1} \Vert^2_2)$ rather than $\Vert \hat{\bm{m}}^{t}_i(I) - \hat{\bm{m}}^{*,\mu,\gamma,\bm{r}}_i(I) \Vert^2_2 \leq O(\Vert \hat{\bm{x}}^{t} - \hat{\bm{x}}^{*,\mu,\gamma,\bm{r}} \Vert^2_2)$. This difference also highlights that our proof approach diverges from the proof employed by OMD algorithms. Then, if $0< \eta \leq {\mu/(2 C_0)}$, we get
\begin{equation}\label{eq:Sec-4-4}
\small
\setlength\abovedisplayskip{3pt}
\setlength\belowdisplayskip{3pt}
% \thinmuskip=0mu
% \medmuskip=0mu
% \thickmuskip=0mu
% \spaceskip=0pt
    \begin{aligned}
        & \frac{\mu\eta}{2} \sum_{t=1}^T \Vert \hat{\bm{x}}^{t}-\hat{\bm{x}}^{*,\mu,\gamma,\bm{r}}\Vert^2_2 
        \leq  
        \sum_{i \in \mathcal{N}} \sum_{I \in \mathcal{I}_i} 
        \zeta_I \bigg(
         D_{\psi}(\sigma_i^{*,\mu,\gamma,\bm{r}}(I), {\bm{\theta}}^{1}_I) 
         \\ & \quad \quad \quad \quad
         + \eta \langle - \hat{\bm{m}}^{*,\mu,\gamma,\bm{r}}_i(I), {\bm{\theta}}^{1}_I \rangle
        - D_{\psi}(\sigma_i^{*,\mu,\gamma,\bm{r}}(I), {\bm{\theta}}^{T+1}_I) - \eta \langle - \hat{\bm{m}}^{*,\mu,\gamma,\bm{r}}_i(I), {\bm{\theta}}^{T+1}_I \rangle 
        \bigg).
    \end{aligned}
\end{equation}

\begin{lemma}
[Proof is in \Cref{subsec:proof:lem:add term is positive}]
\label{lem:add term is positive}
$\forall i \in \mathcal{N}$, $I \in \mathcal{I}_i$, and $\bm{\theta}_I \in \mathbb{R}^{|A(I)|}_{\geq 0}$, $\langle - \hat{\bm{m}}^{*,\mu,\gamma,\bm{r}}_i(I), {\bm{\theta}}_I \rangle \geq 0$.
\end{lemma}

\Cref{lem:add term is positive} is from that an NE is a best response to others at each infoset in perturbed EFGs, \ie, $\forall \sigma_i$, $\small  
\thinmuskip=0mu
\medmuskip=0mu
\thickmuskip=0mu
\spaceskip=0pt
\langle\hat{\bm{v}}^{*,\mu,\gamma,\bm{r}}_i(I),\hat{\sigma}^{*,\mu,\gamma,\bm{r}}_i(I) - \hat{\sigma}_i(I)\rangle \geq 0$ (details are in \Cref{subsec:proof:lem:add term is positive}). By using \Cref{lem:add term is positive}, we get
$\small 
\thinmuskip=0mu
\medmuskip=0mu
\thickmuskip=0mu
\spaceskip=0pt
\forall T \geq 1, \sum_{t=1}^T \Vert \hat{\bm{x}}^{t}-\hat{\bm{x}}^{*,\mu,\gamma,\bm{r}}\Vert^2_2 
        \leq 
        O(1)$,
implying that $\hat{\bm{x}}^{t}$ converges to $\hat{\bm{x}}^{*,\mu,\gamma,\bm{r}}$ with $
\small 
\thinmuskip=0mu
\medmuskip=0mu
\thickmuskip=0mu
\spaceskip=0pt
0< \eta \leq {\mu/(2 C_0)}$.

\citet{farina2021faster} show that when $\bm{\theta}^{1}_I = \bm{0}$, for any $\eta > 0$, the sequence $\small \{ \hat{\bm{x}}^{1}, \hat{\bm{x}}^{2}, \cdots, \hat{\bm{x}}^{t}, \cdots\}$ remains the same. This implies that $\hat{\bm{x}}^{t}$ converges to $\hat{\bm{x}}^{*,\mu,\gamma,\bm{r}}$ for any $\eta > 0$, showing the parameter-free property. In this paper, we further show that for any initial $\bm{\theta}^{1}_I \in \mathbb{R}^{|A(I)|}_{\geq 0}$ and $\eta > 0$, $\hat{\bm{x}}^{t}$ converges to $\hat{\bm{x}}^{*,\mu,\gamma,\bm{r}}$ (see advantages in discussions). This proof is simple yet novel, with the key insights being the linearity of the projection in CFR$^+$ and that $\sum_{t=1}^T \Vert \hat{\bm{x}}^{t} - \hat{\bm{x}}^{*,\mu,\gamma,\bm{r}} \Vert^2_2 \leq O(1)$ holds independently of the value of $\bm{\theta}^{1}_I$. Specifically, from the the linearity of the projection in CFR$^+$, for any accumulated counterfactual regret sequence $\small \{ \bm{\theta}^{1}_I, \bm{\theta}^{2}_I, \dots, \bm{\theta}^{t}_I, \dots \}$ generated by any $\bm{\theta}^{1}_I \in \mathbb{R}^{|A(I)|}_{\geq 0}$ and $\eta > 0$, there exists a corresponding accumulated counterfactual regret sequence $\small \{ {\bm{\theta}^{1}_I}^{\prime}, {\bm{\theta}^{2}_I}^{\prime}, \dots, {\bm{\theta}^{t}_I}^{\prime}, \dots \}$ generated by ${\bm{\theta}^{1}_I}^{\prime}$ and $\eta^{\prime} = \mu / (2 C_0)$, such that the resulting strategy profile sequence $\small \{ \hat{\bm{x}}^{1}, \hat{\bm{x}}^{2}, \dots, \hat{\bm{x}}^{t}, \dots \}$ are identical. Additionally, as the condition $\sum_{t=1}^T \Vert \hat{\bm{x}}^{t} - \hat{\bm{x}}^{*,\mu,\gamma,\bm{r}} \Vert^2_2 \leq O(1)$ holds independently of the value of $\bm{\theta}^{1}_I$ (${\bm{\theta}^{1}_I}^{\prime}$). Based on this analysis, we conclude that for any accumulated counterfactual regret sequence $\small \{ \bm{\theta}^{1}_I, \bm{\theta}^{2}_I, \dots, \bm{\theta}^{t}_I, \dots \}$ generated by any $\bm{\theta}^{1}_I$ and $\eta > 0$, the corresponding strategy profile sequence $\small \{ \hat{\bm{x}}^{1}, \hat{\bm{x}}^{2}, \dots, \hat{\bm{x}}^{t}, \dots \}$ converges to $\hat{\bm{x}}^{*,\mu,\gamma,\bm{r}}$, which indicates the parameter-free property. 

\begin{wrapfigure}{r}{0.35\textwidth}
\vspace{-23pt}
\begin{minipage}{0.35\textwidth}
\begin{algorithm}[H]
    \centering \small
    \caption{RTCFR$^+$}\label{alg:RTRM+}
    \begin{algorithmic}[1]
        \STATE {\bfseries Input:} $N$, $T_u$, $\mu$, $\gamma$, ${\bm{r}}$
        \STATE ${\bm{\theta}}^{1}_I \leftarrow \bm{0}, \eta \leftarrow 1$, $\forall I \in \mathcal{I}$
        \FOR{each $n \in [1,\ 2,\cdots, N]$}
           \STATE Build the perturbed regularized EFGs in (\ref{eq:BSPP-perturbed regularized}) via $\mu$, $\gamma$, and ${\bm{r}}$
           \FOR{each $t \in [1,\ 2,\cdots, T_u]$}
                \STATE Obtain $\hat{\bm{x}}^{t+1}$ and ${\bm{\theta}}^{t+1}_I$ via the update rule in (\ref{eq:update-rule-MPCFR})
            \ENDFOR
            \STATE $\gamma \leftarrow \gamma*0.5$, ${\bm{r}} \leftarrow \hat{\bm{x}}^{T_u+1}$
            \STATE ${\bm{\theta}}^{1}_I \leftarrow {\bm{\theta}}^{T_u+1}_I$, $\forall I \in \mathcal{I}$
        \ENDFOR
        \STATE {\bfseries Return} $\hat{\bm{x}}^{T_u+1}$
    \end{algorithmic}
\end{algorithm}
\end{minipage}
\vspace{-15pt}
\end{wrapfigure}

\textbf{Reward Transformation CFR$^+$ (RTCFR$^+$).} RTCFR$^+$ is the RT algorithm that applies CFR$^+$ to solve perturbed regularized EFGs, whose pseudocode is in \cref{alg:RTRM+}. As analyzed by \citet{abe2023slingshot,bernasconi2024learning}, continuously decreasing $\gamma$ and updating $\bm{r}$ to $\hat{\bm{x}}^{*,\gamma,\mu,\bm{r}}$ allows the sequence of the NEs of the perturbed regularized EFGs to converge to the NE of the set of the original EFG. Specifically, as shown in \cref{alg:RTRM+}, after $T_u$ iterations, RTCFR$^+$ updates $\gamma$ and $\bm{r}$ \footnote{We do not examine the convergence of the sequence of NEs of the perturbed regularized EFGs to the set of NEs of the original EFG when the exact $\hat{\bm{x}}^{*,\gamma,\mu,\bm{r}}$ cannot be learned, but only an approximate $\hat{\bm{x}}^{*,\gamma,\mu,\bm{r}}$ is obtained. For further details on the convergence in this context, refer to \citet{abe2023slingshot,wang2025magnetic}.}, with $N*T_u$ representing the total number of iterations. The implementation of RTCFR$^+$ is in \Cref{sec:Implementation of MPCFR+}.

\textbf{Discussions.} Firstly, to the best of our knowledge, we provide the first parameter-free last-iterate convergence for RM-based CFR algorithms in learning an NE of perturbed regularized EFGs. When considering NFGs, the last-iterate convergence result of CFR$^+$ (RM$^+$) holds even when $\gamma=0$, due to that the smoothness of counterfactual values and \Cref{lem:add term is positive} hold in NFGs with any $\gamma \geq 0$. Secondly, we extend the parameter-free results of CFR$^+$ from \citet{farina2021faster}, demonstrating that CFR$^+$ converges with the parameter-free property for any $\bm{\theta}^{1}_I \in \mathbb{R}^{|A(I)|}_{\geq 0}$, not just when $\bm{\theta}^{1}_I = \bm{0}$ in \citet{farina2021faster}. This new parameter-free result is significant. Specifically, it indicates that after updating $\gamma$ and $\bm{r}$ (line 8 of \Cref{alg:RTRM+}), there is no need to reset $\bm{\theta}^{1}_I$ to $\bm{0}$ to get the parameter-free property (line 9 of \Cref{alg:RTRM+}). This improves the stability of CFR$^+$, i.e., rapid fluctuations in the strategy profiles across iterations, since such stability improves as the lower bound of the 1-norm of $\bm{\theta}^{t}_I$ increases~\citep{farina2023regret} (for CFR$^+$, $\Vert \bm{\theta}^{t}_I \Vert_2 \leq \Vert \bm{\theta}^{t+1}_I \Vert_2$ holds~\citep{liu2021equivalence_icml}, and the 1-norm lower bound can be derived from the 2-norm lower bound). Notably, as shown in \Cref{sec:Additional Experiments}, resetting $\bm{\theta}^{1}_I$ to $\bm{0}$ after updating $\gamma$ and $\bm{r}$ (line 9 of \Cref{alg:RTRM+} becomes ${\bm{\theta}}^{1}_I \leftarrow \bm{0}$, $\forall I \in \mathcal{I}$) causes RTCFR$^+$ to never converge! Also, our proof approach for the parameter-free property can be used to show that CFR$^+$'s average-iterate convergence holds for all $\bm{\theta}^{1}_I \in \mathbb{R}^{|A(I)|}_{\geq 0}$ and $\eta > 0$. As our primary focus is on last-iterate convergence, we discuss the parameter-free average-iterate convergence in \Cref{sec:Parameter-free average-iterate convergence} rather than the main text.

% \vspace{-0.35cm}
\section{Experiments}\label{sec:Experiments}
\begin{figure*}[t]
    \centering %\quad \quad
    \subfigure{
    \includegraphics[width=0.8\linewidth]{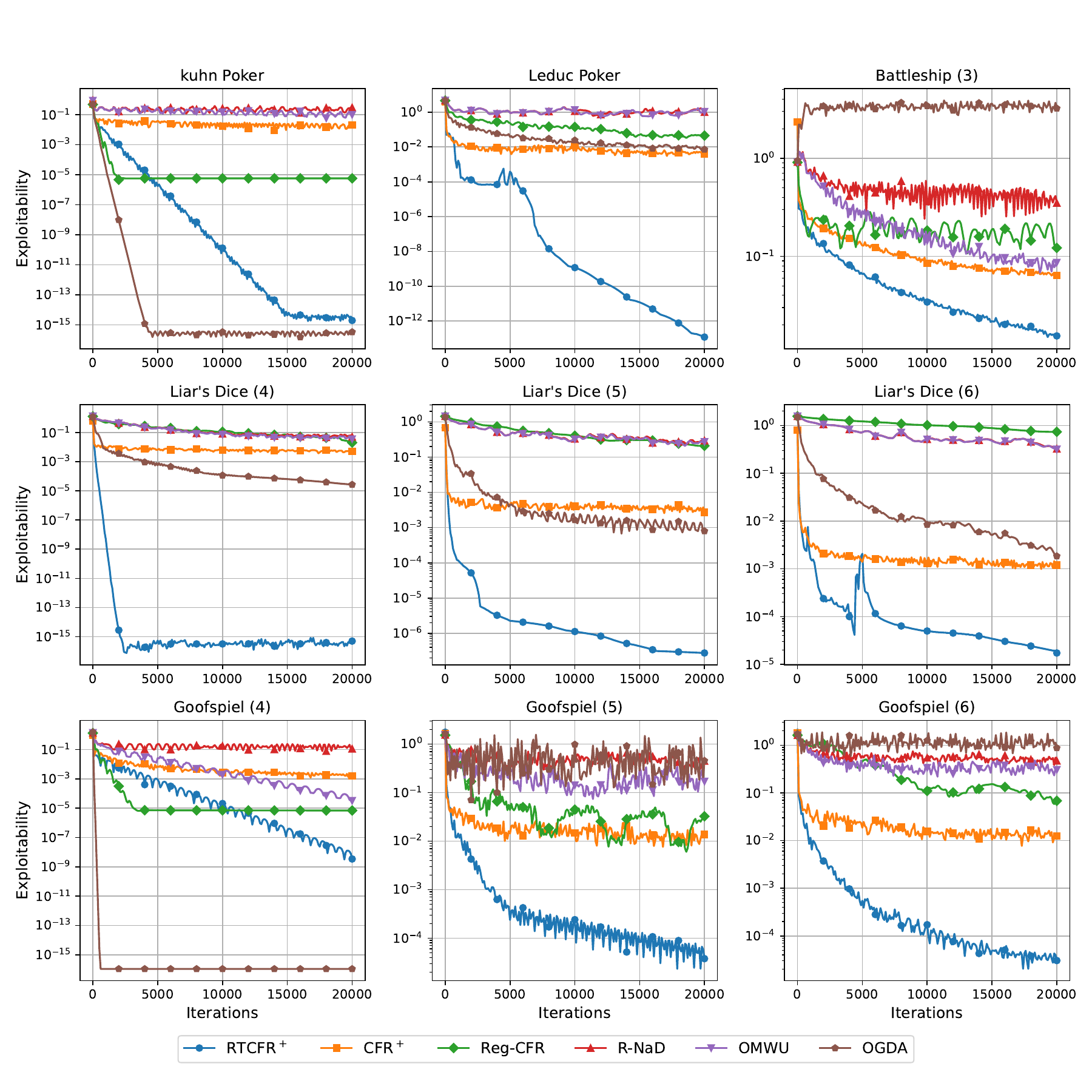}
    }\vspace{-15pt} 
    \caption{Last-iterate convergence rates of different algorithms. Each algorithm runs for 20,000 iterations. In all plots,the x-axis is the number of iteration, and the y-axis represents exploitability,displayed on a logarithmic scale. Liar’s Dice ($x$) represents that every player is given a die with $x$ sides. Goofspiel ($x$) denotes that each player is dealt $x$ cards. Battleship ($x$) implies the size of grids is $x$. The number of infosets of Kuhn Poker, Leduc Poker, Battleship (3), Liar's Dice (3), Liar's Dice (4), Liar's Dice (5), Goofspiel (4), Goofspiel (5), Goofspiel (6) are 12, 936, 81027, 1024, 5120, 24576, 162, 2124, and 34482, respectively. 
}
\label{fig:with-other-algorithms}
\vspace{-0.25cm}
\end{figure*}

\textbf{Configurations.} We now evaluate the empirical convergence rate of RTCFR$^+$. We conduct experiments on five standard EFG benchmarks: Kuhn Poker, Leduc Poker, Goofspiel, Liar's Dice, and Battleship, all implemented using OpenSpiel~\citep{lanctot2019openspiel}. We compare RTCFR$^+$ with classical CFR algorithms, such as CFR$^+$, as well as those with theoretical last-iterate convergence guarantee, including R-NaD~\citep{perolat2021poincare,perolat2022mastering} and Reg-CFR~\citep{liu2022power}. Additionally, we evaluate traditional last-iterate convergence algorithms based on OMD, such as OMWU and OGDA~\citep{wei2020linear,lee2021last}. The algorithm implementations are based on the open-source LiteEFG code~\citep{liu2024liteefgefficientpythonlibrary}, which offers a significant speedup—approximately 100 times faster than OpenSpiel's default implementation for the same number of iterations. For RTCFR$^+$, we set the initial values of $\gamma$ and $\mu$ to $1\mathrm{e}{-10}$ and $1\mathrm{e}{-3}$, respectively. The number of iterations $T_u$ required to update $\gamma$ and $\bm{r}$, is set to $100$. For Reg-CFR, we use the parameters from the original paper. For R-NaD, we initialize $\mu=1\mathrm{e}{-5}$ (R-NaD does not include the parameter $\gamma$), set $T_u = 1000$, and use a learning rate of $\eta = 0.1$. For OMWU and OGDA, we set $\eta$ to $0.5$ and $0.1$, respectively. All algorithms employ alternating updates to enhance empirical convergence rates. Each algorithm is run for 20,000 ($N = 20000/T_u$) iterations to analyze long-term behavior. The experiments are conducted on a machine equipped with a Xeon(R) Gold 6444Y CPU and 256 GB of memory. More experimental results including (i) comparison with more RM-based CFR algorithms, and (ii) performance of RTCFR$^+$ under different hyperparameters, are in \Cref{sec:Additional Experiments}.

\textbf{Results.}
The experimental results are presented in \Cref{fig:with-other-algorithms}. Our proposed algorithm, RTCFR$^+$, demonstrates superior performance compared to all other tested algorithms. Specifically, RTCFR$^+$ exhibits the fastest convergence rate across all games when compared to CFR$^+$. In comparison to existing theoretical last-iterate convergence CFR algorithms, such as Reg-CFR and R-NaD, RTCFR$^+$ is only surpassed by Reg-CFR during the initial stages in small-scale games like Kuhn Poker and Goofspiel (4). Similarly, when compared to traditional last-iterate convergence algorithms, RTCFR$^+$ is only outperformed by OGDA in small-scale games such as Kuhn Poker and Goofspiel (4). It is worth noting that the values of $\gamma$, $\mu$, and $T_u$ for RTCFR$^+$ are not fined tuned at each EFGs. The results of RTCFR$^+$ under the careful tuning of these parameters are presented in \Cref{sec:Additional Experiments}. However, the automatic adjustment of $\gamma$, $\mu$, and $T_u$ remains an open problem. One of our future research directions is to investigate methods for automating the adjustment of these parameters.

% \vspace{-0.35cm}
\section{Conclusions}\label{sec:Conclusions}
In this paper, we explore the last-iterate convergence of parameter-free RM-based CFR algorithms. We establish that a classical parameter-free RM-based CFR algorithm, CFR$^+$, achieves last-iterate convergence  in learning an NE of perturbed regularized EFGs. To the best of our knowledge, this is the first demonstration of parameter-free last-iterate convergence for RM-based CFR algorithms in solving perturbed regularized EFGs. Experimental results indicate that our proposed algorithm, RTCFR$^+$, exhibits a significantly faster empirical convergence rate compared to existing algorithms with theoretical last-iterate convergence guarantee.

% \clearpage
% \newpage

\bibliographystyle{plainnat}
\bibliography{references}

\newpage
\appendix
\section{Related Work}\label{sec:Related Work}
\textbf{Counterfactual Regret Minimization (CFR) algorithms.} CFR algorithms are among the most widely used methods for solving real-world EFGs~\citep{bowling2015heads,moravvcik2017deepstack,brown2018superhuman,brown2019superhuman,perolat2022mastering}. The core idea of CFR is to decompose the problem of regret minimization across the entire game into subproblems within each infoset, employing a regret minimization algorithm as a local regret minimizer. The vanilla CFR algorithm was introduced by \citet{zinkevich2007regret}, which utilize RM~\citep{hart2000simple} as the local regret minimizer. To enhance the performance of CFR, a common approach is to design more effective local regret minimizers, as the choice of local regret minimizer largely determines the overall CFR algorithm's efficiency. Advanced local regret minimizers are typically based on RM, including RM$^+$~\citep{tammelin2014solving}, Discounted RM (DRM)~\citep{brown2019solving}, and Predictive RM$^+$ (PRM$^+$)~\citep{farina2021faster}, which correspond to CFR$^+$~\citep{tammelin2014solving}, Discounted CFR (DCFR)~\citep{brown2019solving}, and Predictive CFR$^+$ (PCFR$^+$)~\citep{farina2021faster}, respectively. However, CFR algorithms typically achieve convergence to the set of NEs of EFGs only through the average of iterates, also be called as average-iterate convergence.

\textbf{Last-iterate convergence results of CFR algorithms.} \citet{perolat2021poincare} provide the first last-iterate convergence result for CFR algorithms in learning an NE of EFGs by transforming the task of learning an NE of the original EFG into finding the NEs of a sequence of regularized EFGs and ensuring the sequence of the NEs of these regularized EFGs converges to the set of NEs of the original EFG. However, their analysis assumes continuous-time feedback, a condition rarely satisfied in practical scenarios. Subsequently, \citet{liu2022power} presents the first last-iterate convergence result for CFR under the discrete-time feedback by transforming the task of learning an NE of the original EFG into finding the NEs of a sequence of perturbed regularized EFGs rather than only regularized EFGs, since the addition of perturbation introduces the smoothness of counterfactual values. Nevertheless, both algorithms do not leverage RM algorithms as the local regret minimizer, leading to a suboptimal empirical last-iterate convergence rate compared to traditional RM-based CFR algorithms that only achieve average-iterate convergence, as demonstrated in our experiments.

\textbf{Last-iterate convergence results of RM algorithms.} 
Except this paper, \citet{cai2023lastiterate,meng2025lastiterate} also investigate the last-iterate convergence of RM algorithms. However, their results mainly focus on non-parameter-free RM algorithms, whereas we considers parameter-free RM algorithms. Specifically, \citet{cai2023lastiterate,meng2025lastiterate} mainly investigate smooth RM$^+$ variants~\citep{farina2023regret}.\footnote{Although \citet{cai2023lastiterate} also investigate RM$^+$ (CFR$^+$ reduces RM$^+$ in NFGs), their proof largely follows from ours in \Cref{sec:prf:thm:convergence results of our algorithm}.} The lack of the parameter-free property in the results of \citet{cai2023lastiterate,meng2025lastiterate} makes them less applicable when solving real-world games. In addition, their results only holds in NFGs while we focus on EFGs in this paper.

We establish the first parameter-free last-iterate convergence for RM-based CFR algorithms in learning an NE of perturbed regularized EFGs. Notably, our parameter-free property holds for any initial accumulated counterfactual regrets not only the zero initialization in previous works~\citep{farina2021faster,grand2021conic,chakrabarti2024extensive}. Experiments show that our algorithms substantially outperform existing algorithms with theoretical last-iterate convergence.

It is important to note that while CFR$^+$'s parameter-free property in its first theoretical convergence result~\citep{bowling2015heads} holds for any initial accumulated counterfactual regrets, this result is exclusively limited to average-iterate convergence. In contrast, our proof technique simultaneously establishes both parameter-free last-iterate and average-iterate convergence for CFR$^+$ under any initial accumulated counterfactual regrets. 
\clearpage
\newpage

\section{Proof of \Cref{thm:convergence results of our algorithm}}\label{sec:prf:thm:convergence results of our algorithm}

\begin{proof} 
To prove the last-iterate convergence of CFR$^+$ in learning an NE of perturbed regularized EFGs defined in (\ref{eq:BSPP-perturbed regularized}), we introduce the following lemmas.

\begin{lemma}
[Adapted from Lemma D.4 in \citet{sokota2022unified}]
For any ${\bm{x}} \in \bm{\mathcal{X}}$, $\mu \geq 0$, and $\gamma \geq 0$,
\begin{equation}
\small
\setlength\abovedisplayskip{0pt}
\setlength\belowdisplayskip{0pt}
\begin{aligned}
\sum_{i \in \mathcal{N}} \langle {\bm{\ell}}^{{\bm{x}}}_i, {\bm{x}}_i - {\bm{x}}^{*,\mu,\gamma,\bm{r}}_i \rangle \geq & \sum_{i \in \mathcal{N}} \langle {\bm{\ell}}^{{\bm{x}}}_i - {\bm{\ell}}^{{\bm{x}}^{*,\mu,\gamma,\bm{r}}}_i, {\bm{x}}_i - {\bm{x}}^{*,\mu,\gamma,\bm{r}}_i \rangle \geq \mu \Vert {\bm{x}}- {\bm{x}}^{*,\mu,\gamma,\bm{r}} \Vert^2_2,
\end{aligned}
\end{equation}
where ${\bm{\ell}}^{{\bm{x}}}_0 = \bm{A}{\bm{x}}_1 + \mu \nabla \psi({\bm{x}}_0) - \mu \nabla \psi({\bm{r}}_0)$ and ${\bm{\ell}}^{\bm{x}}_1 = -\bm{A}^{\text{T}} {\bm{x}}_0 + \mu \nabla \psi({\bm{x}}_1) - \mu \nabla \psi({\bm{r}}_1)$.
\label{lem:strongly monotone condition}
\end{lemma}

\begin{lemma}
[Proof is in \Cref{subsec:proof:lem:maximum value of counterfactual value}]
For any $\bm{x} \in \bm{\mathcal{X}}$, $i \in \mathcal{N}$, $I \in \mathcal{I}_i$, $\mu \geq 0$, and $\gamma \geq 0$,
\begin{equation}
    \Vert \hat{\bm{v}}^{\sigma}_i(I)\Vert_2 \leq  \Vert \hat{\bm{v}}^{\sigma}_i(I)\Vert_1 \leq P + 2\mu D
\end{equation}
where $\hat{\bm{v}}^{\sigma}_i(I)=[\hat{\bm{v}}^{\sigma}_i(I,a)|a \in A(I)]$,  $\hat{\bm{v}}^{\sigma}_i(I,a) = - \hat{\bm{\ell}}^{\bm{x}}_i + \sum_{I^{\prime} \in C_i(I,a)} \langle \hat{\bm{v}}^{\sigma}_i(I^{\prime}), \sigma_i(I^{\prime})\rangle$ with $\hat{\bm{\ell}}^{\bm{x}}_0 = \bm{A}\bm{x}_1 + \mu \nabla \psi(\bm{x}_0) - \mu \nabla \psi({\bm{r}}_0)$ and $\hat{\bm{\ell}}^{\bm{x}}_1 = -\bm{A}^{\text{T}} \bm{x}_0 + \mu \nabla \psi(\bm{x}_1) - \mu \nabla \psi({\bm{r}}_1)$, as well as $\sigma$ is the behavioral strategy profile associated with $\bm{x}$.
\label{lem:maximum value of counterfactual value}
\end{lemma}

\begin{lemma}
[Proof is in \Cref{subsec:proof:lem:smoothness of counterfactual value}]
For any $\bm{x}, \bm{x}^{\prime} \in \bm{\mathcal{X}}$, $i \in \mathcal{N}$, $I \in \mathcal{I}_i$, $\mu \geq 0$, and $\gamma \geq 0$,
\begin{equation}
    \Vert \hat{\bm{v}}^{\sigma}_i(I) - \hat{\bm{v}}^{\sigma^{\prime}}_i(I) \Vert_2 \leq 2( L + \mu )^2 \Vert \bm{x} - \bm{x}^{\prime} \Vert^2_1 + 2(P+2\mu D)^2\Vert \sigma_i  -  \sigma^{\prime}_i  \Vert^2_1,
\end{equation}
where $\hat{\bm{v}}^{\sigma}_i(I)=[\hat{\bm{v}}^{\sigma}_i(I,a)|a \in A(I)]$, $\hat{\bm{v}}^{\sigma}_i(I,a) = - \hat{\bm{\ell}}^{\bm{x}}_i + \sum_{I^{\prime} \in C_i(I,a)} \langle \hat{\bm{v}}^{\sigma}_i(I^{\prime}), \sigma_i(I^{\prime})\rangle$ with $\hat{\bm{\ell}}^{\bm{x}}_0 = \bm{A}\bm{x}_1 + \mu \nabla \psi(\bm{x}_0) - \mu \nabla \psi({\bm{r}}_0)$ and $\hat{\bm{\ell}}^{\bm{x}}_1 = -\bm{A}^{\text{T}} \bm{x}_0 + \mu \nabla \psi(\bm{x}_1) - \mu \nabla \psi({\bm{r}}_1)$, as well as $\sigma$ and $\sigma^{\prime}$ are the behavioral strategy profiles associated with $\bm{x}$ and $ \bm{x}^{\prime}$, respectively.
\label{lem:smoothness of counterfactual value}
\end{lemma}

\begin{lemma}
[Proof is in \Cref{subsec:proof:lem:relationship between behavioral strategy and sequence-form strategy}]
For any $\hat{\bm{x}}, \hat{\bm{x}}^{\prime} \in \bm{\mathcal{X}}^{\gamma}$ with $\gamma > 0$, $i \in \mathcal{N}$, $I \in \mathcal{I}_i$, and $\mu \geq 0$,
\begin{equation}
    \Vert {\hat{\sigma}}_i  -  {\hat{\sigma}}^{\prime}_i  \Vert_1 \leq \frac{2C_{max}}{\gamma^H} \Vert \hat{\bm{x}}_i - \hat{\bm{x}}^{\prime}_i \Vert_1,
\end{equation}
where ${\hat{\sigma}}$ and ${\hat{\sigma}}^{\prime}$ are the behavioral strategy profiles associated with ${\hat{\bm{x}}}$ and ${\hat{\bm{x}}}^{\prime}$, respectively.
\label{lem:relationship between behavioral strategy and sequence-form strategy}
\end{lemma}

By substituting ${\bm{\theta}}_I = \sigma_i^{*,\mu,\gamma,\bm{r}}(I) = \frac{\hat{\sigma}_i^{*,\mu,\gamma,\bm{r}}(I) - \gamma \bm{1}}{1 - \alpha_I}$ into \Cref{lem:update-rule-MPCFR-inequality}, we get
\begin{equation}\label{eq:appendix-A-0}
\small
\thinmuskip=0mu
\medmuskip=0mu
\thickmuskip=0mu
\spaceskip=0pt
    \begin{aligned}
        & \eta \langle \hat{\bm{m}}^{t}_i(I), \sigma_i^{*,\mu,\gamma,\bm{r}}(I) - \bm{\theta}^{t+1}_I \rangle \leq  D_{\psi}(\sigma_i^{*,\mu,\gamma,\bm{r}}(I), {\bm{\theta}}^{t}_I) - D_{\psi}(\sigma_i^{*,\mu,\gamma,\bm{r}}(I), {\bm{\theta}}^{t+1}_I) - D_{\psi}({\bm{\theta}}^{t+1}_I, {\bm{\theta}}^{t}_I).
    \end{aligned}
\end{equation}
Adding $\eta \langle - \hat{\bm{m}}^{*,\mu,\gamma,\bm{r}}_i(I), \bm{\theta}^{t+1}_I - \bm{\theta}^{t}_I \rangle$ to each hand side of (\ref{eq:appendix-A-0}), we have
\begin{equation}\label{eq:appendix-A-1}
\small
% \setlength\abovedisplayskip{2pt}
% \setlength\belowdisplayskip{2pt}
% \thinmuskip=0mu
% \medmuskip=0mu
% \thickmuskip=0mu
% \spaceskip=0pt
    \begin{aligned}
        & \eta \langle \hat{\bm{m}}^{t}_i(I), \sigma_i^{*,\mu,\gamma,\bm{r}}(I) - \bm{\theta}^{t+1}_I \rangle + \eta \langle - \hat{\bm{m}}^{*,\mu,\gamma,\bm{r}}_i(I), \bm{\theta}^{t+1}_I - \bm{\theta}^{t}_I \rangle \\
        \leq  & D_{\psi}(\sigma_i^{*,\mu,\gamma,\bm{r}}(I), {\bm{\theta}}^{t}_I) - D_{\psi}(\sigma_i^{*,\mu,\gamma,\bm{r}}(I), {\bm{\theta}}^{t+1}_I) + \eta \langle - \hat{\bm{m}}^{*,\mu,\gamma,\bm{r}}_i(I), \bm{\theta}^{t+1}_I - \bm{\theta}^{t}_I \rangle - D_{\psi}({\bm{\theta}}^{t+1}_I, {\bm{\theta}}^{t}_I),
    \end{aligned}
\end{equation}
which implies
\begin{equation}\label{eq:appendix-A-2}
\small
% \setlength\abovedisplayskip{2pt}
% \setlength\belowdisplayskip{2pt}
% \thinmuskip=0mu
% \medmuskip=0mu
% \thickmuskip=0mu
% \spaceskip=0pt
    \begin{aligned}
        & \eta \langle \hat{\bm{m}}^{t}_i(I), \sigma_i^{*,\mu,\gamma,\bm{r}}(I) - \bm{\theta}^{t}_I \rangle \\
        \leq  & D_{\psi}(\sigma_i^{*,\mu,\gamma,\bm{r}}(I), {\bm{\theta}}^{t}_I) + \eta \langle - \hat{\bm{m}}^{*,\mu,\gamma,\bm{r}}_i(I), {\bm{\theta}}^{t}_I \rangle
        - D_{\psi}(\sigma_i^{*,\mu,\gamma,\bm{r}}(I), {\bm{\theta}}^{t+1}_I) - \eta \langle - \hat{\bm{m}}^{*,\mu,\gamma,\bm{r}}_i(I), {\bm{\theta}}^{t+1}_I \rangle \\
        & + \eta \langle \hat{\bm{m}}^{t}_i(I) - \hat{\bm{m}}^{*,\mu,\gamma,\bm{r}}_i(I), \bm{\theta}^{t+1}_I - \bm{\theta}^{t}_I \rangle - D_{\psi}({\bm{\theta}}^{t+1}_I, {\bm{\theta}}^{t}_I) \\
        \leq  & D_{\psi}(\sigma_i^{*,\mu,\gamma,\bm{r}}(I), {\bm{\theta}}^{t}_I) + \eta \langle - \hat{\bm{m}}^{*,\mu,\gamma,\bm{r}}_i(I), {\bm{\theta}}^{t}_I \rangle
        - D_{\psi}(\sigma_i^{*,\mu,\gamma,\bm{r}}(I), {\bm{\theta}}^{t+1}_I) - \eta \langle - \hat{\bm{m}}^{*,\mu,\gamma,\bm{r}}_i(I), {\bm{\theta}}^{t+1}_I \rangle \\
        & + \eta^2 \frac{\Vert \hat{\bm{m}}^{t}_i(I) - \hat{\bm{m}}^{*,\mu,\gamma,\bm{r}}_i(I) \Vert^2_2}{2} + \frac{\Vert \bm{\theta}^{t+1}_I - \bm{\theta}^{t}_I \Vert^2_2}{2} - D_{\psi}({\bm{\theta}}^{t+1}_I, {\bm{\theta}}^{t}_I) \\
        \leq  & D_{\psi}(\sigma_i^{*,\mu,\gamma,\bm{r}}(I), {\bm{\theta}}^{t}_I) + \eta \langle - \hat{\bm{m}}^{*,\mu,\gamma,\bm{r}}_i(I), {\bm{\theta}}^{t}_I \rangle
        - D_{\psi}(\sigma_i^{*,\mu,\gamma,\bm{r}}(I), {\bm{\theta}}^{t+1}_I) - \eta \langle - \hat{\bm{m}}^{*,\mu,\gamma,\bm{r}}_i(I), {\bm{\theta}}^{t+1}_I \rangle \\
        & + \eta^2 \frac{\Vert \hat{\bm{m}}^{t}_i(I) - \hat{\bm{m}}^{*,\mu,\gamma,\bm{r}}_i(I) \Vert^2_2}{2} ,
    \end{aligned}
\end{equation}
where the second inequality comes from that $\forall \bm{a}, \bm{b} \in \mathbb{R}^d$, $\rho > 0$, $\langle \bm{a}, \bm{b} \rangle \leq \rho \Vert \bm{a} \Vert^2_2/2 + \Vert \bm{b} \Vert^2_2/(2\rho)$ (in this case, $\bm{a} = \hat{\bm{m}}^{t}_i(I) - \hat{\bm{m}}^{*,\mu,\gamma,\bm{r}}_i(I)$, $\bm{b} = \bm{\theta}^{t+1}_I - \bm{\theta}^{t}_I$, and $\rho = \eta$), and the last inequality is from that $\forall \bm{a}, \bm{b} \in \mathbb{R}^d$, $\Vert  \bm{a} - \bm{b} \Vert^2_2/2 = \Vert  \bm{b} - \bm{a} \Vert^2_2/2 = D_{\psi}(\bm{a}, \bm{b})$ (in this case, $\bm{a} = {\bm{\theta}}^{t+1}_I$, and $\bm{b} = {\bm{\theta}}^{t}_I$).

Arranging the terms in (\ref{eq:appendix-A-2}), we get
\begin{equation}\label{eq:appendix-A-3}
\small
% \setlength\abovedisplayskip{2pt}
% \setlength\belowdisplayskip{2pt}
% \thinmuskip=0mu
% \medmuskip=0mu
% \thickmuskip=0mu
% \spaceskip=0pt
    \begin{aligned}
        & \eta \langle \hat{\bm{m}}^{t}_i(I), \sigma_i^{*,\mu,\gamma,\bm{r}}(I) - \bm{\theta}^{t}_I \rangle - \eta^2 \frac{\Vert \hat{\bm{m}}^{t}_i(I) - \hat{\bm{m}}^{*,\mu,\gamma,\bm{r}}_i(I) \Vert^2_2}{2} \\
        \leq  & D_{\psi}(\sigma_i^{*,\mu,\gamma,\bm{r}}(I), {\bm{\theta}}^{t}_I) + \eta \langle - \hat{\bm{m}}^{*,\mu,\gamma,\bm{r}}_i(I), {\bm{\theta}}^{t}_I \rangle
        - D_{\psi}(\sigma_i^{*,\mu,\gamma,\bm{r}}(I), {\bm{\theta}}^{t+1}_I) - \eta \langle - \hat{\bm{m}}^{*,\mu,\gamma,\bm{r}}_i(I), {\bm{\theta}}^{t+1}_I \rangle. 
    \end{aligned}
\end{equation}
According to the definition of $\hat{\bm{m}}^{t}_i(I)$, we have
\begin{equation}\label{eq:appendix-A-4}
\small
% \setlength\abovedisplayskip{2pt}
% \setlength\belowdisplayskip{2pt}
% \thinmuskip=0mu
% \medmuskip=0mu
% \thickmuskip=0mu
% \spaceskip=0pt
    \begin{aligned}
         \langle \hat{\bm{m}}^{t}_i(I), \sigma_i^{*,\mu,\gamma,\bm{r}}(I) - {\bm{\theta}}^{t}_I \rangle  = & \langle \hat{\bm{v}}^{t}_i(I) - \langle \hat{\bm{v}}^{t}_i(I), \sigma^{t}_i(I)\rangle \bm{1}, \sigma_i^{*,\mu,\gamma,\bm{r}}(I) - {\bm{\theta}}^{t}_I \rangle \\
         = & \langle -\hat{\bm{v}}^{t}_i(I),\sigma^{t}_i(I) - \sigma_i^{*,\mu,\gamma,\bm{r}}(I)\rangle,
    \end{aligned}
\end{equation}
where the second equality comes from the fact that
\begin{equation}\label{eq:appendix-A-4-1}
\small
% \setlength\abovedisplayskip{2pt}
% \setlength\belowdisplayskip{2pt}
% \thinmuskip=0mu
% \medmuskip=0mu
% \thickmuskip=0mu
% \spaceskip=0pt
\begin{aligned}
     \langle \hat{\bm{v}}^{t}_i(I) - \langle \hat{\bm{v}}^{t}_i(I), \sigma^{t}_i(I)\rangle \bm{1}, {\bm{\theta}}^{t}_I \rangle = \langle \hat{\bm{v}}^{t}_i(I) - \langle \hat{\bm{v}}^{t}_i(I), \frac{{\bm{\theta}}^{t}_I}{\langle {\bm{\theta}}^{t}_I, \bm{1} \rangle}\rangle \bm{1}, {\bm{\theta}}^{t}_I \rangle = 0.
\end{aligned}
\end{equation}
Therefore, we have
\begin{equation}\label{eq:appendix-A-5}
\small
% \setlength\abovedisplayskip{2pt}
% \setlength\belowdisplayskip{2pt}
% \thinmuskip=0mu
% \medmuskip=0mu
% \thickmuskip=0mu
% \spaceskip=0pt
    \begin{aligned}
        & \eta \langle -\hat{\bm{v}}^{t}_i(I),\sigma^{t}_i(I) - \sigma_i^{*,\mu,\gamma,\bm{r}}(I)\rangle - \eta^2 \frac{\Vert \hat{\bm{m}}^{t}_i(I) - \hat{\bm{m}}^{*,\mu,\gamma,\bm{r}}_i(I) \Vert^2_2}{2} \\
        \leq  & D_{\psi}(\sigma_i^{*,\mu,\gamma,\bm{r}}(I), {\bm{\theta}}^{t}_I) + \eta \langle - \hat{\bm{m}}^{*,\mu,\gamma,\bm{r}}_i(I), {\bm{\theta}}^{t}_I \rangle
        - D_{\psi}(\sigma_i^{*,\mu,\gamma,\bm{r}}(I), {\bm{\theta}}^{t+1}_I) - \eta \langle - \hat{\bm{m}}^{*,\mu,\gamma,\bm{r}}_i(I), {\bm{\theta}}^{t+1}_I \rangle. 
    \end{aligned}
\end{equation}
Continuing from (\ref{eq:appendix-A-5}), we get
\begin{equation}\label{eq:appendix-A-5-1}
\small
% \setlength\abovedisplayskip{2pt}
% \setlength\belowdisplayskip{2pt}
% \thinmuskip=0mu
% \medmuskip=0mu
% \thickmuskip=0mu
% \spaceskip=0pt
    \begin{aligned}
        & \eta \beta_I \langle -\hat{\bm{v}}^{t}_i(I),(1 - \alpha_I )\sigma^{t}_i(I) - (1 - \alpha_I )\sigma_i^{*,\mu,\gamma,\bm{r}} (I) \rangle - \eta^2 (1 - \alpha_I )\beta_I\frac{\Vert \hat{\bm{m}}^{t}_i(I) - \hat{\bm{m}}^{*,\mu,\gamma,\bm{r}}_i(I) \Vert^2_2}{2}\\
        & \leq (1 - \alpha_I )\beta_I \left( D_{\psi}(\sigma_i^{*,\mu,\gamma,\bm{r}}(I), {\bm{\theta}}^{t}_I) + \eta \langle - \hat{\bm{m}}^{*,\mu,\gamma,\bm{r}}_i(I), {\bm{\theta}}^{t}_I \rangle
        - D_{\psi}(\sigma_i^{*,\mu,\gamma,\bm{r}}(I), {\bm{\theta}}^{t+1}_I) - \eta \langle - \hat{\bm{m}}^{*,\mu,\gamma,\bm{r}}_i(I), {\bm{\theta}}^{t+1}_I \rangle \right) \\
        \Leftrightarrow & 
        \eta \beta_I \langle -\hat{\bm{v}}^{t}_i(I),(1 - \alpha_I )\sigma^{t}_i(I) + \gamma \bm{1} - (1 - \alpha_I )\sigma_i^{*,\mu,\gamma,\bm{r}} (I) - \gamma \bm{1}\rangle - \eta^2(1 - \alpha_I )\beta_I \frac{\Vert \hat{\bm{m}}^{t}_i(I) - \hat{\bm{m}}^{*,\mu,\gamma,\bm{r}}_i(I) \Vert^2_2}{2}\\
        & \leq (1 - \alpha_I )\beta_I \left( D_{\psi}(\sigma_i^{*,\mu,\gamma,\bm{r}}(I), {\bm{\theta}}^{t}_I) + \eta \langle - \hat{\bm{m}}^{*,\mu,\gamma,\bm{r}}_i(I), {\bm{\theta}}^{t}_I \rangle
        - D_{\psi}(\sigma_i^{*,\mu,\gamma,\bm{r}}(I), {\bm{\theta}}^{t+1}_I) - \eta \langle - \hat{\bm{m}}^{*,\mu,\gamma,\bm{r}}_i(I), {\bm{\theta}}^{t+1}_I \rangle \right) \\
        \Leftrightarrow & 
        \eta \beta_I \langle -\hat{\bm{v}}^{t}_i(I),\hat{\sigma}^{t}_i(I)  - \hat{\sigma}_i^{*,\mu,\gamma,\bm{r}} (I) \rangle - \eta^2 (1 - \alpha_I )\beta_I \frac{\Vert \hat{\bm{m}}^{t}_i(I) - \hat{\bm{m}}^{*,\mu,\gamma,\bm{r}}_i(I) \Vert^2_2}{2}\\
        & \leq (1 - \alpha_I )\beta_I \left( D_{\psi}(\sigma_i^{*,\mu,\gamma,\bm{r}}(I), {\bm{\theta}}^{t}_I) + \eta \langle - \hat{\bm{m}}^{*,\mu,\gamma,\bm{r}}_i(I), {\bm{\theta}}^{t}_I \rangle
        - D_{\psi}(\sigma_i^{*,\mu,\gamma,\bm{r}}(I), {\bm{\theta}}^{t+1}_I) - \eta \langle - \hat{\bm{m}}^{*,\mu,\gamma,\bm{r}}_i(I), {\bm{\theta}}^{t+1}_I \rangle \right),
    \end{aligned}
\end{equation}
where $\beta_I = {\pi}^{\hat{\sigma}^{*,\mu,\gamma,\bm{r}}}_i(I)$. By applying \Cref{lem:sum of counterfactual regret}, we have
\begin{equation}\label{eq:appendix-A-6}
\small
\thinmuskip=-0.5mu
\medmuskip=-0.5mu
\thickmuskip=-0.5mu
\spaceskip=-0.5pt
    \begin{aligned}
        & 
        \eta \sum_{t=1}^T \sum_{i \in \mathcal{N}} \langle \hat{\bm{\ell}}^{t}_i, \hat{\bm{x}}^{t}_i - \hat{\bm{x}}^{*,\mu,\gamma,\bm{r}}_i \rangle 
        - \sum_{t=1}^T \sum_{i \in \mathcal{N}} \sum_{I \in \mathcal{I}_i} \eta^2 \zeta_I \frac{\Vert \hat{\bm{m}}^{t}_i(I) - \hat{\bm{m}}^{*,\mu,\gamma,\bm{r}}_i(I) \Vert^2_2}{2} \\
        \leq & 
        \sum_{i \in \mathcal{N}} \sum_{I \in \mathcal{I}_i} 
        \zeta_I \left(
         D_{\psi}(\sigma_i^{*,\mu,\gamma,\bm{r}}(I), {\bm{\theta}}^{1}_I) + \eta \langle - \hat{\bm{m}}^{*,\mu,\gamma,\bm{r}}_i(I), {\bm{\theta}}^{1}_I \rangle
        - D_{\psi}(\sigma_i^{*,\mu,\gamma,\bm{r}}(I), {\bm{\theta}}^{T+1}_I) - \eta \langle - \hat{\bm{m}}^{*,\mu,\gamma,\bm{r}}_i(I), {\bm{\theta}}^{T+1}_I \rangle 
        \right),
    \end{aligned}
\end{equation}
where $\zeta_I = (1 - \alpha_I )\beta_I$. Since $0 \leq \zeta_I \leq 1$ (as $0 \leq \beta_I \leq 1$ and $0 \leq \alpha_I \leq 1$), we get
\begin{equation}\label{eq:appendix-A-7}
\small
\thinmuskip=-0.5mu
\medmuskip=-0.5mu
\thickmuskip=-0.5mu
\spaceskip=-0.5pt
    \begin{aligned}
        & 
        \eta \sum_{t=1}^T \sum_{i \in \mathcal{N}} \langle \hat{\bm{\ell}}^{t}_i, \hat{\bm{x}}^{t}_i - \hat{\bm{x}}^{*,\mu,\gamma,\bm{r}}_i \rangle 
        - \sum_{t=1}^T \sum_{i \in \mathcal{N}} \sum_{I \in \mathcal{I}_i} \eta^2 \frac{\Vert \hat{\bm{m}}^{t}_i(I) - \hat{\bm{m}}^{*,\mu,\gamma,\bm{r}}_i(I) \Vert^2_2}{2} \\
        \leq & 
        \sum_{i \in \mathcal{N}} \sum_{I \in \mathcal{I}_i} 
        \zeta_I \left(
         D_{\psi}(\sigma_i^{*,\mu,\gamma,\bm{r}}(I), {\bm{\theta}}^{1}_I) + \eta \langle - \hat{\bm{m}}^{*,\mu,\gamma,\bm{r}}_i(I), {\bm{\theta}}^{1}_I \rangle
        - D_{\psi}(\sigma_i^{*,\mu,\gamma,\bm{r}}(I), {\bm{\theta}}^{T+1}_I) - \eta \langle - \hat{\bm{m}}^{*,\mu,\gamma,\bm{r}}_i(I), {\bm{\theta}}^{T+1}_I \rangle
        \right).
    \end{aligned}
\end{equation}
By applying \Cref{lem:strongly monotone condition}, we obtain
\begin{equation}\label{eq:appendix-A-8}
\small
\thinmuskip=-0.5mu
\medmuskip=-0.5mu
\thickmuskip=-0.5mu
\spaceskip=-0.5pt
    \begin{aligned}
        & 
        \sum_{t=1}^T \mu \eta \Vert \hat{\bm{x}}^{t}- \hat{\bm{x}}^{*,\mu,\gamma,\bm{r}} \Vert^2_2 
        - \sum_{t=1}^T \sum_{i \in \mathcal{N}} \sum_{I \in \mathcal{I}_i} \eta^2 \frac{\Vert \hat{\bm{m}}^{t}_i(I) - \hat{\bm{m}}^{*,\mu,\gamma,\bm{r}}_i(I) \Vert^2_2}{2} 
        \leq  
        \sum_{i \in \mathcal{N}} \sum_{I \in \mathcal{I}_i} 
        \zeta_I \bigg(
         D_{\psi}(\sigma_i^{*,\mu,\gamma,\bm{r}}(I), {\bm{\theta}}^{1}_I) \\ & \quad \quad \quad \quad \quad  + \eta \langle - \hat{\bm{m}}^{*,\mu,\gamma,\bm{r}}_i(I), {\bm{\theta}}^{1}_I \rangle
        - D_{\psi}(\sigma_i^{*,\mu,\gamma,\bm{r}}(I), {\bm{\theta}}^{T+1}_I) - \eta \langle - \hat{\bm{m}}^{*,\mu,\gamma,\bm{r}}_i(I), {\bm{\theta}}^{T+1}_I \rangle 
        \bigg).
    \end{aligned}
\end{equation}
Now, we use the smoothness of the instantaneous counterfactual regrets to transform $\Vert \hat{\bm{m}}^{t}_i(I) - \hat{\bm{m}}^{*,\mu,\gamma,\bm{r}}_i(I) \Vert^2_2$ into $O(\Vert \hat{\bm{x}}^{t}- \hat{\bm{x}}^{*,\mu,\gamma,\bm{r}} \Vert^2_2)$. Formally, for the term $\Vert \hat{\bm{m}}^{t}_i(I) - \hat{\bm{m}}^{*,\mu,\gamma,\bm{r}}_i(I) \Vert^2_2$, from the definition of $\hat{\bm{m}}^{t}_i(I)$ and $\hat{\bm{m}}^{*,\mu,\gamma,\bm{r}}_i(I)$, we have
\begin{equation}\label{eq:appendix-A-9}
    \small
    % \setlength\abovedisplayskip{2pt}
    % \setlength\belowdisplayskip{2pt}
    % \thinmuskip=0mu
    % \medmuskip=0mu
    % \thickmuskip=0mu
    % \spaceskip=0pt
        \begin{aligned}
            & \Vert \hat{\bm{m}}^{t}_i(I) - \hat{\bm{m}}^{*,\mu,\gamma,\bm{r}}_i(I) \Vert^2_2 \\
            = &  \Vert \hat{\bm{v}}^{t}_i(I) - \langle \hat{\bm{v}}^{t}_i(I), \sigma^{t}_i(I)\rangle \bm{1} 
            - \hat{\bm{v}}^{*,\mu,\gamma,\bm{r}}_i(I) + \langle \hat{\bm{v}}^{*,\mu,\gamma,\bm{r}}_i(I), \sigma^{*,\mu,\gamma,\bm{r}}_i(I)\rangle \bm{1}
            \Vert^2_2 \\
            = &  \Vert \hat{\bm{v}}^{t}_i(I) - \hat{\bm{v}}^{*,\mu,\gamma,\bm{r}}_i(I) - 
            \langle \hat{\bm{v}}^{t}_i(I), \sigma^{t}_i(I)\rangle \bm{1} 
             + \langle \hat{\bm{v}}^{*,\mu,\gamma,\bm{r}}_i(I), \sigma^{*,\mu,\gamma,\bm{r}}_i(I)\rangle \bm{1}
            \Vert^2_2 \\
            \leq & 2 \Vert \hat{\bm{v}}^{t}_i(I)  - \hat{\bm{v}}^{*,\mu,\gamma,\bm{r}}_i(I) \Vert^2_2 + 2 |A(I)|^2 \Vert \langle \hat{\bm{v}}^{t}_i(I), \sigma^{t}_i(I)\rangle  - \langle \hat{\bm{v}}^{*,\mu,\gamma,\bm{r}}_i(I), \sigma^{*,\mu,\gamma,\bm{r}}_i(I)\rangle \Vert^2_2 \\
            \leq & 2 \Vert \hat{\bm{v}}^{t}_i(I)  - \hat{\bm{v}}^{*,\mu,\gamma,\bm{r}}_i(I) \Vert^2_2 \\ & + 2 |A(I)|^2 \Vert \langle \hat{\bm{v}}^{t}_i(I), \sigma^{t}_i(I)\rangle  - \langle \hat{\bm{v}}^{t}_i(I), \sigma^{*,\mu,\gamma,\bm{r}}_i(I)\rangle + \langle \hat{\bm{v}}^{t}_i(I), \sigma^{*,\mu,\gamma,\bm{r}}_i(I)\rangle - \langle \hat{\bm{v}}^{*,\mu,\gamma,\bm{r}}_i(I), \sigma^{*,\mu,\gamma,\bm{r}}_i(I)\rangle \Vert^2_2 \\
            \leq & 2 \Vert \hat{\bm{v}}^{t}_i(I)  - \hat{\bm{v}}^{*,\mu,\gamma,\bm{r}}_i(I) \Vert^2_2  + 2 |A(I)|^2 \Vert \langle \hat{\bm{v}}^{t}_i(I), \sigma^{t}_i(I)\rangle  - \langle \hat{\bm{v}}^{t}_i(I), \sigma^{*,\mu,\gamma,\bm{r}}_i(I)\rangle \Vert^2_2 \\ & + 2 |A(I)|^2 \Vert \langle \hat{\bm{v}}^{t}_i(I), \sigma^{*,\mu,\gamma,\bm{r}}_i(I)\rangle - \langle \hat{\bm{v}}^{*,\mu,\gamma,\bm{r}}_i(I), \sigma^{*,\mu,\gamma,\bm{r}}_i(I)\rangle \Vert^2_2, 
        \end{aligned}
    \end{equation}
    where $\sigma_i^{*,\mu,\gamma,\bm{r}}(I) = \frac{\hat{\sigma}_i^{*,\mu,\gamma,\bm{r}}(I) - \gamma \bm{1}}{1 - \alpha_I}$. By using $A_{max}=\max_{I \in \mathcal{I}} |A(I)|$, we have
    \begin{equation}\label{eq:appendix-A-10}
    \small
    % \setlength\abovedisplayskip{2pt}
    % \setlength\belowdisplayskip{2pt}
    % \thinmuskip=0mu
    % \medmuskip=0mu
    % \thickmuskip=0mu
    % \spaceskip=0pt
        \begin{aligned}
            & \Vert \hat{\bm{m}}^{t}_i(I) - \hat{\bm{m}}^{*,\mu,\gamma,\bm{r}}_i(I) \Vert^2_2 \\
            \leq & 2 \Vert \hat{\bm{v}}^{t}_i(I)  - \hat{\bm{v}}^{*,\mu,\gamma,\bm{r}}_i(I) \Vert^2_2  + 2 A_{max}^2 \Vert \langle \hat{\bm{v}}^{t}_i(I), \sigma^{t}_i(I)\rangle  - \langle \hat{\bm{v}}^{t}_i(I), \sigma^{*,\mu,\gamma,\bm{r}}_i(I)\rangle \Vert^2_2 \\ & + 2 A_{max}^2 \Vert \langle \hat{\bm{v}}^{t}_i(I), \sigma^{*,\mu,\gamma,\bm{r}}_i(I)\rangle - \langle \hat{\bm{v}}^{*,\mu,\gamma,\bm{r}}_i(I), \sigma^{*,\mu,\gamma,\bm{r}}_i(I)\rangle \Vert^2_2. 
        \end{aligned}
    \end{equation}
    For the term $\Vert \langle \hat{\bm{v}}^{t}_i(I), \sigma^{t}_i(I)\rangle  - \langle \hat{\bm{v}}^{t}_i(I), \sigma^{*,\mu,\gamma,\bm{r}}_i(I)\rangle \Vert^2_2$ in (\ref{eq:appendix-A-10}), we have
    \begin{equation}\label{eq:appendix-A-11}
    \small
    % \setlength\abovedisplayskip{2pt}
    % \setlength\belowdisplayskip{2pt}
    % \thinmuskip=0mu
    % \medmuskip=0mu
    % \thickmuskip=0mu
    % \spaceskip=0pt
        \begin{aligned}
            & \Vert \langle \hat{\bm{v}}^{t}_i(I), \sigma^{t}_i(I)\rangle  - \langle \hat{\bm{v}}^{t}_i(I), \sigma^{*,\mu,\gamma,\bm{r}}_i(I)\rangle \Vert^2_2 \\
            = & \Vert \langle \hat{\bm{v}}^{t}_i(I), \sigma^{t}_i(I)  -  \sigma^{*,\mu,\gamma,\bm{r}}_i(I)\rangle \Vert^2_2 \\
            \leq & \Vert \hat{\bm{v}}^{t}_i(I) \Vert^2_2 \Vert \sigma^{t}_i(I)  -  \sigma^{*,\mu,\gamma,\bm{r}}_i(I)\rangle \Vert^2_2 \\
            \leq & (P + 2\mu D )^2 \Vert \sigma^{t}_i(I)  -  \sigma^{*,\mu,\gamma,\bm{r}}_i(I)\rangle \Vert^2_2,
        \end{aligned}
    \end{equation}
    where the last line is from \Cref{lem:maximum value of counterfactual value}.
    For the term $\Vert \langle \hat{\bm{v}}^{t}_i(I), \sigma^{*,\mu,\gamma,\bm{r}}_i(I)\rangle - \langle \hat{\bm{v}}^{*,\mu,\gamma,\bm{r}}_i(I), \sigma^{*,\mu,\gamma,\bm{r}}_i(I)\rangle \Vert^2_2$ in (\ref{eq:appendix-A-10}), we get
    \begin{equation}\label{eq:appendix-A-12}
    \small
    % \setlength\abovedisplayskip{2pt}
    % \setlength\belowdisplayskip{2pt}
    % \thinmuskip=0mu
    % \medmuskip=0mu
    % \thickmuskip=0mu
    % \spaceskip=0pt
        \begin{aligned}
            & \Vert \langle \hat{\bm{v}}^{t}_i(I), \sigma^{*,\mu,\gamma,\bm{r}}_i(I)\rangle - \langle \hat{\bm{v}}^{*,\mu,\gamma,\bm{r}}_i(I), \sigma^{*,\mu,\gamma,\bm{r}}_i(I)\rangle \Vert^2_2 \\
            = & \Vert  \hat{\bm{v}}^{t}_i(I) - \hat{\bm{v}}^{*,\mu,\gamma,\bm{r}}_i(I), \sigma^{*,\mu,\gamma,\bm{r}}_i(I)\rangle\Vert^2_2 \\
            \leq & \Vert  \hat{\bm{v}}^{t}_i(I) - \hat{\bm{v}}^{*,\mu,\gamma,\bm{r}}_i(I)\Vert^2_2  \Vert \sigma^{*,\mu,\gamma,\bm{r}}_i(I)\rangle\Vert^2_2 \\
            \leq & \Vert  \hat{\bm{v}}^{t}_i(I) - \hat{\bm{v}}^{*,\mu,\gamma,\bm{r}}_i(I)\Vert^2_2,
        \end{aligned}
    \end{equation}
    where the last inequality comes from $\Vert \sigma^{*,\mu,\gamma,\bm{r}}_i(I)\Vert^2_2 \leq 1$ as $\sigma^{*,\mu,\gamma,\bm{r}}_i(I)$ is in simplex. 
    By substituting (\ref{eq:appendix-A-11}) and (\ref{eq:appendix-A-12}) into (\ref{eq:appendix-A-10}), as well as using $A_{max} \geq 1$, we obtain
    \begin{equation}\label{eq:appendix-A-13}
    \small
    % \setlength\abovedisplayskip{2pt}
    % \setlength\belowdisplayskip{2pt}
    % \thinmuskip=0mu
    % \medmuskip=0mu
    % \thickmuskip=0mu
    % \spaceskip=0pt
        \begin{aligned}
            & \Vert \hat{\bm{m}}^{t}_i(I) - \hat{\bm{m}}^{*,\mu,\gamma,\bm{r}}_i(I) \Vert^2_2 \\
            \leq & 2 \Vert \hat{\bm{v}}^{t}_i(I)  - \hat{\bm{v}}^{*,\mu,\gamma,\bm{r}}_i(I) \Vert^2_2  + 2 A_{max}^2 (P + 2\mu D )^2 \Vert \hat{\bm{v}}^{t}_i(I)  - \hat{\bm{v}}^{*,\mu,\gamma,\bm{r}}_i(I) \Vert^2_2 \\ & + 2 A_{max}^2 \Vert \hat{\bm{v}}^{t}_i(I)  - \hat{\bm{v}}^{*,\mu,\gamma,\bm{r}}_i(I) \Vert^2_2 \\
            \leq & 4 A_{max}^2 \Vert \hat{\bm{v}}^{t}_i(I)  - \hat{\bm{v}}^{*,\mu,\gamma,\bm{r}}_i(I) \Vert^2_2 + 2 A_{max}^2 (P + 2\mu D )^2 \Vert \sigma^{t}_i(I)  -  \sigma^{*,\mu,\gamma,\bm{r}}_i(I)\rangle \Vert^2_2.
        \end{aligned}
    \end{equation}
    By applying \Cref{lem:smoothness of counterfactual value} into (\ref{eq:appendix-A-13}), we get
    \begin{equation}\label{eq:appendix-A-14}
    \small
    % \setlength\abovedisplayskip{2pt}
    % \setlength\belowdisplayskip{2pt}
    % \thinmuskip=0mu
    % \medmuskip=0mu
    % \thickmuskip=0mu
    % \spaceskip=0pt
        \begin{aligned}
            & \Vert \hat{\bm{m}}^{t}_i(I) - \hat{\bm{m}}^{*,\mu,\gamma,\bm{r}}_i(I) \Vert^2_2 \\
            \leq & 8 A_{max}^2 ( L + \mu )^2 \Vert \hat{\bm{x}}^{t} - \hat{\bm{x}}^{*,\mu,\gamma,\bm{r}} \Vert^2_1 + 8 A_{max}^2 (P+2\mu D)^2\Vert \hat{\sigma}^{t}_i  -  \hat{\sigma}^{*,\mu,\gamma,\bm{r}}_i  \Vert^2_1 \\ & + 2 A_{max}^2 (P + 2\mu D )^2 \Vert \sigma^{t}_i(I)  -  \sigma^{*,\mu,\gamma,\bm{r}}_i(I)\rangle \Vert^2_2 \\
            \leq & 8 A_{max}^2 ( L + \mu )^2 \Vert \hat{\bm{x}}^{t} - \hat{\bm{x}}^{*,\mu,\gamma,\bm{r}} \Vert^2_1 + 10 A_{max}^2 (P+2\mu D)^2\Vert \hat{\sigma}^{t}_i  -  \hat{\sigma}^{*,\mu,\gamma,\bm{r}}_i  \Vert^2_1.
        \end{aligned}
    \end{equation}
By applying \Cref{lem:relationship between behavioral strategy and sequence-form strategy} into (\ref{eq:appendix-A-14}), we get
\begin{equation}\label{eq:appendix-A-15}
\small
% \setlength\abovedisplayskip{2pt}
% \setlength\belowdisplayskip{2pt}
% \thinmuskip=0mu
% \medmuskip=0mu
% \thickmuskip=0mu
% \spaceskip=0pt
    \begin{aligned}
        & \Vert \hat{\bm{m}}^{t}_i(I) - \hat{\bm{m}}^{*,\mu,\gamma,\bm{r}}_i(I) \Vert^2_2 \\
         \leq & 8 A_{max}^2 ( L + \mu )^2 \Vert \hat{\bm{x}}^{t} - \hat{\bm{x}}^{*,\mu,\gamma,\bm{r}} \Vert^2_1 + 10 A_{max}^2 (P+2\mu D)^2\frac{4C^2_{max}}{\gamma^{2H}}\Vert \hat{\bm{x}}^{t}  -  \hat{\bm{x}}^{*,\mu,\gamma,\bm{r}}  \Vert^2_1.
    \end{aligned}
\end{equation}
By substituting (\ref{eq:appendix-A-15}) into (\ref{eq:appendix-A-8}), we have
\begin{equation}\label{eq:appendix-A-16}
\small
\thinmuskip=0mu
\medmuskip=0mu
\thickmuskip=0mu
\spaceskip=0pt
    \begin{aligned}
        & 
        \sum_{t=1}^T \mu \eta \Vert \hat{\bm{x}}^{t}- \hat{\bm{x}}^{*,\mu,\gamma,\bm{r}} \Vert^2_2 
        - \sum_{t=1}^T \sum_{i \in \mathcal{N}} \sum_{I \in \mathcal{I}_i} \eta^2 A_{max}^2 \left( 8 ( L + \mu )^2 + 40 (P+2\mu D)^2\frac{C^2_{max}}{\gamma^{2H}} \right)\frac{\Vert \hat{\bm{x}}^{t}- \hat{\bm{x}}^{*,\mu,\gamma,\bm{r}} \Vert^2_2 }{2} \\
        \leq & 
        \sum_{i \in \mathcal{N}} \sum_{I \in \mathcal{I}_i} 
        \zeta_I \left(
         D_{\psi}(\sigma_i^{*,\mu,\gamma,\bm{r}}(I), {\bm{\theta}}^{1}_I) + \eta \langle - \hat{\bm{m}}^{*,\mu,\gamma,\bm{r}}_i(I), {\bm{\theta}}^{1}_I \rangle
        - D_{\psi}(\sigma_i^{*,\mu,\gamma,\bm{r}}(I), {\bm{\theta}}^{T+1}_I) - \eta \langle - \hat{\bm{m}}^{*,\mu,\gamma,\bm{r}}_i(I), {\bm{\theta}}^{T+1}_I \rangle 
        \right),
    \end{aligned}
\end{equation}
which implies
\begin{equation}\label{eq:appendix-A-17}
\small
\thinmuskip=0mu
\medmuskip=0mu
\thickmuskip=0mu
\spaceskip=0pt
    \begin{aligned}
        & 
        \sum_{t=1}^T \mu \eta \Vert \hat{\bm{x}}^{t}- \hat{\bm{x}}^{*,\mu,\gamma,\bm{r}} \Vert^2_2 
        - \sum_{t=1}^T  \eta^2 |\mathcal{I}| A_{max}^2 \left( 4 ( L + \mu )^2 + 20 (P+2\mu D)^2\frac{C^2_{max}}{\gamma^{2H}} \right)\Vert \hat{\bm{x}}^{t}- \hat{\bm{x}}^{*,\mu,\gamma,\bm{r}} \Vert^2_2  \\
        \leq & 
        \sum_{i \in \mathcal{N}} \sum_{I \in \mathcal{I}_i} 
        \zeta_I \left(
         D_{\psi}(\sigma_i^{*,\mu,\gamma,\bm{r}}(I), {\bm{\theta}}^{1}_I) + \eta \langle - \hat{\bm{m}}^{*,\mu,\gamma,\bm{r}}_i(I), {\bm{\theta}}^{1}_I \rangle
        - D_{\psi}(\sigma_i^{*,\mu,\gamma,\bm{r}}(I), {\bm{\theta}}^{T+1}_I) - \eta \langle - \hat{\bm{m}}^{*,\mu,\gamma,\bm{r}}_i(I), {\bm{\theta}}^{T+1}_I \rangle 
        \right),
    \end{aligned}
\end{equation}
Obviously, if 
\begin{equation}\label{eq:appendix-A-18}
\small
% \setlength\abovedisplayskip{2pt}
% \setlength\belowdisplayskip{2pt}
% \thinmuskip=0mu
% \medmuskip=0mu
% \thickmuskip=0mu
% \spaceskip=0pt
    \begin{aligned}
        & 
        \mu \geq 2 \eta |\mathcal{I}| A_{max}^2  \left( 4 ( L + \mu )^2 + 20 (P+2\mu D)^2\frac{C^2_{max}}{\gamma^{2H}} \right)  > 0\\
        \Leftrightarrow &
        0< \eta \leq \frac{\mu \gamma^{2H}}{2|\mathcal{I}| A_{max}^2  \bigg( 4 \gamma^{2H} ( L + \mu )^2 + 20 C^2_{max} (P+2\mu D)^2 \bigg)},
    \end{aligned}
\end{equation}
we have
\begin{equation}\label{eq:appendix-A-19}
\small
\thinmuskip=0mu
\medmuskip=0mu
\thickmuskip=0mu
\spaceskip=0pt
    \begin{aligned}
        & 
        \sum_{t=1}^T \frac{\mu\eta}{2} \Vert \hat{\bm{x}}^{t}- \hat{\bm{x}}^{*,\mu,\gamma,\bm{r}} \Vert^2_2  \\
        \leq & 
        \sum_{i \in \mathcal{N}} \sum_{I \in \mathcal{I}_i} 
        \zeta_I \left(
         D_{\psi}(\sigma_i^{*,\mu,\gamma,\bm{r}}(I), {\bm{\theta}}^{1}_I) + \eta \langle - \hat{\bm{m}}^{*,\mu,\gamma,\bm{r}}_i(I), {\bm{\theta}}^{1}_I \rangle
        - D_{\psi}(\sigma_i^{*,\mu,\gamma,\bm{r}}(I), {\bm{\theta}}^{T+1}_I) - \eta \langle - \hat{\bm{m}}^{*,\mu,\gamma,\bm{r}}_i(I), {\bm{\theta}}^{T+1}_I \rangle 
        \right),
    \end{aligned}
\end{equation}
By using \Cref{lem:add term is positive}, we have that $\eta \langle - \hat{\bm{m}}^{*,\mu,\gamma,\bm{r}}_i(I), {\bm{\theta}}^{T+1}_I \rangle \geq 0$. As a result, we get $- D_{\psi}(\sigma_i^{*,\mu,\gamma,\bm{r}}(I), {\bm{\theta}}^{T+1}_I) - \eta \langle - \hat{\bm{m}}^{*,\mu,\gamma,\bm{r}}_i(I), {\bm{\theta}}^{T+1}_I \rangle \leq 0$. Then, we conclude that $\forall T \geq 1$
\begin{equation}\label{eq:appendix-A-19}
\small
\thinmuskip=0mu
\medmuskip=0mu
\thickmuskip=0mu
\spaceskip=0pt
    \begin{aligned}
        & 
        \sum_{t=1}^T \frac{\mu\eta}{2} \Vert \hat{\bm{x}}^{t}- \hat{\bm{x}}^{*,\mu,\gamma,\bm{r}} \Vert^2_2 
        \leq \sum_{i \in \mathcal{N}} \sum_{I \in \mathcal{I}_i} 
        \zeta_I \left(
         D_{\psi}(\sigma_i^{*,\mu,\gamma,\bm{r}}(I), {\bm{\theta}}^{1}_I) + \eta \langle - \hat{\bm{m}}^{*,\mu,\gamma,\bm{r}}_i(I), {\bm{\theta}}^{1}_I \rangle
        \right) \\
        \Leftrightarrow &
        \sum_{t=1}^T \Vert \hat{\bm{x}}^{t}-\hat{\bm{x}}^{*,\mu,\gamma,\bm{r}}\Vert^2_2 \leq O(1),
    \end{aligned}
\end{equation}
which implies the asymptotic last-iterate convergence of the sequence $\{ \hat{\bm{x}}^{1}, \hat{\bm{x}}^{2}, \cdots, \hat{\bm{x}}^{t}, \cdots\}$ to NE $\hat{\bm{x}}^{*,\mu,\gamma,\bm{r}}$ of the perturbed regularized EFG since $0 \leq \zeta_I \leq 1$ (as mentioned above). 

As analyzed in \citet{farina2021faster}, if $\bm{\theta}^{1}_I = \bm{0}$, for any $\eta > 0$, the generated sequence $\{ \hat{\bm{x}}^{1}, \hat{\bm{x}}^{2}, \dots, \hat{\bm{x}}^{t}, \dots \}$ remains identical, achieving the parameter-free property. In this paper, we further establish that for any initial $\bm{\theta}^{1}_I \in \mathbb{R}^{|A(I)|}_{\geq 0}$ and $\eta > 0$, the sequence $\hat{\bm{x}}^{t}$ converges to $\hat{\bm{x}}^{*,\mu,\gamma,\bm{r}}$.

We first prove that for the accumulated counterfactual regret sequence $\{ \bm{\theta}^{1}_I, \bm{\theta}^{2}_I, \dots, \bm{\theta}^{t}_I, \dots \}$ generated by $\bm{\theta}^{1}_I \in \mathbb{R}^{|A(I)|}_{\geq 0}$ and $\eta > 0$, there exists a corresponding sequence $\{ {\bm{\theta}^{1}_I}^{\prime}, {\bm{\theta}^{2}_I}^{\prime}, \dots, {\bm{\theta}^{t}_I}^{\prime}, \dots \}$ generated by ${\bm{\theta}^{1}_I}^{\prime} \in \mathbb{R}^{|A(I)|}_{\geq 0}$ and $\eta^{\prime} = \mu / (2 C_0)$, such that the resulting strategy profile sequence $\{ \hat{\bm{x}}^{1}, \hat{\bm{x}}^{2}, \dots, \hat{\bm{x}}^{t}, \dots \}$ is identical. By the update rule of CFR$^+$ defined in (\ref{eq:update-rule-MPCFR}) and the analysis in \citet{farina2021faster}, ${\bm{\theta}}^{t+1}_I \in \argmin_{{\bm{\theta}}_I \in \mathbb{R}^{|A(I)|}_{\geq 0}} \left\{  \langle - \hat{\bm{m}}^{t}_i(I), \bm{\theta}_I\rangle + \frac{1}{\eta} D_{\psi} (\bm{\theta}_I, {\bm{\theta}}^{t}_I) \right\}$ can be expressed as the projection ${\bm{\theta}}^{t+1}_I = [{\bm{\theta}}^{t}_I+\eta \hat{\bm{m}}^{t}_i(I)]^{+}$, where $[\cdot]^{+} = \max(\cdot,\bm{0})$. Setting ${\bm{\theta}^{t}_I}^{\prime} = \eta^{\prime} {\bm{\theta}^{t}_I}/\eta$ for $t \geq 1$, it follows that ${\bm{\theta}^{t+1}_I}^{\prime} = [{\bm{\theta}^{t}_I}^{\prime}+\eta^{\prime} \hat{\bm{m}}^{t}_i(I)]^{+}$ and $\sigma^{t}_i(I) = {{\bm{\theta}}^{t}_I}/{\langle {\bm{\theta}}^{t}_I, \bm{1} \rangle} = {{\bm{\theta}^{t}_I}^{\prime}}/{\langle {\bm{\theta}^{t}_I}^{\prime}, \bm{1} \rangle}$ hold. Furthermore, it is evident that $\sum_{t=1}^T \Vert \hat{\bm{x}}^{t} - \hat{\bm{x}}^{*,\mu,\gamma,\bm{r}} \Vert^2_2 \leq O(1)$ holds independently of the value of $\bm{\theta}^{1}_I$. 

Based on the above analysis, we conclude that (i) for any accumulated counterfactual regret sequence $\{ \bm{\theta}^{1}_I, \bm{\theta}^{2}_I, \dots, \bm{\theta}^{t}_I, \dots \}$ generated by any $\bm{\theta}^{1}_I \in \mathbb{R}^{|A(I)|}_{\geq 0}$ and $\eta > 0$, there exists a corresponding accumulated counterfactual regret sequence $\{ {\bm{\theta}^{1}_I}^{\prime}, {\bm{\theta}^{2}_I}^{\prime}, \dots, {\bm{\theta}^{t}_I}^{\prime}, \dots \}$ generated by ${\bm{\theta}^{1}_I}^{\prime}$ and $\eta^{\prime} = \mu / (2 C_0)$, such that the resulting strategy profile sequence $\{ \hat{\bm{x}}^{1}, \hat{\bm{x}}^{2}, \dots, \hat{\bm{x}}^{t}, \dots \}$ are identical, as well as (ii) the strategy profile sequence $\{ \hat{\bm{x}}^{1}, \hat{\bm{x}}^{2}, \dots, \hat{\bm{x}}^{t}, \dots \}$ generated by the accumulated counterfactual regret sequence $\{ {\bm{\theta}^{1}_I}^{\prime}, {\bm{\theta}^{2}_I}^{\prime}, \dots, {\bm{\theta}^{t}_I}^{\prime}, \dots \}$ converges to $\hat{\bm{x}}^{*,\mu,\gamma,\bm{r}}$. Therefore, we have that for any $\bm{\theta}^{1}_I \in \mathbb{R}^{|A(I)|}_{\geq 0}$ and $\eta > 0$, the generated strategy profile sequence $\{ \hat{\bm{x}}^{1}, \hat{\bm{x}}^{2}, \dots, \hat{\bm{x}}^{t}, \dots \}$ converges to $\hat{\bm{x}}^{*,\mu,\gamma,\bm{r}}$, demonstrating the parameter-free property. We complete the proof.
\end{proof}

\clearpage
\newpage
\section{Proof of Useful Lemmas}\label{sec:Proof of Useful Lemmas}

\subsection{Proof of \Cref{lem:sum of counterfactual regret}}\label{subsec:proof:lem:sum of counterfactual regret}

\begin{proof}
    From the definition of $\sum_{I \in \mathcal{I}_i} \pi^{\sigma^{\prime}}_i(I) \langle -\bm{v}^{\sigma}_i(I,a), \sigma_i(I) - \sigma^{\prime}_i(I) \rangle$, we get
    \begin{equation}\label{eq:0:lem:geq 0 of counterfactual regret}
    \small
    \begin{aligned}
         & \sum_{I \in \mathcal{I}_i} \pi^{\sigma^{\prime}}_i(I) \langle -\bm{v}^{\sigma}_i(I,a), \sigma_i(I) - \sigma^{\prime}_i(I) \rangle\\
         =&\sum_{I \in \mathcal{I}_i }  \pi^{\sigma^{\prime}}_i(I) \langle -\bm{v}^{\sigma}_i(I,a), \sigma_i(I)\rangle - \sum_{I \in \mathcal{I}_i }  \pi^{\sigma^{\prime}}_i(I) \langle -\bm{v}^{\sigma}_i(I,a),\sigma^{\prime}_i(I)\rangle.
    \end{aligned}
    \end{equation}
    For the term $\sum_{I \in \mathcal{I}_i }  \pi^{\sigma^{\prime}}_i(I) \langle -\bm{v}^{\sigma}_i(I,a),\sigma^{\prime}_i(I)\rangle$, we have
    \begin{equation}\label{eq:1:lem:geq 0 of counterfactual regret}
    \small
    \begin{aligned}
         & \sum_{I \in \mathcal{I}_i }  \pi^{\sigma^{\prime}}_i(I) \langle -\bm{v}^{\sigma}_i(I,a),\sigma^{\prime}_i(I)\rangle\\
         =&\sum_{I \in \mathcal{I}_i }  \pi^{\sigma^{\prime}}_i(I) \sum_{a \in A(I)} \sigma^{\prime}_i(I,a)\left( \bm{\ell}_i(I,a)+\sum_{I^{\prime} \in C_i(I,a)} \langle  -\bm{v}^{\sigma}_i(I^{\prime}), \sigma_i(I^{\prime}) \rangle\right) \\
         =&\sum_{I \in \mathcal{I}_i }   \sum_{a \in A(I)} \pi^{\sigma^{\prime}}_i(I) \sigma^{\prime}_i(I,a) \bm{\ell}_i(I,a) + \sum_{I \in \mathcal{I}_i } \sum_{a \in A(I)} \pi^{\sigma^{\prime}}_i(I) \sigma^{\prime}_i(I,a) \sum_{I^{\prime} \in C_i(I,a)} \langle  -\bm{v}^{\sigma}_i(I^{\prime}), \sigma_i(I^{\prime}) \rangle.
    \end{aligned}
    \end{equation}
    Then, by substituting (\ref{eq:1:lem:geq 0 of counterfactual regret}) into (\ref{eq:0:lem:geq 0 of counterfactual regret}), we have
    \begin{equation}\label{eq:2:lem:geq 0 of counterfactual regret}
    \small
    \setlength\abovedisplayskip{2pt}
    \setlength\belowdisplayskip{2pt}
    \thinmuskip=0mu
    \medmuskip=0mu
    \thickmuskip=0mu
    \spaceskip=0pt
    \begin{aligned}
         & \sum_{I \in \mathcal{I}_i} \pi^{\sigma^{\prime}}_i(I) \langle -\bm{v}^{\sigma}_i(I,a), \sigma_i(I) - \sigma^{\prime}_i(I) \rangle \\
         =&\sum_{I \in \mathcal{I}_i }  \pi^{\sigma^{\prime}}_i(I) \langle -\bm{v}^{\sigma}_i(I,a), \sigma_i(I)\rangle -\sum_{I \in \mathcal{I}_i }   \sum_{a \in A(I)} \pi^{\sigma^{\prime}}_i(I) \sigma^{\prime}_i(I,a) \bm{\ell}_i(I,a) -  \sum_{I \in \mathcal{I}_i } \sum_{a \in A(I)} \pi^{\sigma^{\prime}}_i(I) \sigma^{\prime}_i(I,a) \sum_{I^{\prime} \in C_i(I,a)} \langle  -\bm{v}^{\sigma}_i(I^{\prime}), \sigma_i(I^{\prime}) \rangle\\
         =&\sum_{I \in \mathcal{I}_i }  \pi^{\sigma^{\prime}}_i(I) \langle -\bm{v}^{\sigma}_i(I,a), \sigma_i(I)\rangle -\sum_{I \in \mathcal{I}_i }   \sum_{a \in A(I)} \pi^{\sigma^{\prime}}_i(I) \sigma^{\prime}_i(I,a) \bm{\ell}_i(I,a) -  \sum_{I \in \mathcal{I}_i } \sum_{a \in A(I)}  \sum_{I^{\prime} \in C_i(I,a)} \pi^{\sigma^{\prime}}_i(I^{\prime}) \langle  -\bm{v}^{\sigma}_i(I^{\prime}), \sigma_i(I^{\prime}) \rangle.
    \end{aligned}
    \end{equation}
    We denote the initial infosets as $\mathcal{I}_i^{init}$, \ie, for any $I \in \mathcal{I}_i^{init}$, there does not exist $I^{\prime\prime} \in \mathcal{I}_i$ such that $I \in C_i(I^{\prime\prime},a^{\prime\prime})$ holds for a $a^{\prime\prime} \in A(I^{\prime\prime})$. 
    For the term $\sum_{I \in \mathcal{I}_i} \pi^{\sigma^{\prime}}_i(I) \langle -\bm{v}^{\sigma}_i(I,a), \sigma_i(I)\rangle - \sum_{I \in \mathcal{I}_i } \sum_{a \in A(I)}  \sum_{I^{\prime} \in C_i(I,a)} \pi^{\sigma^{\prime}}_i(I^{\prime}) \langle  -\bm{v}^{\sigma}_i(I^{\prime}), \sigma_i(I^{\prime}) \rangle$ in (\ref{eq:2:lem:geq 0 of counterfactual regret}), it follows that
    \begin{equation}\label{eq:3:lem:geq 0 of counterfactual regret}
    \small
    \setlength\abovedisplayskip{2pt}
    \setlength\belowdisplayskip{2pt}
    \thinmuskip=0mu
    \medmuskip=0mu
    \thickmuskip=0mu
    \spaceskip=0pt
    \begin{aligned}
    & \sum_{I \in \mathcal{I}_i} \pi^{\sigma^{\prime}}_i(I) \langle -\bm{v}^{\sigma}_i(I,a), \sigma_i(I)\rangle - \sum_{I \in \mathcal{I}_i } \sum_{a \in A(I)}  \sum_{I^{\prime} \in C_i(I,a)} \pi^{\sigma^{\prime}}_i(I^{\prime}) \langle  -\bm{v}^{\sigma}_i(I^{\prime}), \sigma_i(I^{\prime}) \rangle \\
    =& \sum_{I \in \mathcal{I}_i^{init}} \pi^{\sigma^{\prime}}_i(I) \langle -\bm{v}^{\sigma}_i(I,a), \sigma_i(I)\rangle.
    \end{aligned}
    \end{equation}
    Since the probability of reaching any $I \in \mathcal{I}_i^{init}$ is always 1, regardless of the strategies $\sigma$ or $\sigma^{\prime}$, we have that $\forall \sigma, \sigma^{\prime}$, and $I \in \mathcal{I}_i^{init}$, $\pi^{\sigma^{\prime}}_i(I) = \pi^{\sigma}_i(I)$. Substituting this into (\ref{eq:3:lem:geq 0 of counterfactual regret}), we obtain
    \begin{equation}\label{eq:4:lem:geq 0 of counterfactual regret}
    \small
    \setlength\abovedisplayskip{2pt}
    \setlength\belowdisplayskip{2pt}
    \thinmuskip=0mu
    \medmuskip=0mu
    \thickmuskip=0mu
    \spaceskip=0pt
    \begin{aligned}
    & \sum_{I \in \mathcal{I}_i^{init}} \pi^{\sigma^{\prime}}_i(I) \langle -\bm{v}^{\sigma}_i(I,a), \sigma_i(I)\rangle \\
    =& \sum_{I \in \mathcal{I}_i^{init}} \pi^{\sigma}_i(I) \langle -\bm{v}^{\sigma}_i(I,a), \sigma_i(I)\rangle \\
    =&\sum_{I \in \mathcal{I}_i^{init}}  \pi^{\sigma}_i(I) \sum_{a \in A(I)} \sigma_i(I,a)\left( \bm{\ell}_i(I,a)+\sum_{I^{\prime} \in C_i(I,a)} \langle  -\bm{v}^{\sigma}_i(I^{\prime}), \sigma_i(I^{\prime}) \rangle\right) \\
    =&\sum_{I \in \mathcal{I}_i}  \sum_{a \in A(I)} \pi^{\sigma}_i(I) \sigma_i(I,a) \bm{\ell}_i(I,a) ,
    \end{aligned}
    \end{equation}
    where the last line follows from the recursion. Substituting (\ref{eq:4:lem:geq 0 of counterfactual regret}) into (\ref{eq:2:lem:geq 0 of counterfactual regret}), we obtain
    \begin{equation}\label{eq:5:lem:geq 0 of counterfactual regret}
    \small
    \setlength\abovedisplayskip{2pt}
    \setlength\belowdisplayskip{2pt}
    \thinmuskip=0mu
    \medmuskip=0mu
    \thickmuskip=0mu
    \spaceskip=0pt
    \begin{aligned}
         & \sum_{I \in \mathcal{I}_i} \pi^{\sigma^{\prime}}_i(I) \langle -\bm{v}^{\sigma}_i(I,a), \sigma_i(I) - \sigma^{\prime}_i(I) \rangle \\
         =&\sum_{I \in \mathcal{I}_i }   \sum_{a \in A(I)} \left[ \pi^{\sigma}_i(I) \sigma_i(I,a) \bm{\ell}_i(I,a) - \pi^{\sigma^{\prime}}_i(I) \sigma^{\prime}_i(I,a) \bm{\ell}_i(I,a) \right] 
         =\langle \bm{\ell}_i , \bm{x}_i - \bm{x}^{\prime}_i\rangle.
    \end{aligned}
    \end{equation}
    It finishes the proof.
\end{proof}

\subsection{Proof of \Cref{lem:add term is positive}}\label{subsec:proof:lem:add term is positive}

\begin{proof}
First, when $\bm{\theta}_I = \bm{0}$, we have that $\forall I \in \mathcal{I}_i$, $\langle - \hat{\bm{m}}^{*,\mu,\gamma,\bm{r}}_i(I), {\bm{\theta}}_I \rangle = 0$.

Next, we prove by contradiction that when $\bm{\theta}_I > \bm{0}$, $\forall I \in \mathcal{I}_i$, it holds that $\langle - \hat{\bm{m}}^{*,\mu,\gamma,\bm{r}}_i(I), {\bm{\theta}}_I \rangle \geq 0$. 

Suppose there exists one $I^{\prime} \in \mathcal{I}_i$ and $\bm{\theta}^{\prime}_{I^{\prime}} > \bm{0}$ such that $\langle - \hat{\bm{m}}^{*,\mu,\gamma,\bm{r}}(I^{\prime}), \bm{\theta}^{\prime}_{I^{\prime}} \rangle < 0$. We construct a new strategy ${\sigma}_i^{\prime}$, which matches ${\sigma}_i^{*,\mu,\gamma,\bm{r}}$ (not $\hat{\sigma}_i^{*,\mu,\gamma,\bm{r}}$) except at the infoset $I^{\prime}$, where it is defined as $\bm{\theta}^{\prime}_{I^{\prime}} / \langle \bm{\theta}^{\prime}_{I^{\prime}}, \bm{1}\rangle$. For $\langle - \hat{\bm{m}}^{*,\mu,\gamma,\bm{r}}(I^{\prime}), \bm{\theta}^{\prime}_{I^{\prime}} \rangle$, we have
\begin{equation}\label{eq:appendix-A5-0}
    \small
    % \setlength\abovedisplayskip{2pt}
    % \setlength\belowdisplayskip{2pt}
    % \thinmuskip=0mu
    % \medmuskip=0mu
    % \thickmuskip=0mu
    % \spaceskip=0pt
    \begin{aligned}
    \langle - \hat{\bm{m}}^{*,\mu,\gamma,\bm{r}}(I^{\prime}), \bm{\theta}^{\prime}_{I^{\prime}} \rangle 
    = & - \langle \hat{\bm{v}}^{*,\mu,\gamma,\bm{r}}_i(I^{\prime}) - \langle \hat{\bm{v}}^{*,\mu,\gamma,\bm{r}}_i(I^{\prime}), \sigma^{*,\mu,\gamma,\bm{r}}_i(I^{\prime})\rangle \bm{1}, \bm{\theta}^{\prime}_{I^{\prime}} \rangle \\
    = & - \Vert \bm{\theta}^{\prime}_{I^{\prime}} \Vert_1 \langle \hat{\bm{v}}^{*,\mu,\gamma,\bm{r}}_i(I^{\prime}), {\sigma}_i^{\prime}(I^{\prime}) - \sigma^{*,\mu,\gamma,\bm{r}}_i(I^{\prime})\rangle \\
    = & - \Vert \bm{\theta}^{\prime}_{I^{\prime}} \Vert_1 \langle -\hat{\bm{v}}^{*,\mu,\gamma,\bm{r}}_i(I^{\prime}),\sigma^{*,\mu,\gamma,\bm{r}}_i(I^{\prime}) - {\sigma}_i^{\prime}(I^{\prime})\rangle.
\end{aligned}
\end{equation}
Since $\langle - \hat{\bm{m}}^{*,\mu,\gamma,\bm{r}}(I^{\prime}), \bm{\theta}^{\prime}_{I^{\prime}} \rangle < 0$ and $\Vert \bm{\theta}^{\prime}_{I^{\prime}} \Vert_1 > 0$, we have $\langle -\hat{\bm{v}}^{*,\mu,\gamma,\bm{r}}_i(I^{\prime}),\sigma^{*,\mu,\gamma,\bm{r}}_i(I^{\prime}) - {\sigma}_i^{\prime}(I^{\prime})\rangle > 0$. 

We define $\hat{\sigma}_i^{\prime}(I) = (1 - \alpha_I){\sigma}_i^{\prime}(I) + \gamma \bm{1}$ for all $I \in \mathcal{I}_i$. Additionally, we know that $\hat{\sigma}_i^{*,\mu,\gamma,\bm{r}}(I) = (1 - \alpha_I){\sigma}_i^{*,\mu,\gamma,\bm{r}}(I) + \gamma \bm{1}$ for all $I \in \mathcal{I}_i$, and that $\langle -\hat{\bm{v}}^{*,\mu,\gamma,\bm{r}}_i(I^{\prime}), \sigma^{*,\mu,\gamma,\bm{r}}_i(I^{\prime}) - {\sigma}_i^{\prime}(I^{\prime}) \rangle > 0$. Hence, it follows that $\langle -\hat{\bm{v}}^{*,\mu,\gamma,\bm{r}}_i(I^{\prime}), \hat{\sigma}_i^{*,\mu,\gamma,\bm{r}}(I^{\prime}) - \hat{\sigma}_i^{\prime}(I^{\prime}) \rangle > 0$. 

The correspond sequence-form strategy of $\hat{\sigma}_i^{\prime}$ is represented by $\hat{\bm{x}}_i^{\prime}$. According to \Cref{lem:sum of counterfactual regret} and the definition of NE, we get
\begin{equation}\label{eq:appendix-A5-1}
    \small
    % \setlength\abovedisplayskip{2pt}
    % \setlength\belowdisplayskip{2pt}
    % \thinmuskip=0mu
    % \medmuskip=0mu
    % \thickmuskip=0mu
    % \spaceskip=0pt
    \begin{aligned}
    \langle \hat{\bm{\ell}}^{\bm{x}^{*,\mu,\gamma,\bm{r}}}_i, \hat{\bm{x}}^{*,\mu,\gamma,\bm{r}}_i - \hat{\bm{x}}_i^{\prime}\rangle = \sum_{I \in \mathcal{I}_i} \pi^{\sigma^{\prime}}_i(I) \langle -\hat{\bm{v}}^{*,\mu,\gamma,\bm{r}}_i(I), \hat{\sigma}_i^{*,\mu,\gamma,\bm{r}}(I) - \hat{\sigma}_i^{\prime}(I) \rangle >0.
\end{aligned}
\end{equation}
Since ${\sigma}_i^{\prime}$ matches ${\sigma}_i^{*,\mu,\gamma,\bm{r}}$ except at the infoset $I^{\prime}$, and given that $\hat{\sigma}_i^{\prime}(I) = (1 - \alpha_I){\sigma}_i^{\prime}(I) + \gamma \bm{1}$ for all $I \in \mathcal{I}_i$, as well as $\hat{\sigma}_i^{*,\mu,\gamma,\bm{r}}(I) = (1 - \alpha_I){\sigma}_i^{*,\mu,\gamma,\bm{r}}(I) + \gamma \bm{1}$ for all $I \in \mathcal{I}_i$, we obtain $\hat{\sigma}_i^{*,\mu,\gamma,\bm{r}}(I) - \hat{\sigma}_i^{\prime}(I) = \bm{0}$ holds for all $I \in \mathcal{I}_i$ except $I^{\prime}$. Therefore, we get
\begin{equation}\label{eq:appendix-A5-2}
    \small
    % \setlength\abovedisplayskip{2pt}
    % \setlength\belowdisplayskip{2pt}
    % \thinmuskip=0mu
    % \medmuskip=0mu
    % \thickmuskip=0mu
    % \spaceskip=0pt
    \begin{aligned}
    \langle \hat{\bm{\ell}}^{\bm{x}^{*,\mu,\gamma,\bm{r}}}_i, \hat{\bm{x}}^{*,\mu,\gamma,\bm{r}}_i - \hat{\bm{x}}_i^{\prime}\rangle 
    &= \sum_{I \in \mathcal{I}_i} \pi^{\sigma^{\prime}}_i(I) \langle -\hat{\bm{v}}^{*,\mu,\gamma,\bm{r}}_i(I), \hat{\sigma}_i^{*,\mu,\gamma,\bm{r}}(I) - \hat{\sigma}_i^{\prime}(I) \rangle \\
    &= \langle -\hat{\bm{v}}^{*,\mu,\gamma,\bm{r}}_i(I^{\prime}), \hat{\sigma}_i^{*,\mu,\gamma,\bm{r}}(I^{\prime}) - \hat{\sigma}_i^{\prime}(I^{\prime}) \rangle > 0,
    \end{aligned}
\end{equation}
where $\hat{\bm{x}}_i^{\prime}$ is the sequence-form strategy profile associated with $\hat{\sigma}_i$.
By the definition of $\hat{\bm{x}}_i^{\prime}$, it follows that $\hat{\bm{x}}_i^{\prime} \in \bm{\mathcal{X}}_i^{\gamma}$. From the definition of the NE, we have $\langle \hat{\bm{\ell}}^{\bm{x}^{*,\mu,\gamma,\bm{r}}}_i, \hat{\bm{x}}^{*,\mu,\gamma,\bm{r}}_i - \hat{\bm{x}}_i^{\prime}\rangle \leq 0$, which contradicts the result in (\ref{eq:appendix-A5-2}). Therefore, there exists no $I^{\prime} \in \mathcal{I}_i$ and $\bm{\theta}^{\prime}_{I^{\prime}} > \bm{0}$ such that $\langle - \hat{\bm{m}}^{*,\mu,\gamma,\bm{r}}(I^{\prime}), \bm{\theta}^{\prime}_{I^{\prime}} \rangle < 0$. Consequently, when $\bm{\theta}_I > \bm{0}$ for all $I \in \mathcal{I}_i$, it holds that $\langle - \hat{\bm{m}}^{*,\mu,\gamma,\bm{r}}_i(I), {\bm{\theta}}_I \rangle \geq 0$.

Through the discussion of the above two situations, we complete the proof.
\end{proof}

\subsection{Proof of \Cref{lem:maximum value of counterfactual value}}\label{subsec:proof:lem:maximum value of counterfactual value}

\begin{proof}
    From the definition of $\hat{\bm{v}}^{\sigma}_i(I)$, we get
    \begin{equation}\label{eq:appendix-A2-0}
    \small
    % \setlength\abovedisplayskip{2pt}
    % \setlength\belowdisplayskip{2pt}
    % \thinmuskip=0mu
    % \medmuskip=0mu
    % \thickmuskip=0mu
    % \spaceskip=0pt
    \begin{aligned}
         \Vert \hat{\bm{v}}^{\sigma}_i(I) \Vert_2 \leq \Vert \hat{\bm{v}}^{\sigma}_i(I) \Vert_1  
         = & \sum_{a \in A(I)}\Vert \hat{\bm{v}}^{\sigma}_i(I,a) \Vert_1 \\
         = &\sum_{a \in A(I)}\Vert - \hat{\bm{\ell}}^{\bm{x}}_i(I,a) + \sum_{I^{\prime} \in C_i(I,a)} \langle \hat{\bm{v}}^{\sigma}_i(I^{\prime}), \sigma_i(I^{\prime})\rangle \Vert_1 \\
         \leq & \sum_{a \in A(I)}\Vert - \hat{\bm{\ell}}^{\bm{x}}_i(I,a) \Vert_1 + \sum_{a \in A(I)} \sum_{I^{\prime} \in C_i(I,a)}\Vert  \langle \hat{\bm{v}}^{\sigma}_i(I^{\prime}), \sigma_i(I^{\prime})\rangle \Vert_1 \\
         \leq & \sum_{a \in A(I)}\Vert - \hat{\bm{\ell}}^{\bm{x}}_i(I,a) \Vert_1 + \sum_{a \in A(I)} \sum_{I^{\prime} \in C_i(I,a)}  \Vert \hat{\bm{v}}^{\sigma}_i(I^{\prime}) \Vert_1 \Vert \sigma_i(I^{\prime}) \Vert_1 \\
         \leq & \sum_{a \in A(I)}\Vert - \hat{\bm{\ell}}^{\bm{x}}_i(I,a) \Vert_1 + \sum_{a \in A(I)} \sum_{I^{\prime} \in C_i(I,a)}  \Vert \hat{\bm{v}}^{\sigma}_i(I^{\prime}) \Vert_1\\
         \leq & \sum_{I^{\prime} \in \mathcal{I}_i} \sum_{a^{\prime} \in A(I^{\prime})} \Vert - \hat{\bm{\ell}}^{\bm{x}}_i(I^{\prime}a^{\prime}) \Vert_1,
    \end{aligned}
    \end{equation}
    where the last line is from recursion. Continuing from the above inequality, we get
    \begin{equation}\label{eq:appendix-A2-1}
    \small
    % \setlength\abovedisplayskip{2pt}
    % \setlength\belowdisplayskip{2pt}
    % \thinmuskip=0mu
    % \medmuskip=0mu
    % \thickmuskip=0mu
    % \spaceskip=0pt
    \begin{aligned}
         & \sum_{I^{\prime} \in \mathcal{I}_i} \sum_{a^{\prime} \in A(I^{\prime})} \Vert - \hat{\bm{\ell}}^{\bm{x}}_i(I^{\prime}a^{\prime}) \Vert_1 \\
         = &\Vert \bm{\ell}^{\bm{x}}_i + \mu \nabla \psi(\bm{x}_i)  - \mu \nabla \psi(\bm{r}_i) \Vert_1 \\
         \leq & \Vert \bm{\ell}^{\bm{x}}_i \Vert_1 + \mu \Vert \nabla \psi(\bm{x}_i) \Vert_1 + \mu \Vert \nabla \psi(\bm{r}_i)  \Vert_1 \leq P + 2 \mu D,
    \end{aligned}
    \end{equation}
    where $\bm{\ell}^{\bm{x}}_0 = \bm{A}\bm{x}_1$ and $\bm{\ell}^{\bm{x}}_1 = -\bm{A}^{\text{T}} \bm{x}_0 $. By substituting (\ref{eq:appendix-A2-1}) into (\ref{eq:appendix-A2-0}), we have
    \begin{equation}\label{eq:appendix-A2-2}
    \small
    % \setlength\abovedisplayskip{2pt}
    % \setlength\belowdisplayskip{2pt}
    % \thinmuskip=0mu
    % \medmuskip=0mu
    % \thickmuskip=0mu
    % \spaceskip=0pt
    \begin{aligned}
         \Vert \hat{\bm{v}}^{\sigma}_i(I) \Vert_2 \leq \Vert \hat{\bm{v}}^{\sigma}_i(I) \Vert_1 \leq P + 2 \mu D,
    \end{aligned}
    \end{equation}
    It completes the proof.
\end{proof}

\subsection{Proof of \Cref{lem:smoothness of counterfactual value}}\label{subsec:proof:lem:smoothness of counterfactual value}

\begin{proof}
    From the definition of $\hat{\bm{v}}^{\sigma}_i(I)$ and $\hat{\bm{v}}^{\sigma^{\prime}}_i(I)$, we have
    \begin{equation}\label{eq:appendix-A3-0}
    \small
    % \setlength\abovedisplayskip{2pt}
    % \setlength\belowdisplayskip{2pt}
    % \thinmuskip=0mu
    % \medmuskip=0mu
    % \thickmuskip=0mu
    % \spaceskip=0pt
    \begin{aligned}
         & \Vert \hat{\bm{v}}^{\sigma}_i(I) - \hat{\bm{v}}^{\sigma^{\prime}}_i(I) \Vert_2 \\
         \leq & \Vert \hat{\bm{v}}^{\sigma}_i(I) - \hat{\bm{v}}^{\sigma^{\prime}}_i(I) \Vert_1 \\
         = & \sum_{a \in A(I)}\Vert - \hat{\bm{\ell}}^{\bm{x}}_i(I,a) +  \sum_{I^{\prime} \in C_i(I,a)} \langle \hat{\bm{v}}^{\sigma}_i(I^{\prime}), \sigma_i(I^{\prime})\rangle + \hat{\bm{\ell}}^{\bm{x}^{\prime}}_i(I,a) - \sum_{I^{\prime} \in C_i(I,a)} \langle \hat{\bm{v}}^{\sigma^{\prime}}_i(I^{\prime}), \sigma^{\prime}_i(I^{\prime})\rangle \Vert_1 \\
         \leq & \sum_{a \in A(I)}\Vert - \hat{\bm{\ell}}^{\bm{x}}_i(I,a) + \hat{\bm{\ell}}^{\bm{x}^{\prime}}_i(I,a) \Vert_1 + \sum_{a \in A(I)}  \sum_{I^{\prime} \in C_i(I,a)} \Vert \langle \hat{\bm{v}}^{\sigma}_i(I^{\prime}), \sigma_i(I^{\prime})\rangle - \langle \hat{\bm{v}}^{\sigma^{\prime}}_i(I^{\prime}), \sigma^{\prime}_i(I^{\prime})\rangle\Vert_1 .
    \end{aligned}
    \end{equation}     
    Then, we have
    \begin{equation}\label{eq:appendix-A3-1}
    \small
    % \setlength\abovedisplayskip{2pt}
    % \setlength\belowdisplayskip{2pt}
    % \thinmuskip=0mu
    % \medmuskip=0mu
    % \thickmuskip=0mu
    % \spaceskip=0pt
    \begin{aligned}
         & \Vert \hat{\bm{v}}^{\sigma}_i(I) - \hat{\bm{v}}^{\sigma^{\prime}}_i(I) \Vert_2 \\
         \leq & \sum_{a \in A(I)}\Vert - \hat{\bm{\ell}}^{\bm{x}}_i(I,a) + \hat{\bm{\ell}}^{\bm{x}^{\prime}}_i(I,a) \Vert_1 \\
         & + \sum_{a \in A(I)}  \sum_{I^{\prime} \in C_i(I,a)} \Vert \langle \hat{\bm{v}}^{\sigma}_i(I^{\prime}), \sigma_i(I^{\prime})\rangle  - \langle \hat{\bm{v}}^{\sigma}_i(I^{\prime}), \sigma^{\prime}_i(I^{\prime})\rangle + \langle \hat{\bm{v}}^{\sigma}_i(I^{\prime}), \sigma^{\prime}_i(I^{\prime})\rangle - \langle \hat{\bm{v}}^{\sigma^{\prime}}_i(I^{\prime}), \sigma^{\prime}_i(I^{\prime})\rangle\Vert_1 \\
         \leq & \sum_{a \in A(I)}\Vert - \hat{\bm{\ell}}^{\bm{x}}_i(I,a) + \hat{\bm{\ell}}^{\bm{x}^{\prime}}_i(I,a) \Vert_1 + \sum_{a \in A(I)}  \sum_{I^{\prime} \in C_i(I,a)} \Vert \langle \hat{\bm{v}}^{\sigma}_i(I^{\prime}), \sigma_i(I^{\prime})\rangle  - \langle \hat{\bm{v}}^{\sigma}_i(I^{\prime}), \sigma^{\prime}_i(I^{\prime})\rangle \Vert_1 \\
         & +  \sum_{a \in A(I)}  \sum_{I^{\prime} \in C_i(I,a)} \Vert  \langle \hat{\bm{v}}^{\sigma}_i(I^{\prime}), \sigma^{\prime}_i(I^{\prime})\rangle - \langle \hat{\bm{v}}^{\sigma^{\prime}}_i(I^{\prime}), \sigma^{\prime}_i(I^{\prime})\rangle\Vert_1.
    \end{aligned}
    \end{equation}
    For the term $\Vert \langle \hat{\bm{v}}^{\sigma}_i(I^{\prime}), \sigma_i(I^{\prime})\rangle  - \langle \hat{\bm{v}}^{\sigma}_i(I^{\prime}), \sigma^{\prime}_i(I^{\prime})\rangle \Vert_1$ in (\ref{eq:appendix-A3-1}), we get
    \begin{equation}\label{eq:appendix-A3-2}
    \small
    % \setlength\abovedisplayskip{2pt}
    % \setlength\belowdisplayskip{2pt}
    % \thinmuskip=0mu
    % \medmuskip=0mu
    % \thickmuskip=0mu
    % \spaceskip=0pt
    \begin{aligned}
         \Vert \langle \hat{\bm{v}}^{\sigma}_i(I^{\prime}), \sigma_i(I^{\prime})\rangle  - \langle \hat{\bm{v}}^{\sigma}_i(I^{\prime}), \sigma^{\prime}_i(I^{\prime})\rangle \Vert_1 = & \Vert\langle \hat{\bm{v}}^{\sigma}_i(I^{\prime}), \sigma_i(I^{\prime})  -  \sigma^{\prime}_i(I^{\prime})\rangle  \Vert_1 \\
         \leq & \Vert \hat{\bm{v}}^{\sigma}_i(I^{\prime})\Vert_1 \Vert \sigma_i(I^{\prime})  -  \sigma^{\prime}_i(I^{\prime})  \Vert_1 \\
         \leq & (P+2\mu D) \Vert \sigma_i(I^{\prime})  -  \sigma^{\prime}_i(I^{\prime})  \Vert_1,
    \end{aligned}
    \end{equation}
    where the last line comes from \Cref{lem:maximum value of counterfactual value}. For the term $\Vert  \langle \hat{\bm{v}}^{\sigma}_i(I^{\prime}), \sigma^{\prime}_i(I^{\prime})\rangle - \langle \hat{\bm{v}}^{\sigma^{\prime}}_i(I^{\prime}), \sigma^{\prime}_i(I^{\prime})\rangle\Vert_1$ in (\ref{eq:appendix-A3-1}), we get
    \begin{equation}\label{eq:appendix-A3-3}
    \small
    % \setlength\abovedisplayskip{2pt}
    % \setlength\belowdisplayskip{2pt}
    % \thinmuskip=0mu
    % \medmuskip=0mu
    % \thickmuskip=0mu
    % \spaceskip=0pt
    \begin{aligned}
         \Vert  \langle \hat{\bm{v}}^{\sigma}_i(I^{\prime}), \sigma^{\prime}_i(I^{\prime})\rangle - \langle \hat{\bm{v}}^{\sigma^{\prime}}_i(I^{\prime}), \sigma^{\prime}_i(I^{\prime})\rangle\Vert_1 = & \Vert  \langle \hat{\bm{v}}^{\sigma}_i(I^{\prime}) - \hat{\bm{v}}^{\sigma^{\prime}}_i(I^{\prime}), \sigma^{\prime}_i(I^{\prime})\rangle \Vert_1 \\
         \leq & \Vert  \hat{\bm{v}}^{\sigma}_i(I^{\prime}) - \hat{\bm{v}}^{\sigma^{\prime}}_i(I^{\prime})\Vert_1 \Vert \sigma^{\prime}_i(I^{\prime})\rangle \Vert_1 \\
         \leq & \Vert  \hat{\bm{v}}^{\sigma}_i(I^{\prime}) - \hat{\bm{v}}^{\sigma^{\prime}}_i(I^{\prime})\Vert_1,
    \end{aligned}
    \end{equation}
    where the last line comes from $\Vert \sigma^{\prime}_i(I^{\prime})\rangle \Vert_1 \leq 1$. By substituting (\ref{eq:appendix-A3-2}) and (\ref{eq:appendix-A3-3}) into (\ref{eq:appendix-A3-1}), we obtain
    \begin{equation}\label{eq:appendix-A3-4}
    \small
    % \setlength\abovedisplayskip{2pt}
    % \setlength\belowdisplayskip{2pt}
    % \thinmuskip=0mu
    % \medmuskip=0mu
    % \thickmuskip=0mu
    % \spaceskip=0pt
    \begin{aligned}
         & \Vert \hat{\bm{v}}^{\sigma}_i(I) - \hat{\bm{v}}^{\sigma^{\prime}}_i(I) \Vert_2 \\
         \leq & \sum_{a \in A(I)}\Vert - \hat{\bm{\ell}}^{\bm{x}}_i(I,a) + \hat{\bm{\ell}}^{\bm{x}^{\prime}}_i(I,a) \Vert_1 + \sum_{a \in A(I)}  \sum_{I^{\prime} \in C_i(I,a)} (P+2\mu D) \Vert \sigma_i(I^{\prime})  -  \sigma^{\prime}_i(I^{\prime})  \Vert_1 \\
         & +  \sum_{a \in A(I)}  \sum_{I^{\prime} \in C_i(I,a)} \Vert  \hat{\bm{v}}^{\sigma}_i(I^{\prime}) - \hat{\bm{v}}^{\sigma^{\prime}}_i(I^{\prime})\Vert_1 \\
         \leq & \Vert \hat{\bm{\ell}}^{\bm{x}}_i -\hat{\bm{\ell}}^{\bm{x}^{\prime}}_i \Vert_1 + (P+2\mu D)\Vert \sigma_i  -  \sigma^{\prime}_i  \Vert_1,
    \end{aligned}
    \end{equation}
    where the last line is from recursion. For the term $\Vert \hat{\bm{\ell}}^{\bm{x}}_i -\hat{\bm{\ell}}^{\bm{x}^{\prime}}_i \Vert_1$ in (\ref{eq:appendix-A3-4}), we get
     \begin{equation}\label{eq:appendix-A3-5}
    \small
    % \setlength\abovedisplayskip{2pt}
    % \setlength\belowdisplayskip{2pt}
    % \thinmuskip=0mu
    % \medmuskip=0mu
    % \thickmuskip=0mu
    % \spaceskip=0pt
    \begin{aligned}
         \Vert \hat{\bm{\ell}}^{\bm{x}}_i -\hat{\bm{\ell}}^{\bm{x}^{\prime}}_i \Vert_1 = & \Vert \bm{\ell}^{\bm{x}}_i + \mu \nabla \psi(\bm{x}_i) - \mu \nabla \psi(\bm{r}_i) -\bm{\ell}^{\bm{x}^{\prime}}_i - \mu \nabla \psi(\bm{x}^{\prime}_i) + \mu \nabla \psi(\bm{r}_i)\Vert_1 \\
         = & \Vert \bm{\ell}^{\bm{x}}_i + \mu \bm{x}_i -\bm{\ell}^{\bm{x}^{\prime}}_i - \mu \bm{x}^{\prime}_i \Vert_1 \\
         \leq & L \Vert \bm{x} - \bm{x}^{\prime} \Vert_1 + \mu \Vert \bm{x} - \bm{x}^{\prime} \Vert_1 \\
         \leq & ( L + \mu ) \Vert \bm{x} - \bm{x}^{\prime} \Vert_1,
    \end{aligned}
    \end{equation}
    where $\bm{\ell}^{\bm{x}}_0 = \bm{A}\bm{x}_1$ and $\bm{\ell}^{\bm{x}}_1 = -\bm{A}^{\text{T}} \bm{x}_0$. By substituting (\ref{eq:appendix-A3-5}) into (\ref{eq:appendix-A3-4}), we get
     \begin{equation}\label{eq:appendix-A3-6}
    \small
    \thinmuskip=0mu
    \medmuskip=0mu
    \thickmuskip=0mu
    \spaceskip=0pt
    \begin{aligned}
         & \Vert \hat{\bm{v}}^{\sigma}_i(I) - \hat{\bm{v}}^{\sigma^{\prime}}_i(I) \Vert_2 
         \leq ( L + \mu ) \Vert \bm{x} - \bm{x}^{\prime} \Vert_1 + (P+2\mu D)\Vert \sigma_i  -  \sigma^{\prime}_i  \Vert_1 \\
         \Leftrightarrow & \Vert \hat{\bm{v}}^{\sigma}_i(I) - \hat{\bm{v}}^{\sigma^{\prime}}_i(I) \Vert^2_2  \leq 2( L + \mu )^2 \Vert \bm{x} - \bm{x}^{\prime} \Vert^2_1 + 2(P+2\mu D)^2\Vert \sigma_i  -  \sigma^{\prime}_i  \Vert^2_1,
    \end{aligned}
    \end{equation}
    where the second line is from $\forall b, c \in \mathbb{R}$, $(b+c)^2 \leq 2b^2 + 2c^2$ (in this case, $b = ( L + \mu ) \Vert \bm{x} - \bm{x}^{\prime} \Vert_1$ and $c = (P+2\mu D)\Vert \sigma_i  -  \sigma^{\prime}_i  \Vert_1$).
    It completes the proof.
\end{proof}

\subsection{Proof of \Cref{lem:relationship between behavioral strategy and sequence-form strategy}}\label{subsec:proof:lem:relationship between behavioral strategy and sequence-form strategy}

\begin{proof}
    From the definition of $\Vert {\hat{\sigma}}_i  -  {\hat{\sigma}}^{\prime}_i  \Vert_1$,  we get
    \begin{equation}\label{eq:appendix-A4-0}
    \small
    % \setlength\abovedisplayskip{2pt}
    % \setlength\belowdisplayskip{2pt}
    % \thinmuskip=0mu
    % \medmuskip=0mu
    % \thickmuskip=0mu
    % \spaceskip=0pt
    \begin{aligned}
         & \Vert {\hat{\sigma}}_i  -  {\hat{\sigma}}^{\prime}_i  \Vert_1 \\
         = &\sum_{I \in \mathcal{I}_i} \sum_{a \in A(I)} \Vert \frac{{\hat{\bm{x}}}_i(I,a)}{{\hat{\bm{x}}}_i(\rho_I)} - \frac{{\hat{\bm{x}}}^{\prime}_i(I,a)}{{\hat{\bm{x}}}^{\prime}_i(\rho_I)}\Vert_1 \\
         = & \sum_{I \in \mathcal{I}_i} \sum_{a \in A(I)} \Vert \frac{{\hat{\bm{x}}}_i(I,a){\hat{\bm{x}}}^{\prime}_i(\rho_I)}{{\hat{\bm{x}}}_i(\rho_I){\hat{\bm{x}}}^{\prime}_i(\rho_I)} - \frac{{\hat{\bm{x}}}^{\prime}_i(I,a){\hat{\bm{x}}}_i(\rho_I)}{{\hat{\bm{x}}}_i(\rho_I){\hat{\bm{x}}}^{\prime}_i(\rho_I)}\Vert_1 \\
         = & \sum_{I \in \mathcal{I}_i} \sum_{a \in A(I)} \frac{1}{{\hat{\bm{x}}}_i(\rho_I){\hat{\bm{x}}}^{\prime}_i(\rho_I)} \Vert {\hat{\bm{x}}}_i(I,a){\hat{\bm{x}}}^{\prime}_i(\rho_I) - {\hat{\bm{x}}}^{\prime}_i(I,a){\hat{\bm{x}}}_i(\rho_I)\Vert_1 \\
         = & \sum_{I \in \mathcal{I}_i} \sum_{a \in A(I)} \frac{1}{{\hat{\bm{x}}}_i(\rho_I){\hat{\bm{x}}}^{\prime}_i(\rho_I)} \Vert {\hat{\bm{x}}}_i(I,a){\hat{\bm{x}}}^{\prime}_i(\rho_I) - {\hat{\bm{x}}}_i(I,a){\hat{\bm{x}}}_i(\rho_I) + {\hat{\bm{x}}}_i(I,a){\hat{\bm{x}}}_i(\rho_I) -{\hat{\bm{x}}}^{\prime}_i(I,a){\hat{\bm{x}}}_i(\rho_I)\Vert_1 \\
         = & \sum_{I \in \mathcal{I}_i} \sum_{a \in A(I)} \frac{1}{{\hat{\bm{x}}}_i(\rho_I){\hat{\bm{x}}}^{\prime}_i(\rho_I)} \left(\Vert {\hat{\bm{x}}}_i(I,a){\hat{\bm{x}}}^{\prime}_i(\rho_I) - {\hat{\bm{x}}}_i(I,a){\hat{\bm{x}}}_i(\rho_I) \Vert_1 + \Vert {\hat{\bm{x}}}_i(I,a){\hat{\bm{x}}}_i(\rho_I) -{\hat{\bm{x}}}^{\prime}_i(I,a){\hat{\bm{x}}}_i(\rho_I)\Vert_1 \right).
    \end{aligned}
    \end{equation}
    For the term $\Vert {\hat{\bm{x}}}_i(I,a){\hat{\bm{x}}}^{\prime}_i(\rho_I) - {\hat{\bm{x}}}_i(I,a){\hat{\bm{x}}}_i(\rho_I) \Vert_1$ in (\ref{eq:appendix-A4-0}), we have
    \begin{equation}\label{eq:appendix-A4-1}
    \small
    % \setlength\abovedisplayskip{2pt}
    % \setlength\belowdisplayskip{2pt}
    % \thinmuskip=0mu
    % \medmuskip=0mu
    % \thickmuskip=0mu
    % \spaceskip=0pt
    \begin{aligned}
         \Vert {\hat{\bm{x}}}_i(I,a){\hat{\bm{x}}}^{\prime}_i(\rho_I) - {\hat{\bm{x}}}_i(I,a){\hat{\bm{x}}}_i(\rho_I) \Vert_1 = {\hat{\bm{x}}}_i(I,a) \Vert {\hat{\bm{x}}}^{\prime}_i(\rho_I) - {\hat{\bm{x}}}_i(\rho_I) \Vert_1.
    \end{aligned}
    \end{equation}
    For the term $\Vert {\hat{\bm{x}}}_i(I,a){\hat{\bm{x}}}^{\prime}_i(\rho_I) - {\hat{\bm{x}}}_i(I,a){\hat{\bm{x}}}_i(\rho_I) \Vert_1$ in (\ref{eq:appendix-A4-0}), we have
    \begin{equation}\label{eq:appendix-A4-2}
    \small
    % \setlength\abovedisplayskip{2pt}
    % \setlength\belowdisplayskip{2pt}
    % \thinmuskip=0mu
    % \medmuskip=0mu
    % \thickmuskip=0mu
    % \spaceskip=0pt
    \begin{aligned}
         \Vert {\hat{\bm{x}}}_i(I,a){\hat{\bm{x}}}_i(\rho_I) -{\hat{\bm{x}}}^{\prime}_i(I,a){\hat{\bm{x}}}_i(\rho_I)\Vert_1 = {\hat{\bm{x}}}_i(\rho_I) \Vert {\hat{\bm{x}}}_i(I,a) - {\hat{\bm{x}}}^{\prime}_i(I,a) \Vert_1.
    \end{aligned}
    \end{equation}
    By substituting (\ref{eq:appendix-A4-1}) and (\ref{eq:appendix-A4-2}) into (\ref{eq:appendix-A4-0}), we have
    \begin{equation}\label{eq:appendix-A4-3}
    \small
    % \setlength\abovedisplayskip{2pt}
    % \setlength\belowdisplayskip{2pt}
    % \thinmuskip=0mu
    % \medmuskip=0mu
    % \thickmuskip=0mu
    % \spaceskip=0pt
    \begin{aligned}
         \Vert {\hat{\sigma}}_i  -  {\hat{\sigma}}^{\prime}_i  \Vert_1 
         = & \sum_{I \in \mathcal{I}_i} \sum_{a \in A(I)} \frac{1}{{\hat{\bm{x}}}_i(\rho_I){\hat{\bm{x}}}^{\prime}_i(\rho_I)} \left({\hat{\bm{x}}}_i(I,a) \Vert {\hat{\bm{x}}}^{\prime}_i(\rho_I) - {\hat{\bm{x}}}_i(\rho_I) \Vert_1 + {\hat{\bm{x}}}_i(\rho_I) \Vert {\hat{\bm{x}}}_i(I,a) - {\hat{\bm{x}}}^{\prime}_i(I,a) \Vert_1 \right) \\
         = & \sum_{I \in \mathcal{I}_i} \sum_{a \in A(I)} 
         \left(
        \frac{{\hat{\bm{x}}}_i(I,a)}{{\hat{\bm{x}}}_i(\rho_I){\hat{\bm{x}}}^{\prime}_i(\rho_I)} \Vert {\hat{\bm{x}}}^{\prime}_i(\rho_I) - {\hat{\bm{x}}}_i(\rho_I) \Vert_1 
        + \frac{{\hat{\bm{x}}}_i(\rho_I)}{{\hat{\bm{x}}}_i(\rho_I){\hat{\bm{x}}}^{\prime}_i(\rho_I)} \Vert {\hat{\bm{x}}}_i(I,a) - {\hat{\bm{x}}}^{\prime}_i(I,a) \Vert_1
         \right).
    \end{aligned}
    \end{equation}
    Since ${\hat{\bm{x}}}_i(I,a)/{\hat{\bm{x}}}_i(\rho_I) ={\hat{\sigma}}_i(I,a) \leq 1$, we obtain
    \begin{equation}\label{eq:appendix-A4-4}
    \small
    % \setlength\abovedisplayskip{2pt}
    % \setlength\belowdisplayskip{2pt}
    % \thinmuskip=0mu
    % \medmuskip=0mu
    % \thickmuskip=0mu
    % \spaceskip=0pt
    \begin{aligned}
         \Vert {\hat{\sigma}}_i  -  {\hat{\sigma}}^{\prime}_i  \Vert_1 
         \leq & \sum_{I \in \mathcal{I}_i} \sum_{a \in A(I)} 
         \left(
        \frac{1}{{\hat{\bm{x}}}^{\prime}_i(\rho_I)} \Vert {\hat{\bm{x}}}^{\prime}_i(\rho_I) - {\hat{\bm{x}}}_i(\rho_I) \Vert_1 
        + \frac{1}{{\hat{\bm{x}}}^{\prime}_i(\rho_I)} \Vert {\hat{\bm{x}}}_i(I,a) - {\hat{\bm{x}}}^{\prime}_i(I,a) \Vert_1
         \right) \\
        \leq & \sum_{I \in \mathcal{I}_i} \sum_{a \in A(I)} 
         \frac{1}{\gamma^H} \left(
            \Vert {\hat{\bm{x}}}^{\prime}_i(\rho_I) - {\hat{\bm{x}}}_i(\rho_I) \Vert_1 
            + \Vert {\hat{\bm{x}}}_i(I,a) - {\hat{\bm{x}}}^{\prime}_i(I,a) \Vert_1
             \right),
    \end{aligned}
    \end{equation}
    where the last inequality comes from ${\hat{\bm{x}}}_i(I) \leq 1/\gamma^H$ for all $i \in \mathcal{N}, I \in \mathcal{I}_i$, and ${\hat{\bm{x}}}_i \in \bm{\mathcal{X}}_i^{\gamma}$ (this follows from the facts that $H$ denotes the maximum number of actions taken by all players along any path from the root to a leaf node and the probability of selecting each action is guaranteed to be greater than $\gamma$). Then, if $I \in \mathcal{I}^{init}$ with $\mathcal{I}^{init} = \{ I|i \in \mathcal{N}, I \in  \mathcal{I}^{init}_i\}$, we have ${\hat{\bm{x}}}^{\prime}_i(\rho_I) = {\hat{\bm{x}}}_i(\rho_I)  = 1$. Therefore, continuing from (\ref{eq:appendix-A4-4}), we get
    \begin{equation}\label{eq:appendix-A4-5}
    \small
    % \setlength\abovedisplayskip{2pt}
    % \setlength\belowdisplayskip{2pt}
    % \thinmuskip=0mu
    % \medmuskip=0mu
    % \thickmuskip=0mu
    % \spaceskip=0pt
    \begin{aligned}
         \Vert {\hat{\sigma}}_i  -  {\hat{\sigma}}^{\prime}_i  \Vert_1 
         \leq & \sum_{I \in \mathcal{I}_i} \sum_{a \in A(I)} 
         \frac{C_{max} + 1}{\gamma^H} \Vert {\hat{\bm{x}}}_i(I,a) - {\hat{\bm{x}}}^{\prime}_i(I,a) \Vert_1 \leq \frac{2 C_{max}}{\gamma^H} \Vert {\hat{\bm{x}}}_i - {\hat{\bm{x}}}^{\prime}_i \Vert_1.
    \end{aligned}
    \end{equation}
    It finishes the proof.
\end{proof}

{
% \color{red}
\clearpage
\newpage
\section{Our Parameter-Free Average-Iterate Convergence of CFR$^+$}\label{sec:Parameter-free average-iterate convergence}
Now, we extend the proof of CFR$^+$ in \citet{farina2021faster} via our proof approach in \Cref{sec:prf:thm:convergence results of our algorithm} to demonstrate that for all $\eta > 0$, CFR$^+$'s average-iterate convergence holds for all $\bm{\theta}^{1}_I \in \mathbb{R}^{|A(I)|}_{\geq 0}$ not only for $\bm{\theta}^{1}_I = \bm{0}$. This result is significant because it implies that even when the strategies generated during the initial iterations are discarded, CFR$^+$ remains achieving average-iterate convergence. Specifically, since average-iterate convergence holds for all $\bm{\theta}^{1}_I \in \mathbb{R}^{|A(I)|}_{\geq 0}$,  $\bm{\theta}^{t}_I$ can be treated as a new $\bm{\theta}^{1}_I$, ensuring that CFR$^+$ enjoys average-iterate convergence for all $\eta > 0$ after iteration $t$. Indeed, discarding the initial phase strategies is a common technique to improve the empirical convergence rate of CFR$^+$~\citep{steinberger2019pokerrl}.

\begin{theorem}
Assuming all players follow the update rule of CFR$^+$ with any $\bm{\theta}^{1}_I \in \mathbb{R}^{|A(I)|}_{\geq 0}$ and $\eta > 0$, the average strategy profile $\bar{\bm{x}}^{T} = \frac{\sum_{t=1}^T \bm{x}^t}{T}$ converges to the set of NEs of the perturbed regularized EFGs defined in (\ref{eq:BSPP-perturbed regularized}) with any $\gamma \geq 0$ and $\mu \geq 0$.
\label{thm:average-iterate convergence results of our algorithm}
\end{theorem}

\vspace{-0.35cm}
\begin{proof}
By substituting 
$\small 
\thinmuskip=0.5mu
\medmuskip=0.5mu
\thickmuskip=0.5mu
\spaceskip=0.5pt
{\bm{\theta}}_I = \sigma_i(I) = \frac{\hat{\sigma}_i(I) - \gamma \bm{1}}{1 - \alpha_I} \in \Delta_{\gamma}^{|A(I)|}$ with 
$\small 
\thinmuskip=0.5mu
\medmuskip=0.5mu
\thickmuskip=0.5mu
\spaceskip=0.5pt
\hat{\sigma}_i(I) \in \Delta_{\gamma}^{|A(I)|}$ into \Cref{lem:update-rule-MPCFR-inequality}, we get
\begin{equation}\label{eq:appendix-C-0}
\small
\setlength\abovedisplayskip{2pt}
\setlength\belowdisplayskip{2pt}
% \thinmuskip=0mu
% \medmuskip=0mu
% \thickmuskip=0mu
% \spaceskip=0pt
    \begin{aligned}
        & \eta \langle \hat{\bm{m}}^{t}_i(I), \sigma_i(I) - \bm{\theta}^{t+1}_I \rangle \leq  D_{\psi}(\sigma_i(I), {\bm{\theta}}^{t}_I) - D_{\psi}(\sigma_i(I), {\bm{\theta}}^{t+1}_I) - D_{\psi}({\bm{\theta}}^{t+1}_I, {\bm{\theta}}^{t}_I) \\
        \Leftrightarrow
        & \eta \langle \hat{\bm{m}}^{t}_i(I), \sigma_i(I) - \bm{\theta}^{t}_I \rangle \leq  D_{\psi}(\sigma_i(I), {\bm{\theta}}^{t}_I) - D_{\psi}(\sigma_i(I), {\bm{\theta}}^{t+1}_I) - D_{\psi}({\bm{\theta}}^{t+1}_I, {\bm{\theta}}^{t}_I) + \eta \langle \hat{\bm{m}}^{t}_i(I), \bm{\theta}^{t+1}_I - \bm{\theta}^{t}_I \rangle.
    \end{aligned}
\end{equation}
According to the definition of $\hat{\bm{m}}^{t}_i(I)$, we have
\begin{equation}\label{eq:appendix-C-1}
\small
\setlength\abovedisplayskip{2pt}
\setlength\belowdisplayskip{2pt}
% \thinmuskip=0mu
% \medmuskip=0mu
% \thickmuskip=0mu
% \spaceskip=0pt
    \begin{aligned}
         \langle \hat{\bm{m}}^{t}_i(I), \sigma_i(I) - {\bm{\theta}}^{t}_I \rangle  = & \langle \hat{\bm{v}}^{t}_i(I) - \langle \hat{\bm{v}}^{t}_i(I), \sigma^{t}_i(I)\rangle \bm{1}, \sigma_i(I) - {\bm{\theta}}^{t}_I \rangle \\
         = & \langle -\hat{\bm{v}}^{t}_i(I),\sigma^{t}_i(I) - \sigma_i(I)\rangle,
    \end{aligned}
\end{equation}
where the second equality comes from the fact that
\begin{equation}\label{eq:appendix-C-2}
\small
\setlength\abovedisplayskip{2pt}
\setlength\belowdisplayskip{2pt}
% \thinmuskip=0mu
% \medmuskip=0mu
% \thickmuskip=0mu
% \spaceskip=0pt
\begin{aligned}
     \langle \hat{\bm{v}}^{t}_i(I) - \langle \hat{\bm{v}}^{t}_i(I), \sigma^{t}_i(I)\rangle \bm{1}, {\bm{\theta}}^{t}_I \rangle = \langle \hat{\bm{v}}^{t}_i(I) - \langle \hat{\bm{v}}^{t}_i(I), \frac{{\bm{\theta}}^{t}_I}{\langle {\bm{\theta}}^{t}_I, \bm{1} \rangle}\rangle \bm{1}, {\bm{\theta}}^{t}_I \rangle = 0.
\end{aligned}
\end{equation}
Therefore, we have
\begin{equation}\label{eq:appendix-C-3}
\small
\setlength\abovedisplayskip{2pt}
\setlength\belowdisplayskip{2pt}
\thinmuskip=0mu
\medmuskip=0mu
\thickmuskip=0mu
\spaceskip=0pt
    \begin{aligned}
        & \eta \langle -\hat{\bm{v}}^{t}_i(I),\sigma^{t}_i(I) - \sigma_i(I)\rangle \\
        \leq & D_{\psi}(\sigma_i(I), {\bm{\theta}}^{t}_I) - D_{\psi}(\sigma_i(I), {\bm{\theta}}^{t+1}_I) - D_{\psi}({\bm{\theta}}^{t+1}_I, {\bm{\theta}}^{t}_I) + \eta \langle \hat{\bm{m}}^{t}_i(I), \bm{\theta}^{t+1}_I - \bm{\theta}^{t}_I \rangle.
    \end{aligned}
\end{equation}
Continuing from (\ref{eq:appendix-C-3}), we have
\begin{equation}\label{eq:appendix-C-4}
\small
\setlength\abovedisplayskip{2pt}
\setlength\belowdisplayskip{2pt}
% \thinmuskip=0mu
% \medmuskip=0mu
% \thickmuskip=0mu
% \spaceskip=0pt
    \begin{aligned}
        & \eta \langle -\hat{\bm{v}}^{t}_i(I), (1-\alpha_I)\sigma^{t}_i(I) - (1-\alpha_I) \sigma_i(I)\rangle \\
        \leq & (1-\alpha_I) \left( D_{\psi}(\sigma_i(I), {\bm{\theta}}^{t}_I) - D_{\psi}(\sigma_i(I), {\bm{\theta}}^{t+1}_I) + \eta \langle \hat{\bm{m}}^{t}_i(I), \bm{\theta}^{t+1}_I - \bm{\theta}^{t}_I \rangle \right),
    \end{aligned}
\end{equation}
which implies
\begin{equation}\label{eq:appendix-C-4-1}
\small
\setlength\abovedisplayskip{2pt}
\setlength\belowdisplayskip{2pt}
% \thinmuskip=0mu
% \medmuskip=0mu
% \thickmuskip=0mu
% \spaceskip=0pt
    \begin{aligned}
        & \eta \langle -\hat{\bm{v}}^{t}_i(I), (1-\alpha_I)\sigma^{t}_i(I) + \gamma \bm{1}- (1-\alpha_I) \sigma_i(I) - \gamma \bm{1} \rangle \\
        \leq  & (1-\alpha_I) \left( D_{\psi}(\sigma_i(I), {\bm{\theta}}^{t}_I) - D_{\psi}(\sigma_i(I), {\bm{\theta}}^{t+1}_I) + \eta \langle \hat{\bm{m}}^{t}_i(I), \bm{\theta}^{t+1}_I - \bm{\theta}^{t}_I \rangle \right).
    \end{aligned}
\end{equation}
Therefore, we get
\begin{equation}\label{eq:appendix-C-4-2}
\small
\setlength\abovedisplayskip{2pt}
\setlength\belowdisplayskip{2pt}
% \thinmuskip=0mu
% \medmuskip=0mu
% \thickmuskip=0mu
% \spaceskip=0pt
    \begin{aligned}
        & \eta \langle -\hat{\bm{v}}^{t}_i(I), \hat{\sigma}^{t}_i(I) - \hat{\sigma}_i(I) \rangle 
        \leq  (1-\alpha_I) \left( D_{\psi}(\sigma_i(I), {\bm{\theta}}^{t}_I) - D_{\psi}(\sigma_i(I), {\bm{\theta}}^{t+1}_I)  + \eta \langle \hat{\bm{m}}^{t}_i(I), \bm{\theta}^{t+1}_I - \bm{\theta}^{t}_I \rangle \right).   \end{aligned}
\end{equation}
Continuing from (\ref{eq:appendix-C-3}), we have
\begin{equation}\label{eq:appendix-C-4-3}
\small
\setlength\abovedisplayskip{2pt}
\setlength\belowdisplayskip{2pt}
% \thinmuskip=0mu
% \medmuskip=0mu
% \thickmuskip=0mu
% \spaceskip=0pt
    \begin{aligned}
        & \eta \pi^{\hat{\sigma}}_i(I)\langle -\hat{\bm{v}}^{t}_i(I), \hat{\sigma}^{t}_i(I) - \hat{\sigma}_i(I) \rangle \\
        \leq  & (1-\alpha_I) \pi^{\sigma}_i(I) \left( D_{\psi}(\sigma_i(I), {\bm{\theta}}^{t}_I) - D_{\psi}(\sigma_i(I), {\bm{\theta}}^{t+1}_I)  + \eta \langle \hat{\bm{m}}^{t}_i(I), \bm{\theta}^{t+1}_I - \bm{\theta}^{t}_I \rangle \right).
    \end{aligned}
\end{equation}
By applying \Cref{lem:sum of counterfactual regret}, we get
\begin{equation}\label{eq:appendix-C-5}
\small
\setlength\abovedisplayskip{2pt}
\setlength\belowdisplayskip{2pt}
% \thinmuskip=0mu
% \medmuskip=0mu
% \thickmuskip=0mu
% \spaceskip=0pt
    \begin{aligned}
        & \sum_{t=1}^T \langle \hat{\bm{\ell}}^t_i, \hat{\bm{x}}^t_i - \hat{\bm{x}}_i \rangle \\
        \leq & 
        \sum_{t=1}^T \sum_{i \in \mathcal{N}} \sum_{I \in \mathcal{I}_i} (1-\alpha_I) \pi^{\hat{\sigma}}_i(I) \left( \frac{D_{\psi}(\sigma_i(I), {\bm{\theta}}^{t}_I)}{\eta} - \frac{D_{\psi}(\sigma_i(I), {\bm{\theta}}^{t+1}_I)}{\eta}  + \langle \hat{\bm{m}}^{t}_i(I), \bm{\theta}^{t+1}_I - \bm{\theta}^{t}_I \rangle \right),
    \end{aligned}
\end{equation}
where $\hat{\bm{x}}_i$ is the sequence-form strategy corresponding to $\hat{\sigma}_i$. Using $\xi_I$ to denote $(1-\alpha_I)\pi^{\hat{\sigma}}_i(I)$, we get
\begin{equation}\label{eq:appendix-C-6}
\small
\setlength\abovedisplayskip{2pt}
\setlength\belowdisplayskip{2pt}
\thinmuskip=0.5mu
\medmuskip=0.5mu
\thickmuskip=0.5mu
\spaceskip=0.5pt
    \begin{aligned}
        \sum_{t=1}^T \langle \hat{\bm{\ell}}^t_i, \hat{\bm{x}}^t_i - \hat{\bm{x}}_i \rangle 
        \leq & 
        \sum_{t=1}^T \sum_{i \in \mathcal{N}} \sum_{I \in \mathcal{I}_i}\xi_I \left( \frac{D_{\psi}(\sigma_i(I), {\bm{\theta}}^{t}_I)}{\eta} - \frac{D_{\psi}(\sigma_i(I), {\bm{\theta}}^{t+1}_I)}{\eta}  + \langle \hat{\bm{m}}^{t}_i(I), \bm{\theta}^{t+1}_I - \bm{\theta}^{t}_I \rangle \right) \\
        \leq & 
        \sum_{t=1}^T \sum_{i \in \mathcal{N}} \sum_{I \in \mathcal{I}_i}\xi_I \left( \frac{D_{\psi}(\sigma_i(I), {\bm{\theta}}^{t}_I)}{\eta} - \frac{D_{\psi}(\sigma_i(I), {\bm{\theta}}^{t+1}_I)}{\eta}  + \Vert \hat{\bm{m}}^{t}_i(I)\Vert_2 \Vert \bm{\theta}^{t+1}_I - \bm{\theta}^{t}_I \rangle \Vert_2 \right) .
    \end{aligned}
\end{equation}

\begin{lemma}
[Adapted from Lemma 11 of \citet{wei2020linear}].
If the player $i$ follow the update rule of CFR$^+$, with $\eta > 0$ then for any $I \in \mathcal{I}_i$ and $t \geq 1$, we have
\[
\small
\Vert \bm{\theta}^{t+1}_I - \bm{\theta}^{t}_I \Vert_2 \leq \eta \Vert \hat{\bm{m}}^{t}_i(I) \Vert_2.
\]
\label{thm:bound of two accumulated regret}
\end{lemma}

\vspace{-0.6cm}
By substituting \Cref{thm:bound of two accumulated regret} into (\ref{eq:appendix-C-6}), we get
\begin{equation}\label{eq:appendix-C-7}
\small
\setlength\abovedisplayskip{2pt}
\setlength\belowdisplayskip{2pt}
% \thinmuskip=0mu
% \medmuskip=0mu
% \thickmuskip=0mu
% \spaceskip=0pt
    \begin{aligned}
        & \sum_{t=1}^T \langle \hat{\bm{\ell}}^t_i, \hat{\bm{x}}^t_i - \hat{\bm{x}}_i \rangle 
        \leq 
        \sum_{t=1}^T \sum_{i \in \mathcal{N}} \sum_{I \in \mathcal{I}_i}\xi_I \left( \frac{D_{\psi}(\sigma_i(I), {\bm{\theta}}^{t}_I)}{\eta} - \frac{D_{\psi}(\sigma_i(I), {\bm{\theta}}^{t+1}_I)}{\eta}  + \eta \Vert \hat{\bm{m}}^{t}_i(I)\Vert^2_2 \right) .
    \end{aligned}
\end{equation}
Assuming $\Vert \hat{\bm{m}}^{t}_i(I)\Vert^2_2 \leq M$, we have
\begin{equation}\label{eq:appendix-C-7}
\small
\setlength\abovedisplayskip{2pt}
\setlength\belowdisplayskip{2pt}
% \thinmuskip=0mu
% \medmuskip=0mu
% \thickmuskip=0mu
% \spaceskip=0pt
    \begin{aligned}
        \sum_{t=1}^T \langle \hat{\bm{\ell}}^t_i, \hat{\bm{x}}^t_i - \hat{\bm{x}}_i \rangle 
        \leq &
        \sum_{t=1}^T \sum_{i \in \mathcal{N}} \sum_{I \in \mathcal{I}_i}\xi_I \left( \frac{D_{\psi}(\sigma_i(I), {\bm{\theta}}^{t}_I)}{\eta} - \frac{D_{\psi}(\sigma_i(I), {\bm{\theta}}^{t+1}_I)}{\eta}  + \eta M \right) \\
        \leq &
         \sum_{i \in \mathcal{N}} \sum_{I \in \mathcal{I}_i}\xi_I \left( \frac{D_{\psi}(\sigma_i(I), {\bm{\theta}}^{1}_I)}{\eta} + \sum_{t=1}^T \eta M \right) \\
    \end{aligned}
\end{equation}

According to the analysis in \Cref{sec:prf:thm:convergence results of our algorithm}, we have that for any accumulated counterfactual regret sequence $\{ \bm{\theta}^{1}_I, \bm{\theta}^{2}_I, \dots, \bm{\theta}^{t}_I, \dots \}$ generated by any $\bm{\theta}^{1}_I \in \mathbb{R}^{|A(I)|}_{\geq 0}$ and $\eta > 0$, there exists a corresponding accumulated counterfactual regret sequence $\{ {\bm{\theta}^{1}_I}^{\prime}, {\bm{\theta}^{2}_I}^{\prime}, \dots, {\bm{\theta}^{t}_I}^{\prime}, \dots \}$ generated by ${\bm{\theta}^{1}_I}^{\prime} \in \mathbb{R}^{|A(I)|}_{\geq 0}$ and $\eta^{\prime} > 0$, such that the resulting strategy profile sequence $\{ \hat{\bm{x}}^{1}, \hat{\bm{x}}^{2}, \dots, \hat{\bm{x}}^{t}, \dots \}$ are identical, where ${\bm{\theta}^{t}_I}^{\prime} = \eta^{\prime} {\bm{\theta}^{t}_I}/\eta$. To analysis the convergence rate of the accumulated counterfactual regret sequence $\{ {\bm{\theta}^{1}_I}^{\prime}, {\bm{\theta}^{2}_I}^{\prime}, \dots, {\bm{\theta}^{t}_I}^{\prime}, \dots \}$, from (\ref{eq:appendix-C-7}), we have
\begin{equation}\label{eq:appendix-C-8}
\small
\setlength\abovedisplayskip{2pt}
\setlength\belowdisplayskip{2pt}
% \thinmuskip=0mu
% \medmuskip=0mu
% \thickmuskip=0mu
% \spaceskip=0pt
    \begin{aligned}
        \sum_{t=1}^T \langle \hat{\bm{\ell}}^t_i, \hat{\bm{x}}^t_i - \hat{\bm{x}}_i \rangle 
        \leq 
         \sum_{i \in \mathcal{N}} \sum_{I \in \mathcal{I}_i}\xi_I \left( \frac{D_{\psi}(\sigma_i(I), {\bm{\theta}^{1}_I}^{\prime})}{\eta^{\prime}} + \eta^{\prime} T M \right).
    \end{aligned}
\end{equation}
By substituting ${\bm{\theta}^{t}_I}^{\prime} = \eta^{\prime} {\bm{\theta}^{t}_I}/\eta$ into (\ref{eq:appendix-C-8}), we get
\begin{equation}\label{eq:appendix-C-9}
\small
\setlength\abovedisplayskip{2pt}
\setlength\belowdisplayskip{2pt}
% \thinmuskip=0mu
% \medmuskip=0mu
% \thickmuskip=0mu
% \spaceskip=0pt
    \begin{aligned}
        \sum_{t=1}^T \langle \hat{\bm{\ell}}^t_i, \hat{\bm{x}}^t_i - \hat{\bm{x}}_i \rangle 
        \leq 
         \sum_{i \in \mathcal{N}} \sum_{I \in \mathcal{I}_i}\xi_I \left( \frac{D_{\psi}(\sigma_i(I), \frac{\eta^{\prime} {\bm{\theta}^{t}_I}}{\eta})}{\eta^{\prime}} + \eta^{\prime} T M \right).
    \end{aligned}
\end{equation}
From the fact that $\forall \bm{a}, \bm{b} \in \mathbb{R}^d$, $\Vert  \bm{a} - \bm{b} \Vert^2_2/2 = \Vert  \bm{b} - \bm{a} \Vert^2_2/2 = D_{\psi}(\bm{a}, \bm{b})$, by using $\bm{a} = \sigma_i(I)$ and $\bm{b} = \frac{\eta^{\prime} {\bm{\theta}^{1}_I}}{\eta}$, we get
\begin{equation}\label{eq:appendix-C-10}
\small
\setlength\abovedisplayskip{2pt}
\setlength\belowdisplayskip{2pt}
% \thinmuskip=0mu
% \medmuskip=0mu
% \thickmuskip=0mu
% \spaceskip=0pt
    \begin{aligned}
        \sum_{t=1}^T \langle \hat{\bm{\ell}}^t_i, \hat{\bm{x}}^t_i - \hat{\bm{x}}_i \rangle 
        \leq 
         \sum_{i \in \mathcal{N}} \sum_{I \in \mathcal{I}_i}\xi_I \left(
         \frac{\Vert \sigma_i(I) \Vert^2_2}{\eta^{\prime} } + \frac{(\eta^{\prime} {\bm{\theta}^{1}_I})^2}{ \eta^{\prime} \eta^2} 
        + \eta^{\prime} T M \right).
    \end{aligned}
\end{equation}
As $\sigma_i(I) \in \Delta^{|A(I)|}$, we have $\Vert \sigma_i(I) \Vert^2_2 \leq 1$. In addition, ${\Vert \eta^{\prime} {\bm{\theta}^{1}_I}\Vert_2^2}/{(\eta^{\prime} \eta^2 )} = {\eta^{\prime} \Vert {\bm{\theta}^{1}_I} \Vert^2_2}/{(\eta^2 )}$. Continuing from (\ref{eq:appendix-C-10}), we get
\begin{equation}\label{eq:appendix-C-11}
\small
\setlength\abovedisplayskip{2pt}
\setlength\belowdisplayskip{2pt}
% \thinmuskip=0mu
% \medmuskip=0mu
% \thickmuskip=0mu
% \spaceskip=0pt
    \begin{aligned}
        \sum_{t=1}^T \langle \hat{\bm{\ell}}^t_i, \hat{\bm{x}}^t_i - \hat{\bm{x}}_i \rangle 
        \leq 
         \sum_{i \in \mathcal{N}} \sum_{I \in \mathcal{I}_i}\xi_I \left(
         \frac{1}{\eta^{\prime} } + \eta^{\prime} \frac{\Vert {\bm{\theta}^{1}_I} \Vert^2_2}{\eta^2} 
        + \eta^{\prime} T M \right).
    \end{aligned}
\end{equation}
We use $M^{\bm{\theta}}_{\eta}$ to denote $\max({ \Vert {\bm{\theta}^{1}_I} \Vert^2_2}/{(\eta^2 )},M)$. In addition, as $0 \leq (1-\alpha_I) \leq 1$ and $0 \leq \pi^{\hat{\sigma}}_i(I) \leq 1$, we have $0 \leq \xi_I \leq 1$. Therefore, we get
\begin{equation}\label{eq:appendix-C-12}
\small
\setlength\abovedisplayskip{2pt}
\setlength\belowdisplayskip{2pt}
% \thinmuskip=0mu
% \medmuskip=0mu
% \thickmuskip=0mu
% \spaceskip=0pt
    \begin{aligned}
        \sum_{t=1}^T \langle \hat{\bm{\ell}}^t_i, \hat{\bm{x}}^t_i - \hat{\bm{x}}_i \rangle 
        \leq 
         \sum_{i \in \mathcal{N}} \sum_{I \in \mathcal{I}_i} \left(
         \frac{1}{ \eta^{\prime} } + \eta^{\prime} (T+1)  M^{\bm{\theta}}_{\eta}  \right).
    \end{aligned}
\end{equation}
By setting $\eta^{\prime} = 1/\sqrt{(T+1)M^{\bm{\theta}}_{\eta}}$, we have $\sum_{t=1}^T \langle \hat{\bm{\ell}}^t_i, \hat{\bm{x}}^t_i - \hat{\bm{x}}_i \rangle \leq 2\sqrt{(T+1)M^{\bm{\theta}}_{\eta}} |\mathcal{I}| \leq \sqrt{8T M^{\bm{\theta}}_{\eta}} |\mathcal{I}| $ with any $\bm{\theta}^{1}_I \in \mathbb{R}^{|A(I)|}_{\geq 0}$ and $\eta > 0$. It completes the proof.
\end{proof}
}

\clearpage
\newpage
\section{Additional Experiments}\label{sec:Additional Experiments}

\vspace{-0.25cm}
In this section, we present experimental results for (i) the comparison with additional RM-based CFR algorithms, and (ii) the performance of RTCFR$^+$ under various hyperparameters.

\begin{figure*}[t]
    \centering %\quad \quad
    \subfigure{
    \includegraphics[width=0.8\linewidth]{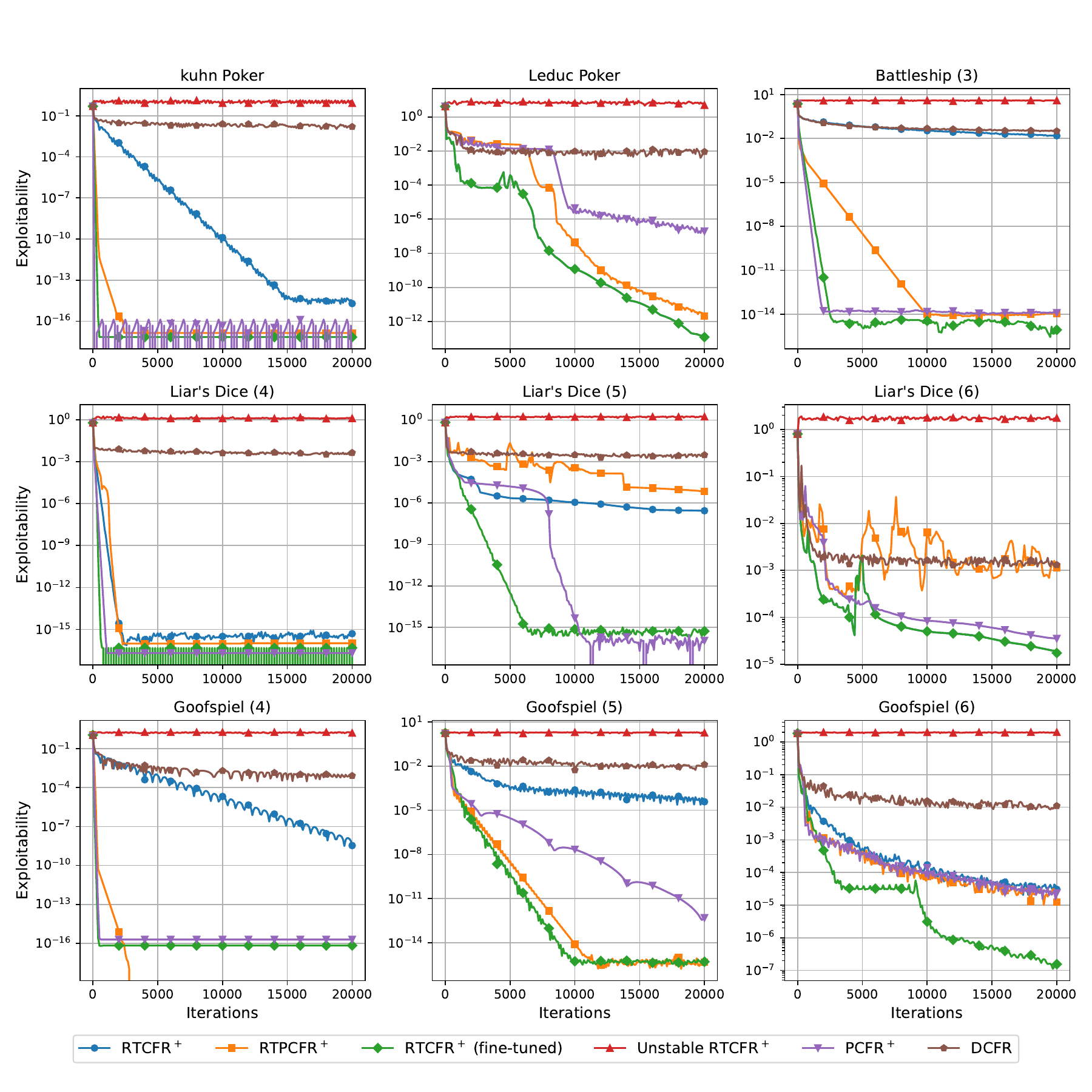}
    }\vspace{-15pt} 
    \caption{Last-iterate convergence rates of more RM-based algorithms. 
}
\label{fig:with-other-algorithms-2}
\vspace{-0.25cm}
\end{figure*}

\textbf{Comparison with additional RM-based CFR algorithms.} We compare RTCFR$^+$ with two advanced RM-based CFR algorithms: PCFR$^+$~\citep{farina2021faster} and DCFR~\citep{brown2019solving}. Additionally, we evaluate RTPCFR$^+$, which employs PCFR$^+$ to solve the perturbed regularized EFG defined in (\ref{eq:BSPP-perturbed regularized}) instead of CFR$^+$. For RTPCFR$^+$, we use the same parameters as RTCFR$^+$ in \Cref{sec:Experiments}, specifically $\gamma = 1\mathrm{e}{-10}$, $\mu = 1\mathrm{e}{-3}$, and $T_u=100$. The experimental results are shown in Figure \ref{fig:with-other-algorithms-2}. Clearly, DCFR is significantly outperformed by the other algorithms. Among RTCFR$^+$, RTPCFR$^+$, and PCFR$^+$, no single algorithm consistently outperforms the others across all EFGs, as their performance varies depending on the specific EFG. This variability may be attributed to the fact that RTCFR$^+$ and RTPCFR$^+$ have not been fine-tuned for individual EFGs. Therefore, we also include a comparison with the fine-tuned RTCFR$^+$, which is denoted as RTCFR$^+$ (fine-tuned) in Figure \ref{fig:with-other-algorithms-2}. Our findings demonstrate that fine-tuning enables RTCFR$^+$ to outperform all tested algorithms. The parameters used for the fine-tuned RTCFR$^+$ are presented in \Cref{tab:RTCFR+ fine-tuned}. In the following, we present the convergence rates of RTCFR$^+$ under various hyperparameters. Lastly, we examine the performance of RTCFR$^+$ that resets $\bm{\theta}^{1}_I$ to $\bm{0}$ at each regularized perturbed EFGs, which is denoted as "Unstable RTCFR$^+$" in Figure \ref{fig:with-other-algorithms-2}. Similarly to RTPCFR$^+$, the parameters of Unstable RTCFR$^+$ are same as RTCFR$^+$ in \Cref{sec:Experiments}. We observe that Unstable RTCFR$^+$ never converges to the set of NEs across all tested games!

\begin{figure*}[t]
    \centering %\quad \quad
    \subfigure{
    \includegraphics[width=0.8\linewidth]{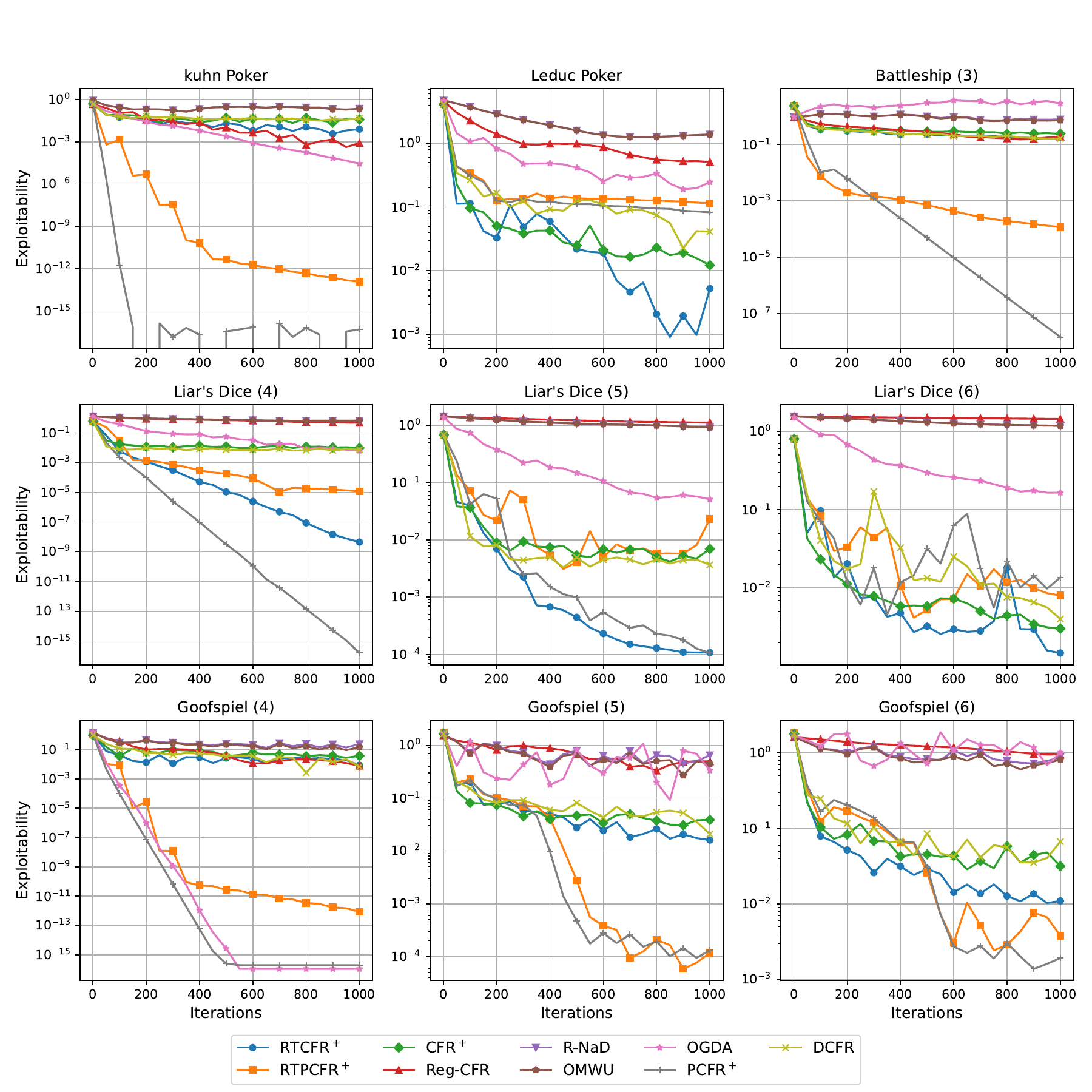}
    }\vspace{-10pt} 
    \caption{Last-iterate convergence rates over the first 1000 iterations.
}
\label{fig:with-other-algorithms-iterations1000}
\vspace{-0.15cm}
\end{figure*}

\textbf{Convergence rates in the initial phase.} We now present the empirical convergence results of our algorithms\textendash RTCFR$^+$ and RTPCFR$^+$, alongside CFR$^+$, R-NaD, Reg-CFR, OMWU, OGDA, PCFR$^+$, and DCFR, over the first 1000 iterations. The experimental results are shown in Figure \ref{fig:with-other-algorithms-iterations1000}. Consistent with the results in Figures \ref{fig:with-other-algorithms} and \ref{fig:with-other-algorithms-2}, RTCFR$^+$, RTPCFR$^+$, and PCFR$^+$ demonstrate superior performance compared to the other algorithms. However, no single algorithm without fine-tuning outperforms all others across all games.

\begin{table}[t]
\centering
\caption{Hyperparameters used in RTCFR$^+$ (fine-tuned).}
\vspace{-3pt}
\small
\renewcommand{\arraystretch}{1.2}
\begin{tabular}{|c|c|c|c|c|c|}
\hline
& \textbf{Kuhn Poker} & \textbf{Leduc Poker} & \textbf{Battleship (3)} & \textbf{Liar's Dice (4)} & \textbf{Liar's Dice (5)} \\
\hline
$\mu$ & 0.1 & 0.001 & 0.1 & 0.01 & 0.0005 \\
\hline
$T_u$ & 10 & 100 & 50 & 10 & 10\\
\hline
& \textbf{Liar's Dice (6)} &\textbf{Goofspiel (4)} & \textbf{Goofspiel (5)} & \textbf{Goofspiel (6)}  & \\
\hline
$\mu$ & 0.005 & 0.1 & 0.05 & 0.005 & \\
\hline
$T_u$ & 100 & 10 & 100 & 50 & \\
\hline
\end{tabular}
\label{tab:RTCFR+ fine-tuned}
\vspace{-0.15cm}
\end{table}

\begin{figure*}[h]
    \centering %\quad \quad
    \subfigure{
    \includegraphics[width=0.8\linewidth]{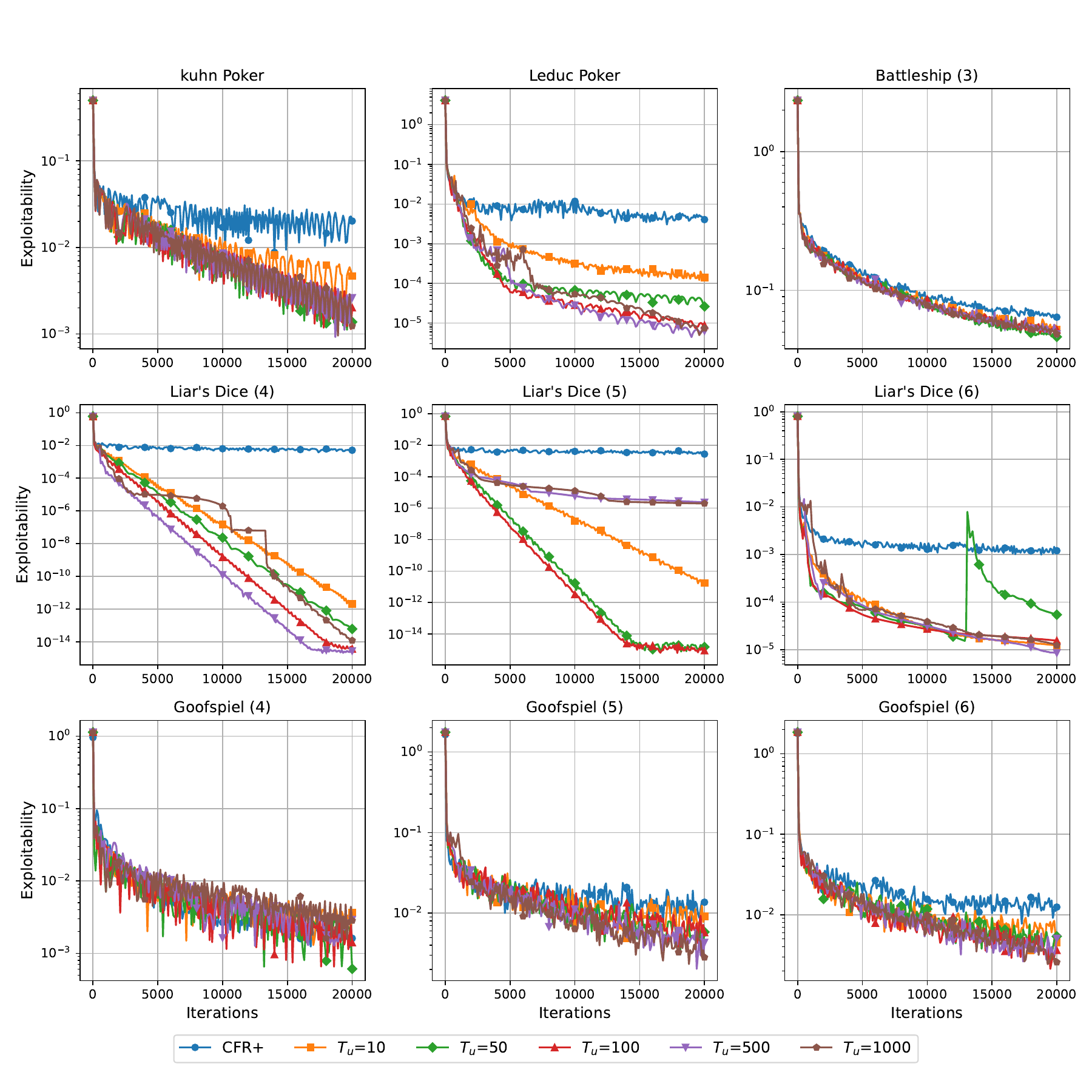}
    }\vspace{-15pt} 
    \caption{Last-iterate convergence rates of RTCFR$^+$ with $\mu = 0.0001$. 
}
\label{fig:RTCFR+-mu=0.0001}
\vspace{-0.25cm}
\end{figure*}

\begin{figure*}[h]
    \centering %\quad \quad
    \subfigure{
    \includegraphics[width=0.8\linewidth]{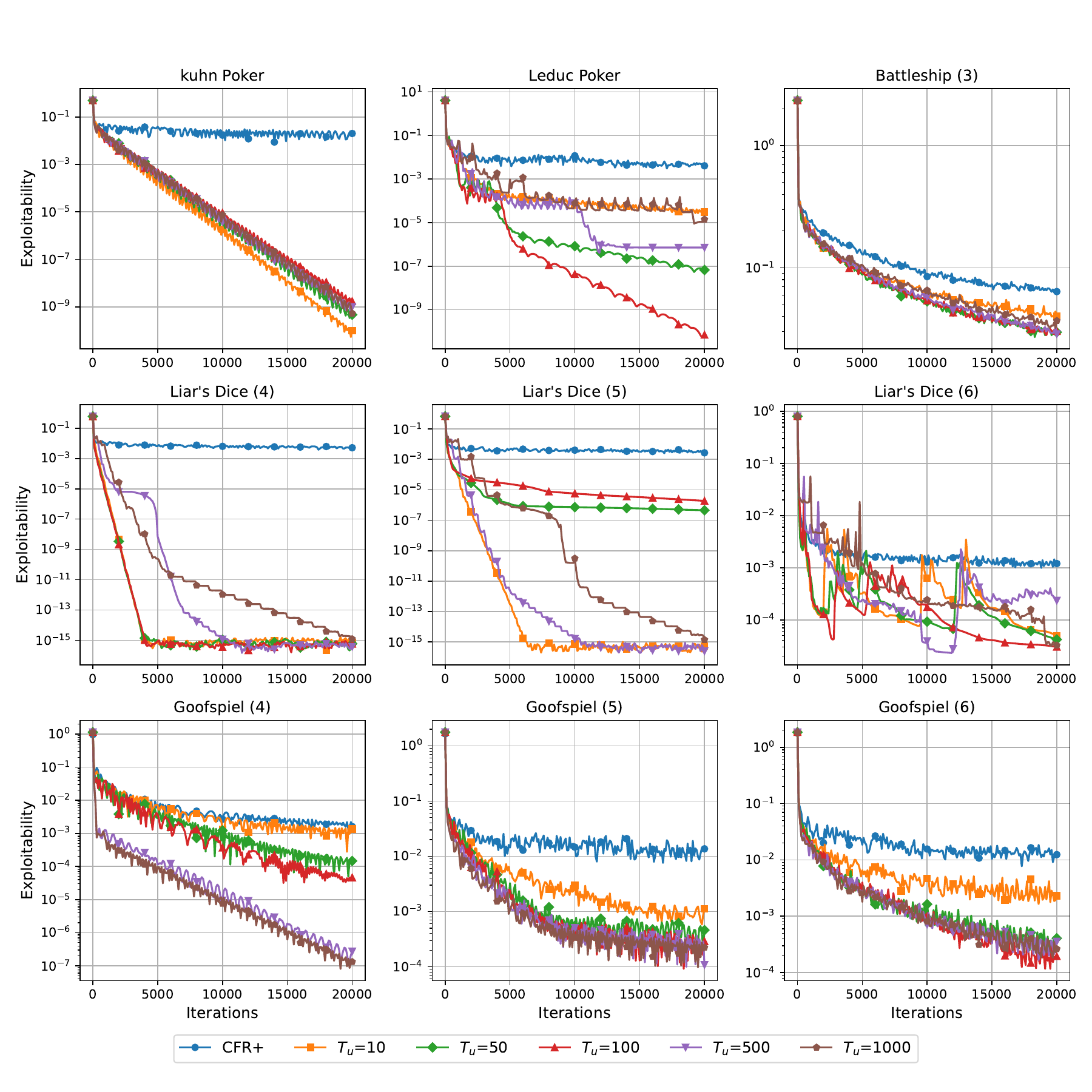}
    }\vspace{-15pt} 
    \caption{Last-iterate convergence rates of RTCFR$^+$ with $\mu = 0.0005$. 
}
\label{fig:RTCFR+-mu=0.0005}
\vspace{-0.25cm}
\end{figure*}

\begin{figure*}[h]
    \centering %\quad \quad
    \subfigure{
    \includegraphics[width=0.8\linewidth]{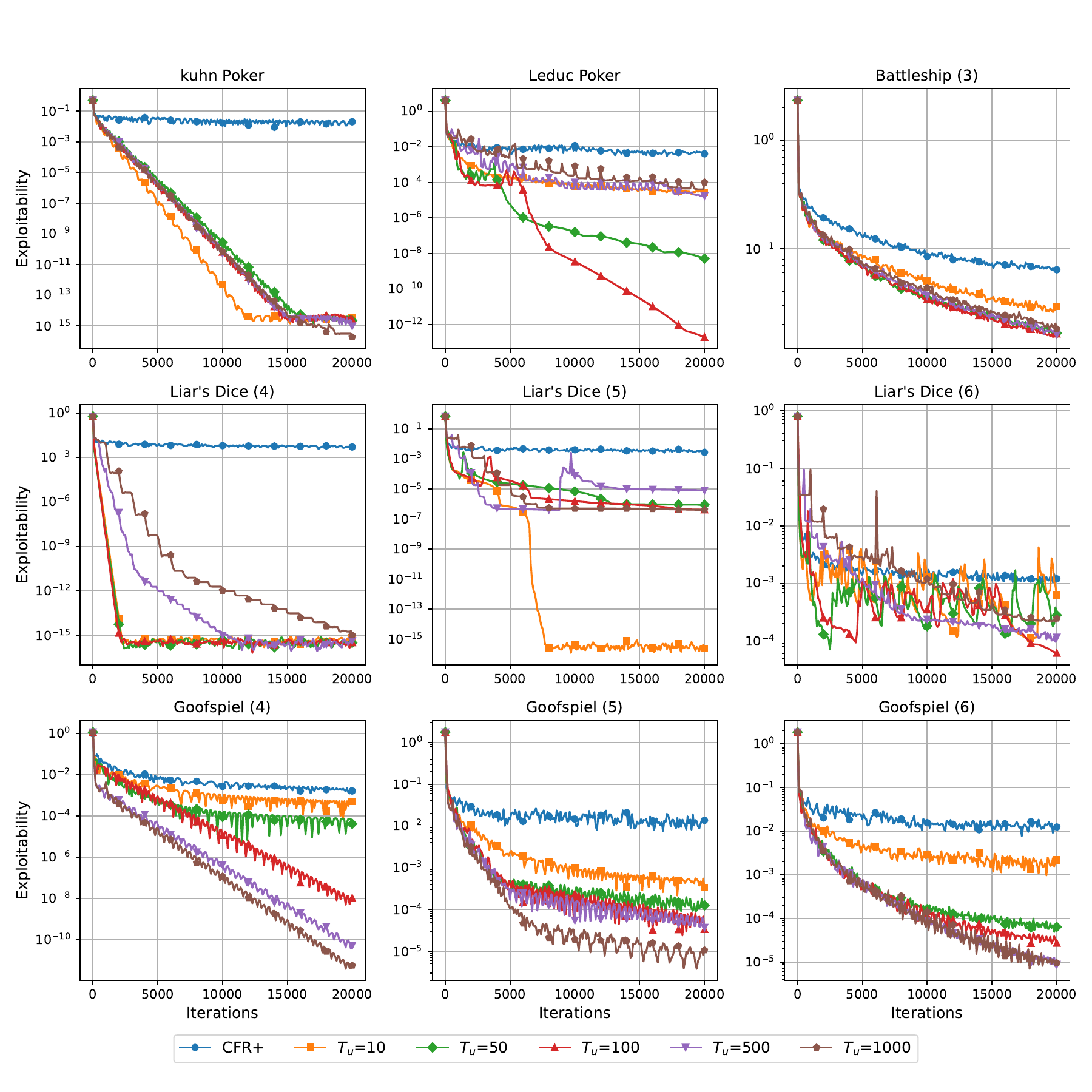}
    }\vspace{-15pt} 
    \caption{Last-iterate convergence rates of RTCFR$^+$ with $\mu = 0.001$. 
}
\label{fig:RTCFR+-mu=0.001}
\vspace{-0.25cm}
\end{figure*}

\begin{figure*}[h]
    \centering %\quad \quad
    \subfigure{
    \includegraphics[width=0.8\linewidth]{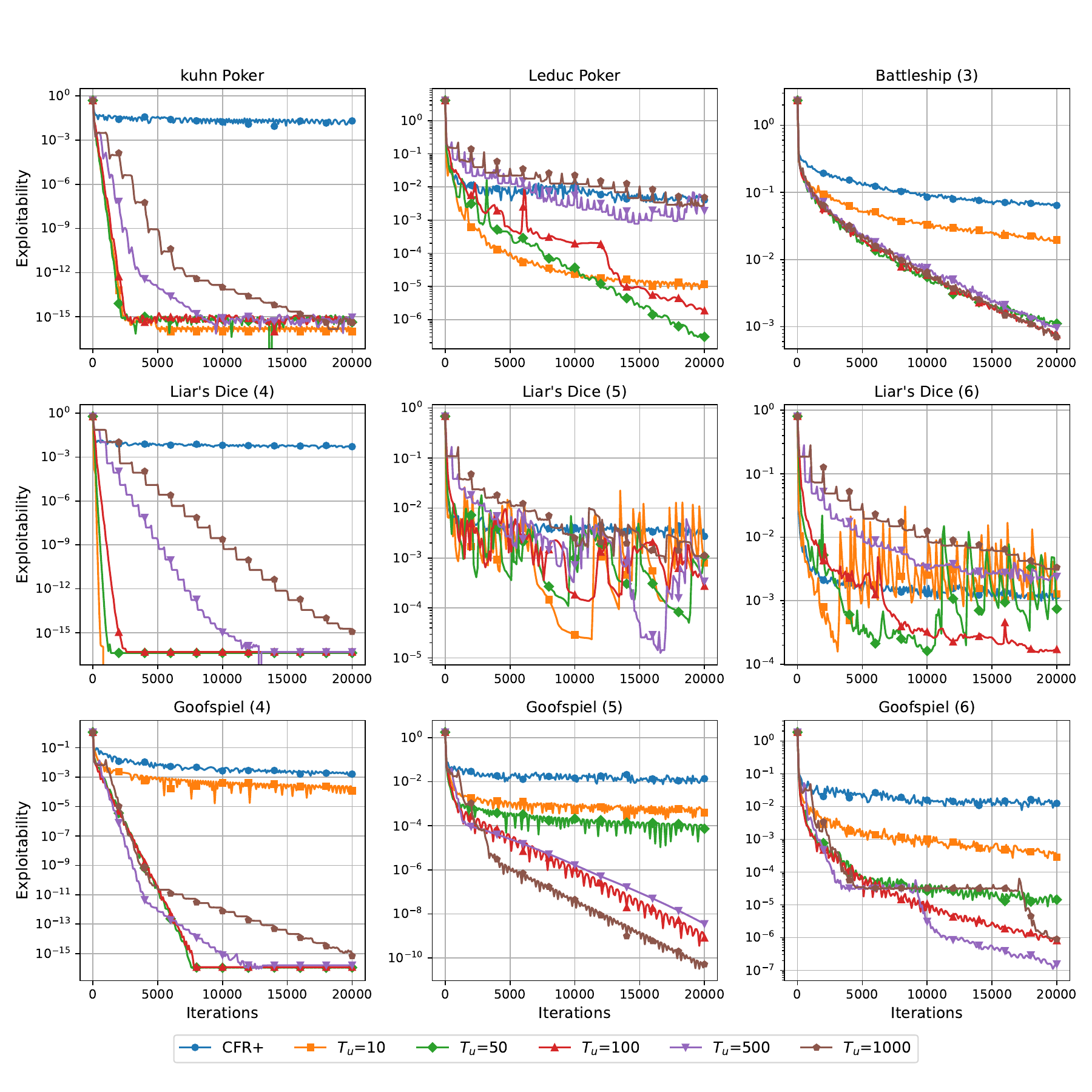}
    }\vspace{-15pt} 
    \caption{Last-iterate convergence rates of RTCFR$^+$ with $\mu = 0.005$. 
}
\label{fig:RTCFR+-mu=0.005}
\vspace{-0.25cm}
\end{figure*}

\begin{figure*}[h]
    \centering %\quad \quad
    \subfigure{
    \includegraphics[width=0.8\linewidth]{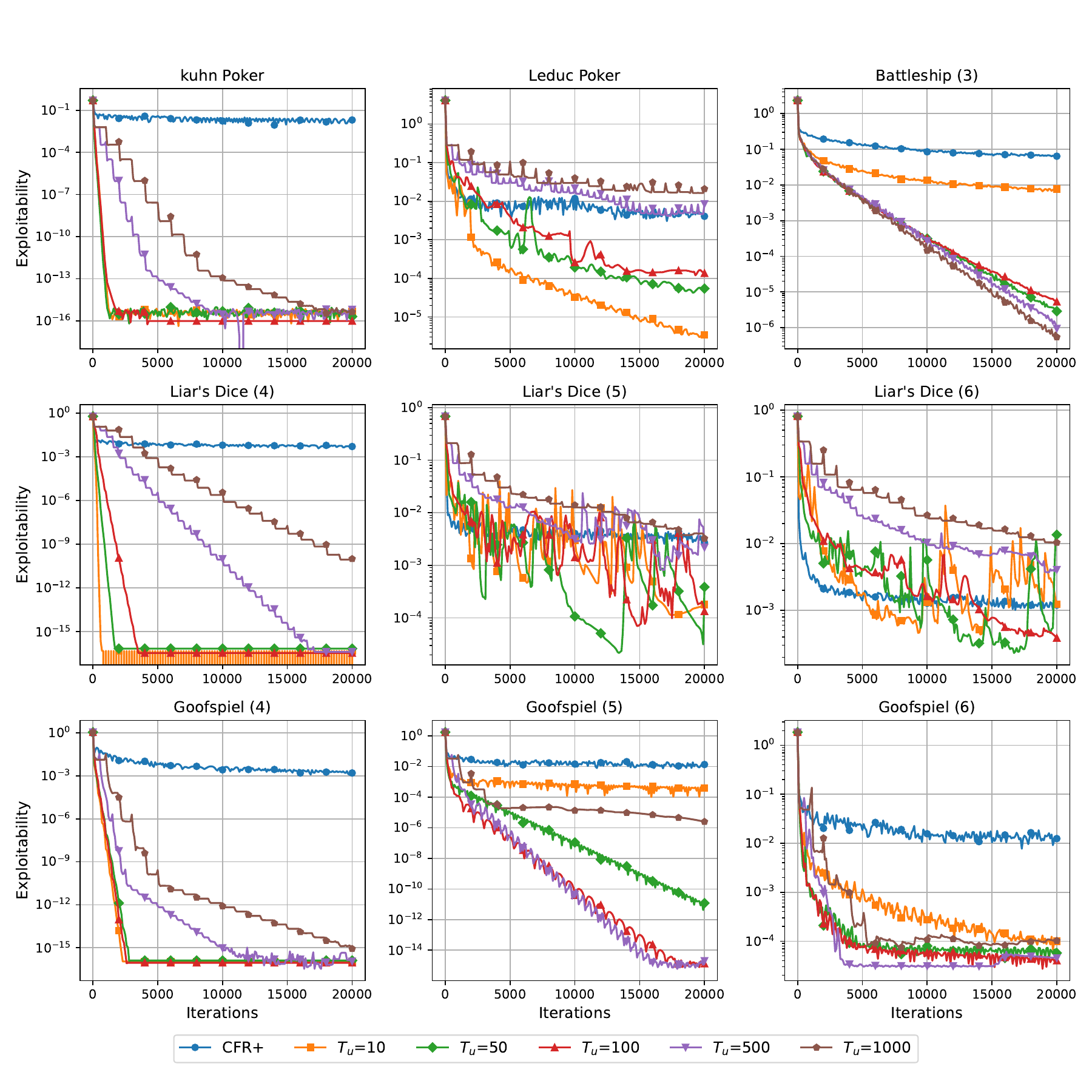}
    }\vspace{-15pt} 
    \caption{Last-iterate convergence rates of RTCFR$^+$ with $\mu = 0.01$. 
}
\label{fig:RTCFR+-mu=0.01}
\vspace{-0.25cm}
\end{figure*}

\begin{figure*}[h]
    \centering %\quad \quad
    \subfigure{
    \includegraphics[width=0.8\linewidth]{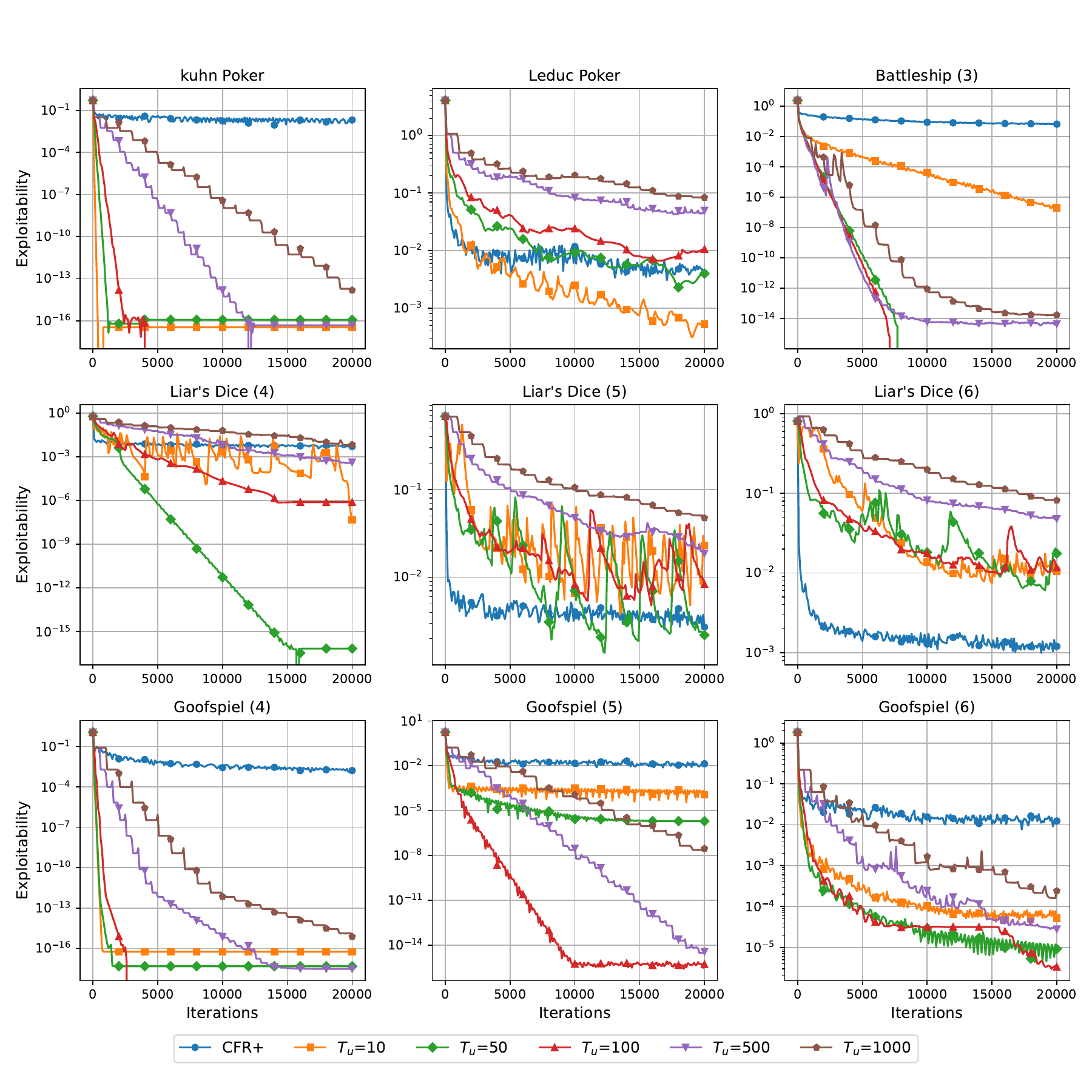}
    }\vspace{-15pt} 
    \caption{Last-iterate convergence rates of RTCFR$^+$ with $\mu = 0.05$. 
}
\label{fig:RTCFR+-mu=0.05}
\vspace{-0.25cm}
\end{figure*}

\begin{figure*}[t!]
    \centering %\quad \quad
    \subfigure{
    \includegraphics[width=0.8\linewidth]{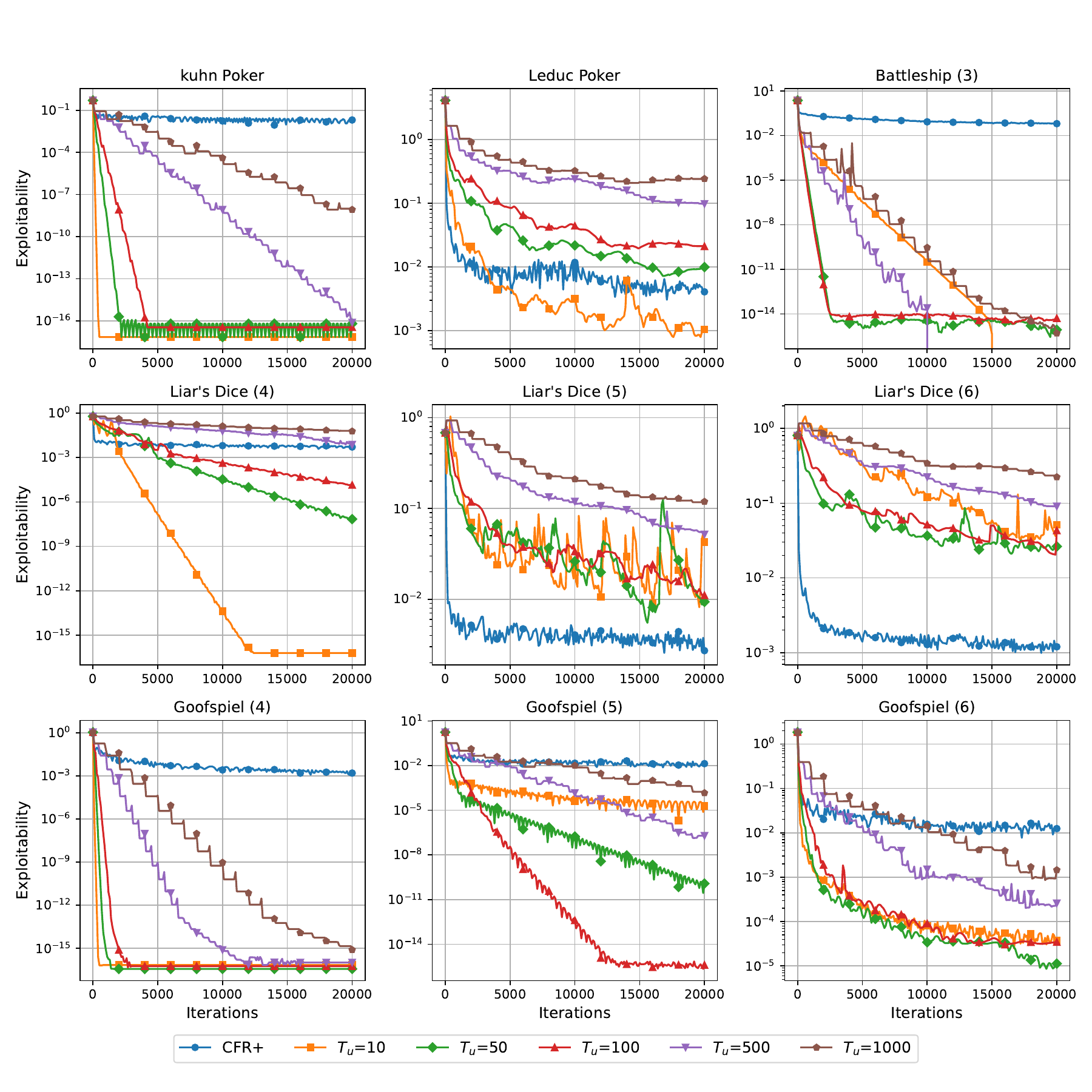}
    }\vspace{-15pt} 
    \caption{Last-iterate convergence rates of RTCFR$^+$ with $\mu = 0.1$. 
}
\label{fig:RTCFR+-mu=0.1}
\vspace{-0.25cm}
\end{figure*}

\begin{figure*}[t!]
    \centering %\quad \quad
    \subfigure{
    \includegraphics[width=0.8\linewidth]{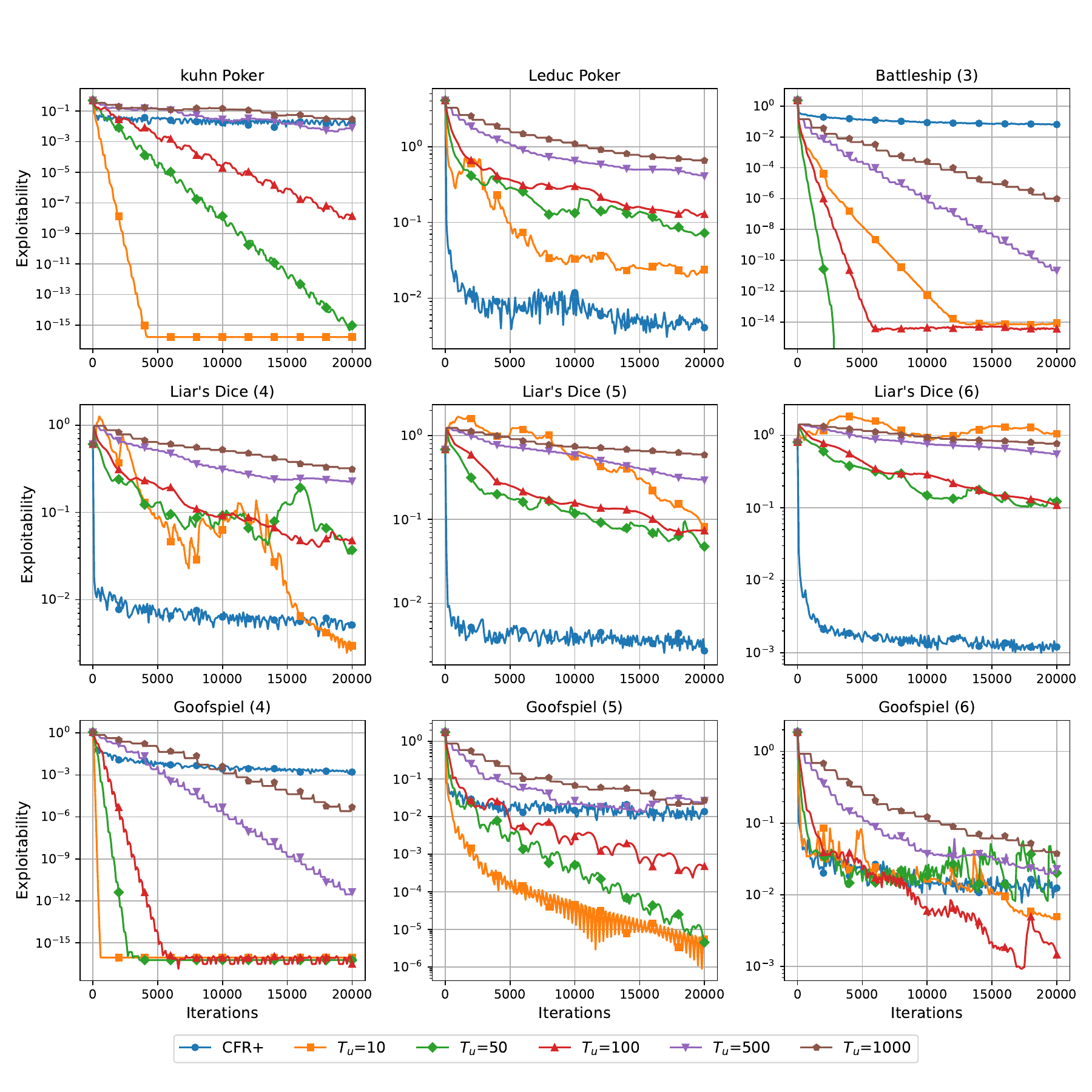}
    }\vspace{-15pt} 
    \caption{Last-iterate convergence rates of RTCFR$^+$ with $\mu = 0.5$. 
}
\label{fig:RTCFR+-mu=0.5}
\vspace{-0.25cm}
\end{figure*}

\textbf{Convergence rates of RTCFR$^+$ under different hyperparameters.} We investigate the convergence rates of RTCFR$^+$ under various hyperparameter settings. Specifically, we focus on the impact of $\mu$ and $T_u$ on the convergence rates, as we observe that $\gamma$ only needs to be set to a sufficiently small value. In fact, \citet{liu2022power} even set $\gamma$ to $0$. The tested ranges for $\mu$ and $T_u$ are $[1\mathrm{e}-4,\ 5\mathrm{e}-4,\ 1\mathrm{e}-3,\ 5\mathrm{e}-3,\ 1\mathrm{e}-2,\ 5\mathrm{e}-2,\ 1\mathrm{e}-1,\ 1\mathrm{e}-5]$ and $[10,\ 50,\ 100,\ 500,\ 1000]$, respectively. Experimental results are shown in Figures \ref{fig:RTCFR+-mu=0.0001}, \ref{fig:RTCFR+-mu=0.0005}, \ref{fig:RTCFR+-mu=0.001}, \ref{fig:RTCFR+-mu=0.005}, \ref{fig:RTCFR+-mu=0.01}, \ref{fig:RTCFR+-mu=0.05}, \ref{fig:RTCFR+-mu=0.1}, and \ref{fig:RTCFR+-mu=0.5}. Results reveal that the performance of RTCFR$^+$ is primarily contingent upon the value of $\mu$. To elucidate this dependency, we discuss the performance implications of varying $\mu$ values. Specifically, for small $\mu$ values, CFR$^+$ encounters difficulties in accurately learning an NE of perturbed regularized EFGs. Consequently, this challenge persists irrespective of the value of $T_u$, enabling that learning an NE of perturbed regularized EFGs becomes impossible. As a result, attaining an NE of the original game becomes impracticable for any $T_u$ value, which is also consistent with the experimental results. Conversely, when $\mu$ is optimal, neither too small nor too large, this condition enables CFR$^+$ to learn sufficiently accurate approximate an NE of perturbed regularized EFGs. These allow RTCFR$^+$ to achieve commendable performance. However, for large $\mu$ values, although CFR$^+$ are capable of learning the exact NE of perturbed regularized EFGs, the requisite number of reference strategy updates becomes excessively large. Hence, we observe that with large $\mu$ values, a smaller $T_u$ yields better performance. Based on these analyses, we advocate for the prioritization of determining $\mu$'s value, followed by the value of $T_u$, when practically applying our algorithm.

\clearpage
\newpage
\section{Implementation of RTCFR$^+$}\label{sec:Implementation of MPCFR+}

In this section, we present a detailed description of the implementation of RTCFR$^+$, which is adapted from the open-source implementation of CFR$^+$ by LiteEFG~\citep{liu2024liteefgefficientpythonlibrary}.

\begin{lstlisting}
import LiteEFG
class RTCFRPlusGraph(LiteEFG.Graph):
    def __init__(self, gamma=1e-10, mu=1e-3, shrink_iter=100): #default parameters
        super().__init__()
        self.timestep = 0
        self.shrink_iter = shrink_iter # shrink_iter is T_u

        # Initialization of RTCFR+
        with LiteEFG.backward(is_static=True):
            ev = 1.0 * LiteEFG.const(1, 0.0)
            # unperturbed_strategy is \sigma
            self.unperturbed_strategy = LiteEFG.const(self.action_set_size, 1.0 / self.action_set_size)
            # perturbed_strategy is \hat{\sigma}
            self.strategy = LiteEFG.const(self.action_set_size, 1.0 / self.action_set_size)
            # regret_buffer is \bm{\theta}
            self.regret_buffer = LiteEFG.const(self.action_set_size, 0.0)

            # ref_strategy is \bm{r}
            self.ref_strategy = LiteEFG.const(self.action_set_size, 1.0 / self.action_set_size)
            # the following three variables are used to compute \nabla \psi(\bm{r}), note that self.ref_reach_prob(I) = \nabla \psi(\bm{r})(I)
            self.ref_reach_prob = LiteEFG.const(self.action_set_size, 1.0)
            self.parent_reach_prob = LiteEFG.const(self.action_set_size, 1.0)
            self.parent_to_child_prob = LiteEFG.const(self.action_set_size, 1.0)

            self.iteration = LiteEFG.const(1, 0)
            self.mu = LiteEFG.const(1, mu)
            self.gamma = LiteEFG.const(1, gamma)
            self.alpha_I = self.gamma*self.action_set_size

        with LiteEFG.backward(color=0):
            self.iteration.inplace(self.iteration+1)
            # to compute the \hat{\bm{v}}_i^t(I) defined in (4)
            gradient = LiteEFG.aggregate(ev, aggregator="sum") + self.utility - self.mu*(self.reach_prob*self.strategy - self.ref_reach_prob*self.ref_strategy)
            # to compute the \langle \hat{\bm{v}}_i^t(I), \sigma^t_i(I) \rangle defined in (4)
            ev.inplace(LiteEFG.dot(gradient, self.unperturbed_strategy))
            # gradient - ev is the instantaneous counterfactual regret \hat{\bm{m}}_i^t(I ) defined in (4)
            self.regret_buffer.inplace(LiteEFG.maximum(self.regret_buffer + gradient - ev, 0.0))
            
            # to get \sigma^{t+1}_i(I)
            self.unperturbed_strategy.inplace(LiteEFG.normalize(self.regret_buffer, p_norm=1.0, ignore_negative=True))
            # to employ PCFR+ to solve the perturbed regularized EFGs, please use the following line
            # self.unperturbed_strategy.inplace(LiteEFG.normalize(self.regret_buffer + gradient - ev, p_norm=1.0, ignore_negative=True))
            # to get \hat{\sigma}^{t+1}_i(I)
            self.strategy.inplace(LiteEFG.normalize((1 - self.alpha_I)*self.unperturbed_strategy + self.gamma, p_norm=1.0, ignore_negative=True))

        # update gamma and the reference strategy profile
        with LiteEFG.backward(color=1):
            self.gamma.inplace(self.gamma * 0.5)
            self.ref_strategy.inplace(self.strategy * 1.0)
        
        with LiteEFG.forward(color=2):
            # to compute \nabla \psi(\bm{r}) after updating the reference strategy profile
            self.parent_reach_prob.inplace(LiteEFG.aggregate(self.ref_reach_prob, "sum", object="parent", player="self", padding=1))
            self.parent_to_child_prob.inplace(LiteEFG.aggregate(self.ref_strategy, "sum", object="parent", player="self", padding=1))
            self.ref_reach_prob.inplace(self.parent_reach_prob*self.parent_to_child_prob)

        
        print("===============Graph is ready for RTCFR+===============")

    def update_graph(self, env : LiteEFG.Environment) -> None:
        self.timestep += 1
        if self.timestep==1:
            env.update(self.strategy, upd_color=[2])
        if self.timestep % self.shrink_iter == 0:
            env.update(self.strategy, upd_color=[1])
            env.update(self.strategy, upd_color=[2])
            env.update(self.strategy, upd_color=[0], upd_player=1)
            env.update(self.strategy, upd_color=[0], upd_player=2)
        else:
            env.update(self.strategy, upd_color=[0], upd_player=1)
            env.update(self.strategy, upd_color=[0], upd_player=2)
        
    def current_strategy(self, type_name="last-iterate") -> LiteEFG.GraphNode:
        return self.strategy
\end{lstlisting}

\end{document}